\documentclass[letterpaper,11pt]{article}
\pdfoutput=1
\usepackage{hyperref}
\hypersetup{
    colorlinks=true,       
    linkcolor=blue,          
    citecolor=blue,        
    filecolor=blue,      
    urlcolor=blue           
}
\usepackage{float}                 
\usepackage{braket,slashed,bm}
\usepackage{array,multirow}
\usepackage[normalem]{ulem}
\usepackage{xcolor,cancel,youngtab}
 
\usepackage[T1]{fontenc} 
\usepackage{mathrsfs}
\usepackage{booktabs}
\usepackage{adjustbox}
\usepackage{mathtools}
\usepackage{soul}
\definecolor{labelkey}{rgb}{0,0.5,0.0}
\usepackage{xspace}
\usepackage{slashed}   
\usepackage{array,multirow}
\usepackage{graphicx}
\usepackage{graphics}
\usepackage{epsfig}
\usepackage{amsfonts}
\usepackage{amsmath}
\usepackage{amssymb}
\usepackage{dsfont}
\usepackage{color}
\usepackage{xcolor}
\usepackage{cite}
\usepackage{pdflscape}
\usepackage{pifont}
\usepackage{pgfplots}
\usepackage{tikz} 
\usepackage{pgfplotstable}
\usepackage{bbold}
\usepackage{subcaption}

\newcommand{\hc}{\mathrm{H.c.}}

\newcommand{\al}{\alpha}
\newcommand{\bt}{\beta}
\newcommand{\g}{\gamma}

\newcommand{\dt}{\delta}

\newcommand{\la}{\lambda}
\newcommand{\simu}{\sigma^{\mu\nu}}

\newcommand{\vL}{\ensuremath{\mathcal{L}}}    
    \newcommand{\Dt}{\Delta}
\newcolumntype{P}[1]{>{\centering\arraybackslash}p{#1}}

\newcommand{\eV}{{\rm ~eV}}

\newcommand{\ga}{\gamma}

\newcommand{\Tr}[1]{{{\rm Tr}\left( #1\right)}} 

\usepackage{bm}  %

\newcommand{\beq}{\begin{equation}}
\newcommand{\eeq}{\end{equation}}
\newcommand{\be}{\begin{equation}}
\newcommand{\ee}{\end{equation}}
\newcommand{\bea}{\begin{eqnarray}}
\newcommand{\eea}{\end{eqnarray}}
\newcommand{\ben}{\begin{eqnarray*}}
\newcommand{\een}{\end{eqnarray*}}

\newcommand{\boldsigma}{\mbox{\boldmath $\sigma$}}

\renewcommand{\vec}[1]{{\mathbf #1}} 

\newcommand{\bma}{\begin{pmatrix}}
\newcommand{\ema}{\end{pmatrix}}

\renewcommand{\O}{\mathcal{O}}
\newcommand{\op}[3]{\O^{#2,#3}_{#1}}

\def\lixo#1{}

\def\slashchar#1{\setbox0=\hbox{$#1$}           
  \dimen0=\wd0                                    
  \setbox1=\hbox{/} \dimen1=\wd1                  
  \ifdim\dimen0>\dimen1                           
    \rlap{\hbox to \dimen0{\hfil/\hfil}}            
    #1                                             
  \else                                          
    \rlap{\hbox to \dimen1{\hfil$#1$\hfil}}        
    /                                           
 \fi}                                           %

\newcommand{\Or}{\mathcal O}

\newcommand{\vp}{\varphi}
\newcommand{\sq}{^{2}}

\newcommand{\dslash}[1]{#1 \llap{/\kern-0.5pt}}
\newcommand{\Dslash}[1]{#1 \llap{/\kern+1.5pt}}
\newcommand{\DDslash}[1]{#1 \llap{/\kern+2.3pt}}
\newcommand{\dslashh}[1]{#1 \llap{/\kern+1pt}}
\newcommand{\abs}[1]{|#1|}

\newcommand{\Ex}[1]{\cdot 10^{#1}}

\newcommand{\blue}[1]{{\color{blue}#1}}

\newcommand{\textoverline}[1]{$\overline{\mbox{#1}}$}

\definecolor{cadmiumgreen}{rgb}{0.0, 0.42, 0.24}
\definecolor{darkpastelgreen}{rgb}{0.01, 0.75, 0.24}
\definecolor{darkspringgreen}{rgb}{0.09, 0.45, 0.27}
\definecolor{forestgreen(web)}{rgb}{0.13, 0.55, 0.13}
\definecolor{forestgreen(traditional)}{rgb}{0.0, 0.27, 0.13}
\definecolor{cobalt}{rgb}{0.0, 0.28, 0.67}
\definecolor{darkblue}{rgb}{0.0, 0.0, 0.75}
\definecolor{darkred}{rgb}{0.55, 0.0, 0.0}
\definecolor{palatinatepurple}{rgb}{0.41, 0.16, 0.38}
\definecolor{burntorange}{rgb}{0.8, 0.33, 0.0}

\newcommand{\nn}{\nonumber}

\def\12{\frac{1}{2}}
\def\0nbb{$0\nu\beta\beta$}

\newcommand{\eot}{E\"{o}tv\"{o}s parameter }
\newcommand{\ten}[1]{\cdot 10^{#1}}

\begin{document}
	
\begin{titlepage}

\begin{flushright}
INT-PUB-22-011
\end{flushright}

\vspace{2.0cm}

\begin{center}
	{\LARGE  \bf
CP-violating axion interactions in effective field theory
	}
	\vspace{1cm}
	
	{\large \bf  Wouter Dekens$^{a}$, Jordy de Vries$^{b,c}$, Sachin Shain$^{b,c}$}
	\vspace{0.5cm}
	
	\vspace{0.25cm}
	
	\vspace{0.25cm}

{\large 
$^a$ 
{\it Institute for Nuclear Theory, University of Washington, \\Seattle WA 91195-1550, USA}}
	
{\large 
$^b$ 
{\it Institute for Theoretical Physics Amsterdam and Delta Institute for Theoretical Physics, University of Amsterdam, Science Park 904,\\ 1098 XH Amsterdam, The Netherlands}}

{\large 
$^c$ 
{\it Nikhef, Theory Group, Science Park 105, 1098 XG, Amsterdam, The Netherlands}}
\end{center}

\vspace{0.2cm}
\begin{abstract}
\vspace{0.1cm}
Axions are introduced to explain the observed smallness of the $\bar \theta $ term of QCD. Standard Model extensions typically contain new sources of CP violation, for instance to account for the baryon asymmetry of the universe. In the presence of additional CP-violating sources a Peccei-Quinn mechanism does not remove all CP violation, leading to CP-odd interactions among axions and Standard Model fields. In this work, we use effective field theory to parametrize generic sources of beyond-the-Standard-Model CP violation. We systematically compute the resulting CP-odd couplings of axions to leptons and hadrons by using chiral perturbation theory. We discuss in detail the phenomenology of the CP-odd axion couplings and compare limits from axion searches, such as fifth force and monopole-dipole searches and astrophysics, to direct limits on the CP-violating operators from electric dipole moment experiments. While limits from electric dipole moment searches are tight, the proposed ARIADNE experiment can potentially improve the existing constraints in a window of axion masses. 
\vspace{2cm}
\looseness-1
\end{abstract}
\vfill
\end{titlepage}
\tableofcontents

\section{Introduction}

The main motivation for a Peccei-Quinn (PQ) mechanism and the associated axion has been the potential resolution of the strong CP problem. In the Standard Model (SM) it is not clear why the strong interactions seem to conserve CP symmetry to very high accuracy, while there has been no experimental sign of a possible CP-odd interaction $\vL_{\bar\theta}\sim \bar \theta G\tilde G$. The current limits on the electric dipole moment (EDM) of the neutron and the ${}^{199}$Hg atom constrain the CP-violating vacuum angle $\bar \theta \lesssim 10^{-10}$ \cite{Graner:2016ses,Dragos:2019oxn, Abel:2020pzs}. The QCD axion \cite{Peccei:1977ur,Peccei:1977hh,Wilczek:1977pj,Weinberg:1977ma} $a$ realizes a $U(1)_{PQ}$ symmetry and modifies the CP-violating gluonic interaction $\mathcal L_{\bar\theta} \rightarrow (\bar \theta + a/f_a)G\tilde G$, where $f_a$ is the axion decay constant. The $U(1)_{PQ}$ symmetry is broken by non-perturbative QCD effects inducing an effective axion potential. Minimizing this potential leads to a vacuum expectation value (vev) for the axion field that sets the effective CP-violating phase $\bar \theta + \langle a \rangle/f_a =0$, resolving the strong CP problem. Even more compelling is that the axion could be the dark matter in our universe in certain scenarios \cite{Abbott:1982af,Dine:1982ah,Marsh:2015xka}. For these reasons the search for the axion has grown into a giant endeavor on both the experimental and theoretical front, see Refs.~\cite{Sikivie:2020zpn,DiLuzio:2020wdo} for recent reviews. So far, roughly 45 years after the initial proposal, without success.

If the only source of CP violation is the QCD $\bar \theta$ term, then the above mechanism removes all CP violation from the theory after the axion field takes its vev. In the presence of additional CP-odd interactions the minimum of the axion potential is shifted, which leaves a remnant of CP violation behind in the form of an induced $\bar \theta$ term and higher-dimensional operators. The CKM phase is one such source of CP violation, but due to its flavor properties it leads to a highly suppressed value of the induced $\bar \theta$ term that cannot be detected with present or expected experiments. However, it is very possible that additional sources of CP violation arise from beyond-the-SM (BSM) physics at a scale $\Lambda$ well above the electroweak scale. In fact, generic extensions of the SM have additional CP phases that cannot be rotated away, something which is reflected by the large number of CP invariants in the SM effective field theory (SM-EFT) \cite{Bonnefoy:2021tbt} In fact, the number of CP invariants is sizable even if one only considers the operators that would result from a minimal seesaw scenario \cite{Yu:2022nxj,Yu:2022ttm}. In addition, the generation of the matter-antimatter asymmetry of the universe requires additional sources of CP violation.

A curious property of the SM is that small values of $\bar \theta$ are technically natural. That is, radiative corrections to $\bar\theta$ start at high loop order~\cite{Ellis:1978hq} and lie well below the experimental limit. Once $\bar \theta$ is chosen small at some scale, it remains small. This is no longer true in generic BSM extensions. For example, in certain supersymmetric scenarios the phases of the soft parameters induce large threshold corrections to $\bar \theta$ \cite{Dine:2015jga}. In left-right symmetric models (LRSMs) \cite{Pati:1974yy, Mohapatra:1974hk, Senjanovic1975,Senjanovic:1978ev,Deshpande:1990ip} parity can be conserved in the UV so that $\bar \theta=0$ by symmetry. However, after electroweak symmetry breaking a new $\bar \theta$ is induced by phases in the scalar sector of the model \cite{Maiezza:2014ala}. Even models that are specifically constructed to solve the strong CP problem in the UV \cite{Babu:1989rb}, have severe trouble keeping $\bar \theta$ small enough after electroweak symmetry breaking \cite{deVries:2021pzl}. It was argued from an EFT viewpoint \cite{deVries:2018mgf} that the presence of higher-dimensional sources of CP violation essentially requires a PQ mechanism as otherwise it is hard to understand why a large correction to $\bar \theta$ is not induced. 

A natural question is then what the presence of additional sources of CP violation implies for the interactions of the axion. The main consequence is that the pseudoscalar axion field will obtain CP-violating scalar couplings to leptons and quarks (and thus nucleons and atoms) as was already proposed in Ref.~\cite{Moody:1984ba}. The scalar axion-fermion couplings lead to an axion-mediated scalar-scalar (monopole-monopole) potential between atoms, while they induce a scalar-pseudoscalar (monopole-dipole) potential when combined with the conventional CP-conserving pseudoscalar axion-fermion interactions. The resulting forces can be looked for in dedicated experiments, see e.g.\ Refs.\ \cite{OHare:2020wah,Irastorza:2018dyq} for an overview. 

As the PQ mechanism acts in the infrared, BSM sources of CP violation can be parametrized in terms of effective higher-dimensional operators. Various studies have computed the scalar axion-nucleon interactions for specific CP-odd dimension-six operators such as the quark electric and chromo-electric dipole moments \cite{Pospelov:1997uv, Okawa:2021fto} and, more recently, certain four-quark operators \cite{Bigazzi:2019hav,Bertolini:2020hjc,DiLuzio:2021jfy}. The main goal of this work is to generalize, extend, and systemize these results. Our starting point is the general set of CP-violating effective operators among light SM fields (light quarks, electrons and muons, and photons and gluons). We then compute the axion vev by minimizing the axion potential, align the vacuum to eliminate mesonic tadpoles, and use chiral perturbation theory ($\chi$PT) to compute the resulting CP-odd axion couplings to mesons, baryons, and leptons. 

The second goal of this paper is to determine the prospects of measuring the resulting CP-odd axion interactions. We therefore compare the experimental limits on the original CP-odd EFT interactions (mainly coming from EDM experiments) to limits on CP-odd axion interactions. The latter arise for instance from fifth-force searches, violations of the weak equivalence principle, monopole-dipole searches, rare decays, and various astrophysical processes. We find that EDM experiments set very stringent constraints and the prospects of detecting CP-violating axion interactions are slim, especially when the CP violation is sourced by a single SM-EFT operator. A nonzero signal of CP-odd axion couplings would then imply significant cancellations between the contributions to EDMs of multiple operators, or a scenario not captured by the EFT involving new light degrees of freedom. We also consider several projected experiments and show that the proposed ARIADNE experiment could detect signs of CP-odd axion couplings in parts of parameter space without coming into conflict with current EDM limits.

Here we do not assume that axions make up dark matter, but note that this assumption would lead to additional interesting signatures including oscillating EDMs \cite{Graham:2011qk,Graham:2013gfa, Budker:2013hfa,Stadnik:2013raa,Flambaum:2019acd,Janish:2020knz}, which can be searched for in a wide range of experiments \cite{Semertzidis:2021rxs,JacksonKimball:2017elr,Mitridate:2020kly,Graham:2020kai}.

This paper is organized as follows. We start by introducing the relevant axion interactions and higher-dimensional operators, derive the chiral rotations that are needed to align the vacuum, and minimize the axion potential in Section \ref{sec:EFT}. The resulting Lagrangian is subsequently matched onto chiral perturbation theory in Section \ref{sec:ChiLag}, where the induced CP-odd lepton-nucleon and pion-nucleon interactions as well as the  axion-nucleon and axion-lepton couplings are derived. The contributions of the former to EDM experiments are discussed in Section \ref{sec:EDMs}, while Sections \ref{sec:5thforce} and \ref{sec:ariadne} are dedicated to the effects of the latter in fifth-force, monopole-dipole, and astrophysical searches. We subsequently apply the derived framework to several specific BSM scenarios involving CP-violating interactions and a PQ mechanism in Section \ref{sec:pheno}.  We conclude in Section \ref{sec:conclusions}, while several technical details are relegated to several appendices.

\section{The effective Lagrangian}\label{sec:EFT}

In this section we introduce the interactions that can arise from BSM scenarios in which additional sources of CP violation originate at a scale $\Lambda$, while a PQ mechanism is active at the same time. Assuming that the BSM scale lies well above the electroweak scale, $\Lambda\gg v\simeq 246$ GeV, any new heavy fields can be integrated out, leading to higher-dimensional operators made up of SM fields and the axion. Just below the scale $\Lambda$, the resulting interactions between SM fields can be described by the SM-EFT  \cite{Buchmuller:1985jz,Grzadkowski:2010es}, while the possible axion interactions are given by its coupling to the SM fermions and to the $SU(3)_c$,  $SU(2)_L$, and  $U(1)_Y$ theta terms. 

To describe their effects on EDM experiments and searches for axion-mediated forces, these interactions need to be evolved to the QCD scale, $\mu\simeq 2$ GeV, after which they can be matched to chiral perturbation theory ($\chi$PT) in terms of leptons, nucleons, and pions instead of leptons, quarks, and gluons. This would require the evolution of the $SU(3)_c\times SU(2)_L\times U(1)_Y$-invariant SM-EFT to the electroweak scale and its subsequent matching onto an $SU(3)_c\times U(1)_{\rm em}$-invariant EFT, sometimes called LEFT \cite{Jenkins:2017jig}.  The resulting LEFT interactions could then be evolved to the QCD scale, where they can finally be matched onto $\chi$PT. Many of the ingredients needed to perform the steps above the QCD scale are in principle available at the one-loop level. For example, the running and matching of the SM-EFT and LEFT operators was computed in \cite{Jenkins:2013zja,Jenkins:2013wua,Alonso:2013hga,Dekens:2019ept,Jenkins:2017dyc}, while the renormalization of the axion couplings  \cite{Bauer:2020jbp,Chala:2020wvs} and the running due to axion loops \cite{Galda:2021hbr} were discussed more recently. However, as we will mainly be concerned with low-energy measurements, in this section we start directly with the $SU(3)_c\times U(1)_{\rm em}$-invariant effective theory involving three flavors of quarks ($u$, $d$, and $s$), at a scale of $\mu\sim2$ GeV, and  only briefly comment on the connection to $SU(3)_c\times SU(2)_L\times U(1)_Y$-invariant operators. Nevertheless, the assumption that the EFT below the electroweak scale originates from an $SU(2)_L$-invariant theory will prove useful as it provides additional information about the scaling of certain operators with respect to $\Lambda$, as we  discuss below. 
\subsection{The interactions}

The quark-level interactions of the axion-like particle (ALP), $a$, and the higher-dimensional CP-odd sources that we consider can be split into three parts consisting of the SM, the ALP, and the EFT operators
\bea\label{eq:lag0}
\vL = \vL_{\rm SM} + \vL_{\rm axion} + \vL_{\rm LEFT}\,.
\eea
\subsubsection*{The Standard Model terms}
The SM terms that will be relevant for our discussion are
\bea
\vL_{\rm SM} = \bar q i \slashed{D} q-\bar q_L M_0q_R-\bar q_R M_0^\dagger q_L-\bar\theta\frac{\al_s}{8\pi}\tilde G^A_{\mu\nu} G^{A\,\mu\nu}+\dots\,,
\eea
where the dots stand for the lepton sector and the kinetic terms of the gauge fields, while $q=(u,d,s)^T$, $D_\mu = \partial_\mu +ig_s T^A G^A_\mu +i e Q  A_\mu$, and we work in the quark mass basis $M_0 = {\rm diag}(m_u,m_d,m_s) $. Furthermore, $Q={\rm diag}(2/3,-1/3,-1/3)$ is the matrix of electromagnetic quark charges, $e=|e|$ is the charge of the proton, and $\tilde G^A_{\mu\nu} = \frac{1}{2} \varepsilon_{\mu\nu\al\bt}G^{A\,\al\bt}$ with $\varepsilon^{0123}=+1$, the Gell-Mann matrices $T^A$, and color index $A$.

\subsubsection*{The axion Lagrangian}

The second term in Eq.\ \eqref{eq:lag0} consists of all possible ALP interactions up to dimension five that are invariant under a shift symmetry, $a\to a+c$, up to total derivatives. These can be written as
\bea\label{eq:Laxion}
\vL_{\rm axion} &=&  \frac{1}{2}\partial_\mu a\partial^\mu a-\frac{\al_s}{8\pi}\frac{a}{f_a} \tilde G^A_{\mu\nu} G^{A\,\mu\nu}-\frac{1}{4} g_{a\gamma}^{(0)} \frac{a}{f_a}\tilde F_{\mu\nu}F^{\mu\nu}\nn\\
&&+\sum_{f = \nu,e,q} \frac{\partial_\mu a}{2f_a}\left[\bar f_L c^{f}_{L}\gamma^\mu f_L+\bar f_R c^{f}_{R}\gamma^\mu f_R\right]\,,
\eea
where $c^{f}_{L,R}$ are hermitian matrices in flavor space and $f_a$ is the ALP decay constant, indicating the scale related to the PQ mechanism, $\Lambda_a\sim 4\pi f_a$, which we will assume to be above the scale at which the EFT operators are generated, $\Lambda_a\gg\Lambda$. We do not consider the possibility of light right-handed neutrinos~\footnote{The effective theory that systematically includes light right-handed neutrinos is called the $\nu$SM-EFT \cite{delAguila:2008ir,Liao:2016qyd}.}, so that the term $\sim c^{\nu}_R$ vanishes.

The interactions in Eq.\ \eqref{eq:Laxion} respect a PQ symmetry, $a\to a+c$, at the classical level. The terms involving derivatives are manifestly invariant, while the shifts of the $a\tilde FF$ and $a\tilde GG$ terms lead to total derivatives which, in the case of the $\tilde GG$ coupling, gives rise to non-perturbative effects that break the PQ symmetry at the quantum level. The Lagrangian of Eq.\ \eqref{eq:Laxion} describes all the interactions that are generally induced when the axion arises as the phase of a complex scalar field. Note, however, that the form of the axion-fermion interactions is not unique and one can trade the $c_{L,R}^{f} $ couplings for non-derivative interactions of the form $a \bar f_L f_R$ along with a shift of the $a\tilde FF$ and/or $a\tilde GG$ terms, through a redefinition of the fermion fields~\footnote{Such terms do not lead to additional independent operators as long as we assume (classical) invariance under the PQ shift symmetry.}.

Finally, as alluded to above, these interactions in principle arise from an $SU(2)_L$-invariant Lagrangian, such as the one discussed in  \cite{Bauer:2020jbp}. At tree level, the matching of the above axion couplings to $F\tilde F$ and $G\tilde G$ and the fermions is given by 
\bea
\frac{1}{f_a}&=&-2\frac{c_{GG}}{f}\,,\qquad \frac{g_{a\g}^{(0)}}{f_a}= -\frac{\al}{\pi f}\left(c_{WW}+c_{BB}\right)\,, \nn\\
\frac{c_L^{(e)}}{f_a} &=& 2\frac{\boldsymbol{c}_L}{f}\,, \qquad \frac{c_R^{(e)}}{f_a} = 2\frac{\boldsymbol{c}_e}{f}\,,\qquad c_L^{(\nu)}= U_{\rm PMNS}^\dagger c_L^{(e)}U_{\rm PMNS}\,,\nn\\
\frac{c_L^{(q)}}{f_a} &=& \frac{2}{f}
\bma \left[\boldsymbol{c}_Q\right]_{1\times1} &\boldsymbol{0}_{1\times 2}\\
\boldsymbol{0}_{2\times 1} & \left[V_{\rm CKM}^\dagger \boldsymbol{c}_{Q}V_{\rm CKM}\right]_{2\times 2}
\ema
\,, \qquad \frac{c_R^{(q)}}{f_a} = \frac{2}{f}
\bma \left[\boldsymbol{c}_u\right]_{1\times1} &\boldsymbol{0}_{1\times 2}\\
\boldsymbol{0}_{2\times 1} & \left[ \boldsymbol{c}_{d}\right]_{2\times 2}
\ema\,,
\eea
where the right-hand sides correspond to the $SU(2)_L$-invariant couplings in the notation of \cite{Bauer:2020jbp}, while $U_{\rm PMNS}$ and $V_{\rm CKM}$ are the PMNS and CKM matrices.

\subsubsection*{The higher-dimensional operators}

Finally, the third term in Eq.\ \eqref{eq:lag0} involves operators of up to dimension-six, that consist of SM fields. At energies above the electroweak scale such operators are described by the SM-EFT \cite{Buchmuller:1985jz,Grzadkowski:2010es}, while for processes at energies below $\mu\sim m_W$, where the $SU(2)_L$ gauge group of the SM has been broken, these interactions make up the so-called LEFT. 
The operators in this EFT are invariant under $SU(3)_c\times U(1)_{\rm em}$ and a complete basis up to dimension-six has been derived in Ref.\ \cite{Jenkins:2017jig}, its Lagrangian can be written as
\bea\label{eq:lagLEFT}
\vL_{\rm LEFT} = \sum_i L_i \Or_i\,,
\eea
where the sum extends over all operators in Table \ref{tab:oplist1}, their flavor indices, as well as their hermitian conjugates, when applicable. 

Here we do not consider the complete set of operators derived in Ref.\ \cite{Jenkins:2017jig}, as we are interested in the CP-violating ones only. We focus on purely hadronic operators that give unsuppressed contributions to the chiral Lagrangian, i.e.\  their chiral representations come without derivatives. We also take into account LEFT operators that couple the photon or lepton fields to quark currents, which can straightforwardly be included as source terms in the chiral Lagrangian. The hadronic operators give rise to CP-odd interactions between nucleons and pions, while the semi-leptonic operators induce couplings of nucleons to leptons. Both types of interactions are probed by EDM measurements. As we will see in the upcoming sections, in the presence of a PQ mechanism, the same operators also induce CP-odd couplings of the axion to hadrons and leptons, which can be constrained by searches for axion-mediated forces.  The operators that satisfy the above conditions are collected in Table \ref{tab:oplist1}, while the derivation of this list is discussed in more detail in App.\ \ref{app:operators}.

One possible complication arises due to the fact that Table \ref{tab:oplist1} involves both dimension-five and -six operators. For our purposes the relevant dimension-five operators are the dipole interactions in the $\boldsymbol{ (\bar LR)X}$ class, which, within the LEFT, scale as $\Lambda^{-1}$. As we include operators up to dimension-six, scaling as $\Lambda^{-2}$,  terms such as $ L_{q\g}^2$, $L_{qG}^2$, or $L_{q\g}L_{qG}$ would need to be considered as well, since they enter at the same order. However, whenever the LEFT operators originate from an $SU(2)_L$-invariant EFT, the dipole operators are generated by dimension-six operators and scale as $L_{q\g,qG}\sim \frac{m_q}{\Lambda^2}$. This is what we will be assuming in what follows, such that all the operators in Table \ref{tab:oplist1} scale as $\Lambda^{-2}$. The complete tree-level matching of the LEFT interactions to the SM-EFT is given in \cite{Jenkins:2017dyc}.

\begin{table}[t!]
\vspace{-0.75cm}
\begin{adjustbox}{width=0.5\textwidth,center}
\begin{minipage}[t]{3cm}
\renewcommand{\arraystretch}{1.51}
\small
\begin{align*}
\begin{array}[t]{c|c}
\multicolumn{2}{c}{\boldsymbol{(\overline L R ) X+\hc}} \\
\hline
\blue{\O_{u \gamma} }& \bar u_{Lp}   \sigma^{\mu \nu}  u_{Rr}\, F_{\mu \nu}   \\
\blue{\O_{d \gamma} }& \bar d_{Lp}  \sigma^{\mu \nu} d_{Rr}\, F_{\mu \nu}  \\
\O_{u G} & \bar u_{Lp}   \sigma^{\mu \nu}  T^A u_{Rr}\,  G_{\mu \nu}^A  \\
\O_{d G} & \bar d_{Lp}   \sigma^{\mu \nu} T^A d_{Rr}\,  G_{\mu \nu}^A \\
\end{array}
\end{align*}
\end{minipage}
\begin{minipage}[t]{3cm}
\renewcommand{\arraystretch}{1.51}
\small
\begin{align*}
\begin{array}[t]{c|c}
\multicolumn{2}{c}{\boldsymbol{X^3}} \\
\hline
\O_{\widetilde G} & f^{ABC} \widetilde G_\mu^{A\nu} G_\nu^{B\rho} G_\rho^{C\mu}   \\
\end{array}
\end{align*}
\end{minipage}
\end{adjustbox}
\mbox{}\\[-0.75cm]
\begin{adjustbox}{width=1.05\textwidth,center}
\begin{minipage}[t]{3cm}
\renewcommand{\arraystretch}{1.51}
\small
\begin{align*}
\begin{array}[t]{c|c}
\multicolumn{2}{c}{\boldsymbol{(\overline L L)(\overline L  L)}} \\
\hline
\blue{\op{\nu u}{V}{LL}  }     & (\bar \nu_{Lp} \gamma^\mu \nu_{Lr}) (\bar u_{Ls}  \gamma_\mu u_{Lt})  \\
\blue{\op{\nu d}{V}{LL}   }    & (\bar \nu_{Lp} \gamma^\mu \nu_{Lr})(\bar d_{Ls} \gamma_\mu d_{Lt})     \\
\blue{\op{eu}{V}{LL}  }    & (\bar e_{Lp}  \gamma^\mu e_{Lr})(\bar u_{Ls} \gamma_\mu u_{Lt})   \\
\blue{\op{ed}{V}{LL}   }    & (\bar e_{Lp}  \gamma^\mu e_{Lr})(\bar d_{Ls} \gamma_\mu d_{Lt})  \\
\blue{\op{\nu edu}{V}{LL} }     & (\bar \nu_{Lp} \gamma^\mu e_{Lr}) (\bar d_{Ls} \gamma_\mu u_{Lt})  + \hc   \\
[-0.5cm]
\end{array}
\end{align*}
\renewcommand{\arraystretch}{1.51}
\small
\begin{align*}
\begin{array}[t]{c|c}
\multicolumn{2}{c}{\boldsymbol{(\overline R  R)(\overline R R)}} \\
\hline
\blue{\op{eu}{V}{RR}    }   & (\bar e_{Rp}  \gamma^\mu e_{Rr})(\bar u_{Rs} \gamma_\mu u_{Rt})   \\
\blue{\op{ed}{V}{RR}   }  & (\bar e_{Rp} \gamma^\mu e_{Rr})  (\bar d_{Rs} \gamma_\mu d_{Rt})   \\
\end{array}
\end{align*}
\end{minipage}

\begin{minipage}[t]{3cm}
\renewcommand{\arraystretch}{1.51}
\small
\begin{align*}
\begin{array}[t]{c|c}
\multicolumn{2}{c}{\boldsymbol{(\overline L  L)(\overline R  R)}} \\
\hline
\blue{\op{\nu u}{V}{LR}   }      & (\bar \nu_{Lp} \gamma^\mu \nu_{Lr})(\bar u_{Rs}  \gamma_\mu u_{Rt})    \\
\blue{\op{\nu d}{V}{LR}    }     & (\bar \nu_{Lp} \gamma^\mu \nu_{Lr})(\bar d_{Rs} \gamma_\mu d_{Rt})   \\
\blue{\op{eu}{V}{LR}   }     & (\bar e_{Lp}  \gamma^\mu e_{Lr})(\bar u_{Rs} \gamma_\mu u_{Rt})   \\
\blue{\op{ed}{V}{LR}     }   & (\bar e_{Lp}  \gamma^\mu e_{Lr})(\bar d_{Rs} \gamma_\mu d_{Rt})   \\
\blue{\op{ue}{V}{LR}  }      & (\bar u_{Lp} \gamma^\mu u_{Lr})(\bar e_{Rs}  \gamma_\mu e_{Rt})   \\
\blue{\op{de}{V}{LR}     }    & (\bar d_{Lp} \gamma^\mu d_{Lr}) (\bar e_{Rs} \gamma_\mu e_{Rt})   \\
\blue{\op{\nu edu}{V}{LR}  }      & (\bar \nu_{Lp} \gamma^\mu e_{Lr})(\bar d_{Rs} \gamma_\mu u_{Rt})  +\hc \\
\op{uu}{V1}{LR}        & (\bar u_{Lp} \gamma^\mu u_{Lr})(\bar u_{Rs} \gamma_\mu u_{Rt})   \\
\op{uu}{V8}{LR}       & (\bar u_{Lp} \gamma^\mu T^A u_{Lr})(\bar u_{Rs} \gamma_\mu T^A u_{Rt})    \\ 
\op{ud}{V1}{LR}       & (\bar u_{Lp} \gamma^\mu u_{Lr}) (\bar d_{Rs} \gamma_\mu d_{Rt})  \\
\op{ud}{V8}{LR}       & (\bar u_{Lp} \gamma^\mu T^A u_{Lr})  (\bar d_{Rs} \gamma_\mu T^A d_{Rt})  \\
\op{du}{V1}{LR}       & (\bar d_{Lp} \gamma^\mu d_{Lr})(\bar u_{Rs} \gamma_\mu u_{Rt})   \\
\op{du}{V8}{LR}       & (\bar d_{Lp} \gamma^\mu T^A d_{Lr})(\bar u_{Rs} \gamma_\mu T^A u_{Rt}) \\
\op{dd}{V1}{LR}      & (\bar d_{Lp} \gamma^\mu d_{Lr})(\bar d_{Rs} \gamma_\mu d_{Rt})  \\
\op{dd}{V8}{LR}   & (\bar d_{Lp} \gamma^\mu T^A d_{Lr})(\bar d_{Rs} \gamma_\mu T^A d_{Rt}) \\
\op{uddu}{V1}{LR}   & (\bar u_{Lp} \gamma^\mu d_{Lr})(\bar d_{Rs} \gamma_\mu u_{Rt})  + \hc  \\
\op{uddu}{V8}{LR}      & (\bar u_{Lp} \gamma^\mu T^A d_{Lr})(\bar d_{Rs} \gamma_\mu T^A  u_{Rt})  + \hc \\
\end{array}
\end{align*}
\end{minipage}

\begin{minipage}[t]{3cm}
\renewcommand{\arraystretch}{1.51}
\small
\begin{align*}
\begin{array}[t]{c|c}
\multicolumn{2}{c}{\boldsymbol{(\overline L R)(\overline L R)+\hc}} \\
\hline
\blue{\op{eu}{S}{RR} } & (\bar e_{Lp}   e_{Rr}) (\bar u_{Ls} u_{Rt})   \\
\blue{\op{eu}{T}{RR} }& (\bar e_{Lp}   \sigma^{\mu \nu}   e_{Rr}) (\bar u_{Ls}  \sigma_{\mu \nu}  u_{Rt})  \\
\blue{\op{ed}{S}{RR} } & (\bar e_{Lp} e_{Rr})(\bar d_{Ls} d_{Rt})  \\
\blue{\op{ed}{T}{RR} }& (\bar e_{Lp} \sigma^{\mu \nu} e_{Rr}) (\bar d_{Ls} \sigma_{\mu \nu} d_{Rt})   \\
\blue{\op{\nu edu}{S}{RR}} & (\bar   \nu_{Lp} e_{Rr})  (\bar d_{Ls} u_{Rt} ) \\
\blue{\op{\nu edu}{T}{RR} }&  (\bar  \nu_{Lp}  \sigma^{\mu \nu} e_{Rr} )  (\bar  d_{Ls}  \sigma_{\mu \nu} u_{Rt} )   \\
\op{uu}{S1}{RR}  & (\bar u_{Lp}   u_{Rr}) (\bar u_{Ls} u_{Rt})  \\
\op{uu}{S8}{RR}   & (\bar u_{Lp}   T^A u_{Rr}) (\bar u_{Ls} T^A u_{Rt})  \\
\op{ud}{S1}{RR}   & (\bar u_{Lp} u_{Rr})  (\bar d_{Ls} d_{Rt})   \\
\op{ud}{S8}{RR}  & (\bar u_{Lp} T^A u_{Rr})  (\bar d_{Ls} T^A d_{Rt})  \\
\op{dd}{S1}{RR}   & (\bar d_{Lp} d_{Rr}) (\bar d_{Ls} d_{Rt}) \\
\op{dd}{S8}{RR}  & (\bar d_{Lp} T^A d_{Rr}) (\bar d_{Ls} T^A d_{Rt})  \\
\op{uddu}{S1}{RR} &  (\bar u_{Lp} d_{Rr}) (\bar d_{Ls}  u_{Rt})   \\
\op{uddu}{S8}{RR}  &  (\bar u_{Lp} T^A d_{Rr}) (\bar d_{Ls}  T^A u_{Rt})  \\[-0.5cm]
\end{array}
\end{align*}
\renewcommand{\arraystretch}{1.51}
\small
\begin{align*}
\begin{array}[t]{c|c}
\multicolumn{2}{c}{\boldsymbol{(\overline L R)(\overline R L) +\hc}} \\
\hline
\blue{\op{eu}{S}{RL} } & (\bar e_{Lp} e_{Rr}) (\bar u_{Rs}  u_{Lt})  \\
\blue{\op{ed}{S}{RL}} & (\bar e_{Lp} e_{Rr}) (\bar d_{Rs} d_{Lt}) \\
\blue{\op{\nu edu}{S}{RL} } & (\bar \nu_{Lp} e_{Rr}) (\bar d_{Rs}  u_{Lt})  \\
\end{array}
\end{align*}
\end{minipage}
\end{adjustbox}
\setlength{\abovecaptionskip}{0.15cm}
\caption{The $B$- and $L$-conserving operators of the LEFT of dimension five and six that contribute to CP-violating effects in the meson sector at leading order. Only the hadronic operators that contribute to the non-derivative meson interactions and the semi-leptonic operators that can be written as external sources (shown in blue) are listed.}
\label{tab:oplist1}
\end{table}

\subsection*{Chiral representations}

In order to build the chiral Lagrangian in the upcoming section, it is convenient to group the above described interactions by their transformation properties under the chiral symmetry group $SU(3)_L\times SU(3)_R$. 
The kinetic terms for the quarks and the axion remain unchanged,
\bea\label{eq:LagIrreps}
\vL &=&\bar q (i\slashed \partial -g_s\ga^\mu G^A_\mu T^A)q-\bar \theta\frac{\al_s}{8\pi}\tilde G^A_{\mu\nu} G^{A\,\mu\nu}\nn\\
&&+\frac{1}{2}\partial_\mu a\partial^\mu a-\frac{\al_s}{8\pi}\frac{a}{f_a} \tilde G^A_{\mu\nu} G^{A\,\mu\nu}-\frac{1}{4} g_{a\gamma}^{(0)} \frac{a}{f_a}\tilde F_{\mu\nu}F^{\mu\nu}\nn\\
&&+\vL_{\rm sources}+\vL_{6}+\dots\,,
\eea
while we collect the couplings of quark bilinears to leptons, axions, or photons to quark in $\vL_{\rm sources}$,
\bea\label{eq:sourcesLag}
\vL_{\rm sources}&=&\bar q\Bigg[\ga^\mu l_\mu P_L+\ga^\mu r_\mu P_R-MP_R-M^\dagger P_L+t_R^{\mu\nu}\sigma_{\mu\nu}P_R+t_L^{\mu\nu}\sigma_{\mu\nu}P_L\Bigg]q\,.
\eea
The electromagnetic gauge couplings, the $c_{L,R}^{q}$ derivative axion couplings, and semileptonic vector operators  are now contained in the $l_\mu$ and $r_\mu$ currents. The quark masses as well as the scalar and pseudoscalar sources $s$ and $p$, which contain semileptonic scalar interactions, are collected in $M = M_0+s-ip$, while the tensor sources are denoted by $t_{L,R}^{\mu\nu}$ and capture semileptonic tensor interactions as well as the quark EDMs. All of these sources form $3\times 3$ matrices in flavor space and depend on the axion, photon, and lepton fields. Their explicit expressions are given in App.\ \ref{app:sources}.

Finally, the remaining LEFT operators, those that cannot be written as the couplings of quark bilinears, transform under $SU(3)_L\times SU(3)_R$ in several different ways. In particular, the quark color-EDM operators transform as $\bf \bar 3_L\times  3_R$, while the four-quark operators transform as the irreps $\bf{8_{L}}\times 8_{R}$, $\bf3_L\times {\bar 3}_R$, and $\bf \bar6 _L\times { 6}_R$. This allows us to write
\bea\label{eq:LagLEFT}
\vL_{6}&=&\Bigg[\bar q_L L_5 T^AG^A_{\mu\nu}\sigma^{\mu\nu}q_R+{\rm h.c.}\Bigg]\nn\\
&&+ L_{\bf 8\times 8}^{ijkl}(\bar q_L^i\ga^\mu q_L^j)\, ( \bar q_R^k\ga_\mu q_R^l)+\bar L_{\bf 8\times 8}^{ijkl}(\bar q_L^i\ga^\mu T^A q_L^j)\, ( \bar q_R^k\ga_\mu T^A q_R^l)\nn\\
&&+\Bigg[L_{\bf 3\times 3}^{ijkl} (\bar q_L^iq_R^j)\,(  \bar q_L^k q_R^l)+ L_{\bf 6\times 6}^{ijkl}(\bar q_L^i q_R^j)\,  (\bar q_L^k q_R^l)\nn\\
&&+\bar L_{\bf 3\times 3}^{ijkl} (\bar q_L^iT^A q_R^j)\,(  \bar q_L^k T^A q_R^l)+\bar L_{\bf 6\times 6}^{ijkl}(\bar q_L^i T^A q_R^j)\,  (\bar q_L^k T^A q_R^l)+{\rm h.c.}\Bigg]\,,
\eea
where $i,j,k,l\in \{1,2,3\}$ are flavor indices which are summed over when repeated. Here $L_{\bf 6\times 6}^{ijkl}$ ($L_{\bf 3\times 3}^{ijkl}$) are (a)symmetric in $i\leftrightarrow k$ and $j\leftrightarrow l$, while the $L_{\bf 8\times 8}^{ijkl}$ couplings are traceless in $(i,j)$ and $(k,l)$ and satisfy $L_{\bf 8\times 8}^{ijkl}=L_{\bf 8\times 8}^{jilk\,*}$, so that they project out their particular representations.
The explicit expressions for the $L_\al$ couplings in terms of the Wilson coefficients of the LEFT are given in App.\ \ref{app:dim56}. After collecting the operators either in the external sources, or by their chiral representations, we can use Eq.\ \eqref{eq:LagIrreps} as a starting point to derive the chiral Lagrangian. 

\subsection{Vacuum alignment}\label{sec:vacAlign}

As the higher-dimensional operators violate CP and chiral symmetry they generally lead to a misalignment of the vacuum. In this case the $SU(3)$ subgroup of $SU(3)_L\times SU(3)_R$ that is left unbroken by chiral symmetry breaking does not necessarily correspond to the diagonal subgroup, $SU(3)_V$, under which $q_{L,R}$ transform as $q_{L,R}\to U q_{L,R}$. Diagrammatically this corresponds to the appearance of so-called tadpole vertices that allow for meson-vacuum transitions. It is convenient to remove such tadpole diagrams by a non-anomalous chiral rotation, which, at the same time, aligns the unbroken subgroup with $SU(3)_V$. In addition, we will find it useful to perform an anomalous $U(1)$ chiral rotation that removes the $\tilde GG$ and $a\tilde GG$ couplings from the quark-level Lagrangian and trade them for $\bar qq$ and $a\bar qq$ terms. Here we briefly describe these field redefinitions, as well as how the needed angles of rotation can be determined, before constructing the chiral Lagrangian in the next section.

\subsubsection*{Chiral rotation}

All in all we perform the following unitary basis transformation 
\bea\label{eq:transf}
q_L = A^\dagger q_L'\,,\qquad  q_R = A q_R'\,, \qquad A = {\rm exp}(i\left[\al_0/3 +\al\cdot t\right])\,,
\eea
where $t$ are the Gell-Mann matrices in flavor space and we will allow the $\al_i$ to depend on the axion field $a(x)$. Here $\al_0$ is the anomalous chiral rotation that removes the $\tilde GG$ and $a\tilde GG$ terms, while the $\al_i$ are chosen to eliminate the tadpoles.
This rotation leads to a transformed Lagrangian
\bea\label{eq:LagTransf}
\vL' &=&\bar q (i\slashed \partial -g_s\ga^\mu G^A_\mu T^A)q+\frac{1}{2}\partial_\mu a\partial^\mu a-\frac{1}{4} g_{a\gamma} \frac{a}{f_a}\tilde F_{\mu\nu}F^{\mu\nu}+\vL_{\rm sources}'+\vL_{6}'+\dots\,,
\eea
where the $(a)\tilde GG$ terms have been removed by choosing $2\al_0=\theta_a\equiv  \bar \theta + \frac{a}{f_a}$
while the 
 $a\tilde FF$ coupling is given by
\bea\label{eq:theta_a}
g_{a\ga} = g_{a\ga}^{(0)} -2N_c\frac{\al}{\pi}{\rm Tr}\left[H_a Q^2\right]\,,
\eea
and $H_a$ is determined by the axion-dependent part of $A={\rm exp}(iH)$, with $H = H_0 + \frac{a}{f_a} H_a$ where $H_{0,a}$ are $a$-independent matrices in flavor space.
  $\vL_{\rm sources}'$ and $\vL_{6}'$ can be obtained from the original Lagrangians by the following replacements
\bea\label{eq:rotatedCouplings}
&l_\mu\to l_\mu' = A\left[l_\mu +i\partial_\mu \right]A^{-1}\,,\qquad &r_\mu\to r_\mu' = A^{-1}\left[r_\mu +i\partial_\mu\right] A\,,\nn\\
&S\to S' = ASA\,,\hspace{2.6cm}&S\in \{M\,, t_R^{\mu\nu}\,,L_5\}\,,\nn\\
&L_{ijkl}\to L_{ijkl} '= A_{ia}A_{jb}^*L_{abcd} A_{ck}^*A_{dl}\,,\qquad &L\in\{L_{\bf 8\times 8}\,,\bar L_{\bf 8\times 8}\}\,,\nn\\
&L_{ijkl}\to L_{ijkl} '= A_{ia}A_{bj}L_{abcd} A_{kc}A_{dl}\,,\qquad &L\in\{L_{\bf 3\times 3}\,,\bar L_{\bf 3\times 3}\,,L_{\bf 6\times 6}\,,\bar L_{\bf 6\times 6}\}\,.
\eea
This rotation leads to an effective quark mass term that now depends on the CP-violating Wilson coefficients, while the higher-dimensional operators generally obtain a dependence on $\theta_a$.

The angles of rotation in $A$ can be determined by requiring that the unbroken subgroup corresponds to $SU(3)_V$.  This can be achieved by demanding that the potential is in a minimum 
\bea\label{eq:Vmin}
\frac{\partial V}{\partial \al_i }=0\,, \qquad V = -\langle 0 |\vL'_{int} |0\rangle\,,
\eea
and solving for the $\al_i$. 
One can show that the above condition on the $\al_i$ also ensures that the chiral Lagrangian will not induce any tadpole terms. In our particular case we have $\al_{1,2}=\al_{4,5}=0$, as these angles would allow one to remove tadpole terms for the charged mesons, $\pi^\pm$ and $K^\pm$, which are never induced thanks to $U(1)_{\rm em}$ invariance. The remaining $\al_i$ are generally nonzero and become functions of the Wilson coefficients, $L_\al$ and the matrix elements of the higher-dimensional operators, while $\langle \theta_a \rangle = \langle a\rangle/f_a$ enters through the axion dependence of the Lagrangian. The hadronic matrix elements are related to the low-energy constants (LECs) that will be introduced in the next section, we give the explicit relations in App.\ \ref{app:LECs}. The vev of the axion field, $\langle a\rangle$, is similarly obtained through minimization of the axion potential.
Explicit expressions for the $\al_i$ and $\langle a\rangle$ are discussed in App.~\ref{app:rotations}.

Although the solutions obtained from Eq.\ \eqref{eq:Vmin} lead to a chiral Lagrangian without tadpole interactions, it will generally mix the pion and axion fields. One can show that such mass-mixing terms can be removed by allowing the $\al_i$ to depend on the axion field. The needed modification of these angles can then be obtained by including the physical axion field, $a_{\rm ph}(x)$, whenever the axion vacuum expectation value (vevs) would otherwise appear in the solutions of Eq.\ \eqref{eq:Vmin}. I.e.\ we replace $\langle a\rangle\to a\equiv \langle a\rangle+a_{\rm ph}(x)$. Finally, the chiral Lagrangian in principle allows the kinetic terms of the axions and pions fields to become mixed. Such mixings can be removed by a redefinition of the axion and pion fields \cite{DiLuzio:2020wdo}, which only modifies the axion-pion interactions beyond the precision we work at.

\section{Chiral Lagrangian}\label{sec:ChiLag}

\subsection{Mesonic Lagrangian}

After performing the basis transformations discussed in the previous section, the mesonic part of the chiral Lagrangian becomes
\bea\label{eq:mesonLag}
\vL_\pi &=& \frac{F_0^2}{4}\Tr{D_\mu U D^\mu U^\dagger} + \frac{F_0^2}{4}\Tr{\chi^\dagger U+\chi U^\dagger}- F_0^2 \bar B\Tr{L_5^{\prime\,\dagger} U+L_5' U^\dagger}\nn\\
&&-\frac{F_0^4}{4}\left[\mathcal A_{\bf 8\times 8}L_{\bf 8\times 8}^{\prime\, ijkl}+\mathcal {\bar A}_{\bf 8\times 8}\bar L_{\bf 8\times 8}^{\prime\, ijkl}\right]U_{jk}U^*_{il}\nn\\
&&-\frac{F_0^4}{8}\Bigg\{\left[\mathcal A_{\bf 3\times 3}L_{\bf 3\times 3}^{\prime\, ijkl}+\mathcal {\bar A}_{\bf 3\times 3}\bar L_{\bf 3\times 3}^{\prime\, ijkl}\right]\left(U_{ij}^* U^*_{kl}-U^*_{il}U^*_{kj}\right)\nn\\
&&\hspace{.9cm} +\left[\mathcal A_{\bf 6\times 6}L_{\bf 6\times 6}^{\prime\, ijkl}+\mathcal {\bar A}_{\bf 6\times 6}\bar L_{\bf 6\times 6}^{\prime\, ijkl}\right]\left(U^*_{ij}U^*_{kl}+U^*_{il}U^*_{kj}\right)+{\rm h.c.}\Bigg\}\nn\\
&&-\frac{1}{4} g_{a\gamma} \frac{a}{f_a}\tilde F_{\mu\nu}F^{\mu\nu}\,,
\eea
where $F_0$ is the pion decay constant in the chiral limit,  $\chi = 2B M'$, the covariant derivative is given by $D_\mu U = \partial_\mu U-il_\mu' U+iUr_\mu'$, and the matrix of pseudo-Goldstone fields is 
\bea
U =u^2= {\rm exp}\left(\frac{2i\pi\cdot t}{F_0}\right)\,,\qquad \pi\cdot t = \frac{1}{\sqrt{2}}
\bma
\frac{\pi_3}{\sqrt{2}}+\frac{\pi_8}{\sqrt{6}} & \pi^+ & K^+ \\
\pi^- & -\frac{\pi_3}{\sqrt{2}}+\frac{\pi_8}{\sqrt{6}} &  K^0 \\
K^-&\overline{K}^0 & -2\frac{\pi_8}{\sqrt{6}} 
\ema
\,.
\eea
 Furthermore, $B$, $\bar B$, $\mathcal A_i$, and  $\bar{\mathcal A}_i$ are LECs, which are defined in terms of matrix elements of the corresponding operators in App.\ \ref{app:LECs}.

The axion field and $\bar \theta$ enter through the couplings $M'$ and $L_\al'$ and their dependence on the rotation angles, $\al_i$, which are obtained by solving Eq.\ \eqref{eq:Vmin}. These solutions ensure that the chiral Lagrangian is free of tadpole terms and the mass matrices do not mix the pion and axion fields. We checked explicitly that the above Lagrangian, together with our solutions of the $\al_i$, satisfies these conditions. By expanding the first two terms of the above Lagrangian one can obtain the axion and meson interactions induced by dimension-four operators. Apart from the usual $\chi$PT Lagrangian, this gives rise to the  expression for the axion mass
\beq\label{axionmass}
m_a =\frac{\sqrt{m_*B} F_0}{f_a}
=\sqrt{\frac{m_*}{m_u+m_d}}\frac{F_0}{f_a} m_\pi\simeq 5.9\cdot 10^{6}\left(\frac{{\rm eV}}{f_a}\right){\rm GeV}\,,
\eeq
where $m_* = (\frac{1}{m_u}+\frac{1}{m_d}+\frac{1}{m_s})^{-1}$.
By taking into account the solutions for the $\al_i$, as well as the remaining terms in the above Lagrangian, we can determine the non-standard axion-meson interactions induced by higher-dimensional operators. 

\subsection{Nucleon-pion sector}

The $\pi N$ Lagrangian can be built from the baryon fields
\begin{equation}\label{eq:3.0}
B = \left( \begin{array}{c c c}
\frac{1}{\sqrt{2}}\Sigma^0 + \frac{1}{\sqrt{6}}\Lambda & \Sigma^+ 							& p \\
\Sigma^-					       & -\frac{1}{\sqrt{2}}\Sigma^0 + \frac{1}{\sqrt{6}}\Lambda 	& n \\
\Xi^-						       & \Xi^0								& -\frac{2}{\sqrt{6}} \Lambda
\end{array} \right)\, ,
\end{equation}
and several combinations of the Wilson coefficients and meson fields, $\chi_+ = 2B(u^\dagger M' u^\dagger+u M^{\prime \dagger} u)$, $\bar \chi_+ = 2\bar B(u^\dagger L_5'u^\dagger+u L_5^{\prime \dagger} u)$, and
\bea
l^{ijkl}_{\bf 8\times 8} &=& u^*_{ai} u_{bj} L^{\prime abcd}_{\bf 8\times 8}u_{kc}u^*_{ld}\,,\nn\\
l^{ijkl}_{\bf 3\times 3,6\times 6} &=& u^*_{ai} u^*_{jb} L^{\prime abcd}_{\bf 3\times 3,6\times 6}u^*_{ck}u^*_{ld}\,,
\eea
with analogous definitions for the color-octet operators with Wilson coefficients, $\bar L_{\al}$.
The different parts of the Lagrangian then take the form
\bea
\vL^{\pi N} _{M}&=& 
\langle \bar B iv\cdot D B\rangle + F\langle \bar BS_\mu\left[\hat u^\mu,B\right]\rangle+ D\langle \bar BS_\mu\big\{\hat u^\mu,B\big\}\rangle + g_A^0\langle u^\mu\rangle \langle  \bar BS_\mu B\rangle\label{eq:LagpiNindirect}\\
&&+b_0\langle \bar BB\rangle\langle  \chi_+\rangle+b_D\langle \bar B\{B,\,\chi_+\}\rangle+b_F\langle \bar B[\chi_+,\, B]\rangle\,,\nn\\
\vL^{\pi N}_{L_5} &=& \bar b_0\langle \bar BB\rangle\langle  \bar\chi_+\rangle+\bar b_D\langle \bar B\{B,\,\bar\chi_+\}\rangle+\bar b_F\langle \bar B[\bar\chi_+,\, B]\rangle\,,\nn\\
\vL^{\pi N}_{\bf 8\times 8} &=&a_{\bf 8\times 8}^{(1)}\langle \bar BB\rangle l^{ ijji}_{\bf 8\times 8}+\bar B_{ji} B_{lk}b_{\bf 8\times 8}^{(27)}\left[l^{ ijkl}_{\bf 8\times 8} \right]^{\bf 27} \nn\\
&&+a_{\bf 8\times 8}^{(8)}\left[\left(\bar BB\right)_{ij}-\frac{\dt_{ij}}{3}\langle B\bar B\rangle\right]  \left[ l^{ j kk i}_{\bf 8\times 8} + l^{ kijk}_{\bf 8\times 8} \right] +b_{\bf 8\times 8}^{(8)}\left[\left( B \bar B\right)_{ij}-\frac{\dt_{ij}}{3}\langle B\bar B\rangle\right] \left[ l^{ j kk i}_{\bf 8\times 8} +l^{kijk}_{\bf 8\times 8} \right]\,,\nn\\
\vL^{\pi N}_{\bf 3\times 3} &=&b_{\bf 3\times 3}^{(1)}\langle \bar BB\rangle l_{\bf 3 \times3}^{iijj}+a_{\bf 3\times 3}^{(8)}\left[ (\bar BB)_{ij}-\frac{\dt_{ij}}{3}\langle \bar BB\rangle\right] l_{\bf 3 \times3}^{ kk ji}+b_{\bf 3\times 3}^{(8)}\left[ (B\bar B)_{ij}-\frac{\dt_{ij}}{3}\langle \bar BB\rangle\right] l_{\bf 3 \times3}^{ kk ji}+{\rm h.c.}\,,\nn\\
\vL^{\pi N}_{\bf 6\times 6} &=&b_{\bf 6\times 6}^{(1)}\langle \bar BB\rangle l_{\bf 6 \times6}^{ iijj}+a_{\bf 6\times 6}^{(8)}\left[ (\bar BB)_{ij}-\frac{\dt_{ij}}{3}\langle \bar BB\rangle\right]l_{\bf 6 \times6}^{ kk ji}+b_{\bf 6\times 6}^{(8)}\left[ (B\bar B)_{ij}-\frac{\dt_{ij}}{3}\langle \bar BB\rangle\right]l_{\bf 6 \times6}^{ kk ji}\nn\\
&&+b_{\bf 6\times 6}^{(27)} 
\bar B_{ji} B_{lk}\left[l^{ ijkl}_{\bf 6\times 6}\right]^{\bf 27}+{\rm h.c.}\,,
\label{eq:LagpiNdirect}
\eea
where $v^\mu$ is the baryon velocity and $S^\mu$ is its spin ($v^\mu =(1,\boldsymbol 0)$ and $S^\mu = (0,\, \boldsymbol
\sigma/2 )$ in the nucleon rest frame), while $\hat u_\mu = u_\mu-\langle u_\mu\rangle/3$ with $u_\mu =-i(u^\dagger \partial_\mu u-u \partial_\mu u^\dagger-i u^\dagger l'_\mu u+i ur'_\mu u^\dagger)$ and $D_\mu B= \partial_\mu B+ \left[V_\mu, B\right]$, with $V_\mu = (u^\dagger \partial_\mu u+u \partial_\mu u^\dagger-i u^\dagger l'_\mu u-i ur'_\mu u^\dagger)/2$. 
Furthermore,
$\langle ..\rangle$ denotes a trace in flavor space and $\left[l^{ ijkl}_{\bf r}\right]^{\bf 27}$ stands for the combination of couplings that is symmetric in both $(i \leftrightarrow k)$ and $(j \leftrightarrow l)$ as well as traceless in $(i,j)$, $(k,l)$. 
The LECs $D$, $F$, $g_A^{0}$, and $b_{0,D,F}$ determine the axial and mass terms of the baryons, while the terms generated by the quark color-EDM terms are proportional to $\bar b_{0,D,F}$. 
Finally, $a_{\bf r}^{(r')},b_{\bf r}^{(r')}$ are LECs whose subscripts denote the chiral representation of the corresponding quark-level operator, while the superscripts indicate the representation the baryon fields appear in. For example, $b_{\bf 8\times 8}^{(1)}$ indicates a singlet and therefore appears with the trace of baryon fields, while $b_{\bf 8\times 8,6\times 6}^{(27)}$ indicates the (symmetric) $\bf 27$ representation and thus appears with a symmetric combination of baryon fields.

The total $\pi N$ Lagrangian is then given by
\bea
\vL^{\pi N} = \vL^{\pi N} _{M}+\vL^{\pi N}_{L_5}+\left[\vL^{\pi N}_{\bf 8\times 8} +\vL^{\pi N}_{\bf 3\times 3} +\vL^{\pi N}_{\bf 6\times 6} 
+\bma l_{\bf r}&& \bar l_{\bf r}\\ a_{\bf r}&\to&\bar a_{\bf r}\\b_{\bf r}& &\bar b_{\bf r}\ema\right]\,,
\eea
with $\bf r = \{\bf8\times 8,3\times 3,6\times 6\}$.

\subsection{Interactions from semi-leptonic operators}

The semi-leptonic operators of Table \ref{tab:oplist1} enter the meson and nucleon Lagrangians through the source terms. The most important CP-violating axion couplings arise from the scalar operators, which contribute to the sources $s$ and $p$ and enter the chiral Lagrangian through $\chi$. In the nucleon sector, these terms induce  $\bar NN \bar ei\ga_5e$ interactions, while $a\bar ee$ and $\pi \bar ee$ couplings appear in the mesonic Lagrangian. At low energies, the latter of these give rise to CP-odd spin-dependent nucleon-lepton couplings, through pion exchange. The axion-lepton couplings can be probed by searches for axion-mediated forces, while EDMs are sensitive to the CP-odd hadron-lepton interactions.
\subsubsection*{Axion-lepton couplings}
The relevant interactions are
\bea
\vL_{a\bar ll} &=& g_S^{(e)} a\bar e e\,,\nn\\
g_S^{(e)} &= &\frac{m_*BF_0^2 }{2f_a}{\rm Im}\left[\frac{1}{m_u}L_{\substack{e u\\eeuu}}^{\rm S,RR}+\frac{1}{m_d}L_{\substack{e d\\eedd}}^{\rm S,RR}+\frac{1}{m_s}L_{\substack{e d\\eess}}^{\rm S, RR}-\left(RR\to RL\right)\right]\,.
\label{eq:lepAxion}\eea
Here we have focused on couplings to electrons but by replacing $e\rightarrow \{\mu,\,\tau\}$ in the above expressions we also obtain scalar couplings to muons and taus. While the experimental limits on these couplings are less stringent than for electron-axion couplings, it should be kept in mind that the indirect limits, arising from the $\mu$ and $\tau$ EDMs \cite{Muong-2:2008ebm,Belle:2002nla}, are also weaker. By allowing for lepton flavor-violating dimension-six couplings, e.g. $L_{\substack{e u\\e\mu uu}}^{\rm S,RR}$, we also induce couplings of the form $a\,\bar e \mu$ that can be probed in $\mu \rightarrow e +a $ searches.  However, the $L_{\substack{e u\\e\mu uu}}^{\rm S,RR}$ dimension-six operators are stringently constrained by muon-to-electron conversion ($\mu + N \rightarrow e +N$) experiments \cite{Cirigliano:2009bz,Crivellin:2017rmk}, and we leave a detailed study of these couplings to future work.

In principle the CP-violating phase in the CKM matrix, $\dt_{\rm CKM}$, in the SM can also induce $g_S^{(e)}$ although we are not aware of estimates in the literature (the estimates for the coupling to nucleons is discussed in Sect.~\ref{sec:HadrAxionCouplings}). It is clear these couplings must be proportional to $m_e$ due to the scalar nature of $g_S^{(e)}$. 
Furthermore, at least two insertions of the weak interactions, $\sim G_F^2$, are needed to obtain the CP-violating combination of CKM elements, given by the Jarlskog invariant $J_{\mathrm{CP}} \simeq 3 \cdot 10^{-5}$ \cite{Jarlskog:1985ht}.
While there are many ways to combine these interactions, using naive dimensional analysis \cite{Manohar:1983md,Gavela:2016bzc} to estimate one of the possibilities leads to, 
\bea
g_S^{(e)}(\dt_{\rm CKM})f_a\sim m_e\left(\frac{\al}{4\pi}\right)^2\left(G_FF_0^2\right)^2J_{\mathrm{CP}}\sim 10^{-25}\,{\rm MeV}\,,
\label{eq:gSeSM}\eea
with $J_{\rm CP}$ the Jarlskog invariant $J_{\mathrm{CP}} \simeq 3 \cdot 10^{-5}$ \cite{Jarlskog:1985ht}.
This contributions can be seen to be induced by a $\Delta S=1$ Fermi interaction, which, together with electromagnetism, can induce couplings $\sim K^0 F_{\mu\nu}F^{\mu\nu}$ in the chiral Lagrangian. Such terms can subsequently generate a $\sim m_e K^0 \bar e e$ coupling through a one-loop diagram. Finally, 
an additional insertion of a $\Delta S = 1$ Fermi interaction can generate mixing between the kaon and the axion, thereby giving rise to the axion-electron coupling. While it is certainly possible that other contributions are enhanced with respect to the above estimate, Eq.\ \eqref{eq:gSeSM} should give a rough lower limit on $g_S^{(e)}(\dt_{\rm CKM})$.

\subsubsection*{Semi-leptonic couplings}

The induced nucleon-lepton interactions that contribute to EDMs can be written as,
\bea  \label{CSP}
\vL & =&-\frac{G_F}{\sqrt{2}}\bigg\{\bar e i\g_5 e\, \bar N\left(C_S^{(0)}+\tau_3 C_S^{(1)}\right) N + \bar e e\, \frac{\partial_\mu}{m_N} \left[\bar N\left(C_P^{(0)}+\tau_3 C_P^{(1)}\right)S^\mu N\right]
\bigg\}+\dots \,,
\label{eq:eEDM_interaction}
\eea
where $G_F$ is the Fermi constant and $N
= (p\,\,n)^T$  is the non-relativistic nucleon doublet with mass $m_N$. The matching coefficients are given by
\begin{align}
C_S^{(0)} &= -v_H^2\frac{\sigma_{\pi N}}{m_u+m_d}{\rm Im}\left[L_{\substack{eu\\eeuu}}^{\rm S,RR}+L_{\substack{ed\\eedd}}^{\rm S,RR}
\right]-v^2\frac{\sigma_{s}}{m_s}{\rm Im}\left[L_{\substack{ed\\eess}}^{\rm S,RR}
\right]+(RR\to RL) \,\nn\\
C_S^{(1)} &= -v_H^2\frac{1}{2}\frac{\delta m_N}{m_d-m_u}{\rm Im}\left[L_{\substack{eu\\eeuu}}^{\rm S,RR}-L_{\substack{ed\\eedd}}^{\rm S,RR}+(RR\to RL)
\right]\,,\nn\\
C_P^{(0)} &= -v_H^2\frac{m_N B(D-3F)}{3m_\eta^2}{\rm Im}\left[L_{\substack{eu\\eeuu}}^{\rm S,RL}+L_{\substack{ed\\eedd}}^{\rm S,RL}-2L_{\substack{ed\\eess}}^{\rm S,RL}-(RL\to RR)
\right]\,,\nn\\
C_P^{(1)} &= v_H^2\frac{m_N B g_A}{m_\pi^2}{\rm Im}\left[L_{\substack{eu\\eeuu}}^{\rm S,RL}-L_{\substack{ed\\eedd}}^{\rm S,RL}-(RL\to RR)\right]\,,
\end{align}
where $v_H$ is the vev of the Higgs field, at tree level $v_H^2 =\left(\sqrt{2}G_F\right)^{-1} \simeq (246\, {\rm GeV})^2$, while 
$C_{P}^{(0)}$ and $C_{P}^{(1)}$ arise from the exchange of an $\eta$ and $\pi^0$, respectively.
Here $C_S^{(0,1)}$ induce CP-odd effects in ThO and the mercury EDM, while $C_P^{(0,1)}$ only contribute to the latter. Furthermore, $g_A = D+F$ is the axial charge of the nucleon, $\delta m_N = (m_n-m_p)_{QCD}$ is the strong nucleon mass splitting, while the nucleon sigma terms are given by $\sigma_q = m_q\frac{\partial \Delta m_N}{\partial m_q}$, where $\Dt m_N = \frac{m_n+m_p}{2}$, and $\sigma_{\pi N} = \sigma_u+\sigma_d$. 
The input for these hadronic matrix elements can be summarized as \cite{Airapetian:2006vy,Hoferichter:2015dsa,Abdel-Rehim:2016won,Borsanyi:2014jba,Brantley:2016our} 
\bea\label{sigma}
\sigma_{\pi N} &=& (59.1 \pm 3.5)\,\mathrm{MeV}\ ,\qquad \sigma_s = (41.1_{-10.0}^{+11.3})\,\mathrm{MeV} \, ,\nn\\ 
\delta m_N &=&( 2.32\pm0.17)\, \mathrm{MeV}\, ,\qquad
  g_A = 1.27\pm0.002\,.
\eea

\subsection{Interactions from hadronic operators}\label{sec:HadrAxionCouplings}

\subsubsection*{Axion-meson-meson couplings}

Although we will mainly focus on the couplings of the axion to nucleons in what follows, here we briefly discuss the CP-odd interactions between axions and mesons that arise from Eq.\ \eqref{eq:mesonLag}. These interactions contribute to the axion-mediated potential between nuclei, through the second and third diagram in Fig.\ \ref{fig:feynman_diagrams}, but they are typically subleading with respect to direct axion-nucleon interactions discussed in the next subsection. 

Apart from contributing to axion-mediated forces between nucleons, the meson-axion interactions can also give rise to rare decays of kaons and the $\eta$. In particular, the strangeness-violating operators induce the decay $K\to \pi a$. For example, the strangeness-changing elements of the $L_{\bf 3\times 3}$ Wilson coefficients  induce the vertex
\bea
\vL_{aK\pi} &=& g_{aK\pi} \,a K^+ \pi^-+{\rm h.c.}\nn\\
&=&im_*F_0^2{\cal A_{\bf 3\times 3}}\frac{4m_d m_s(m_d+m_s)+(3m_d+m_s)(m_d+3m_s)m_u}{16 m_d m_u m_s(m_d+m_s)}\left[\left(L_{\bf 3\times 3}^{2311}\right)^*-L_{\bf 3\times 3}^{3211}\right] \nn\\
&&\times \frac{a}{f_a}K^+ \pi^- +{\rm h.c.}\,.
\label{eq:lag_aKpi}
\eea
The general form of the meson-meson-axion interactions is discussed in App.\ \ref{app:pipia}. 

For axion masses $m_a < 2 m_e$, the only available decay channel is $a \rightarrow \gamma \gamma$, which is induced by the model-dependent CP-even $g_{a\gamma}$ coupling of Eq.\ \eqref{eq:LagTransf}~\footnote{The derivative axion couplings to fermions, $c_{R}^{(f)}-c_{L}^{(f)}$, also contribute to $a\to\ga\ga$. However, these contributions can be captured by an effective $g_{a\ga}^{\rm eff}\sim g_{a\ga}+ \frac{\al}{\pi}(c_{R}^{(f)}-c_{L}^{(f)})$, after a chiral rotation.}. In this case the axion lifetime is given by~\cite{DiLuzio:2020wdo} 
\bea 
\tau = \frac{64 \pi}{g_{a\gamma}^2} \frac{f_a^2}{m_a^3}\simeq \frac{5\cdot 10^{18}}{g_{a\gamma}^2}\left(\frac{\rm eV}{m_a}\right)^5\,{\rm s}\,.
\eea
Since we focus on the range $m_a< 2 m_e$ and $g_{a\gamma}=\Or(\al/\pi)$ we can treat the axion as stable in meson decays, implying
$K^+ \rightarrow \pi^+ a$ and $K^+ \rightarrow \pi^+ \nu \bar \nu$ will have the same experimental signature, i.e.\ $K^+\to\pi^+$ with missing energy and momentum. 
We use the upper limit on BR($K^+ \rightarrow \pi^+a$) set by the NA62 experiment~\cite{NA62:2021zjw,NA62:2018ctf,NA62:2020fhy} to constrain $g_{aK\pi}$
\be
{\rm BR}(K^+ \rightarrow \pi^+a) \simeq \frac{ \tau_{K} }{16 \pi m_{K}}\frac{m_K^2-m_\pi^2}{m_K^2}|g_{aK\pi}|^2  < 5 \ten{-11} \, ,
\label{eq:Kpia}
\ee
for $m_a \ll m_\pi$. 
For the couplings of Eq.\ \eqref{eq:lag_aKpi}, this implies
\bea
\left(\frac{m_a}{\rm eV}\right)^2 \bigg| \left(L_{\bf 3\times 3}^{2311}\right)^*-L_{\bf 3\times 3}^{3211}\bigg|^2 {\rm TeV}^4\leq 6.4\ten{5} \,,
\eea
implying TeV-level constraints on $\Lambda$ for axion masses in the keV range. 

At the same time, the $\Delta S=1$ CP-odd operators induce CP violation in kaon decays thereby contributing to $\varepsilon'$. Using the expressions and estimates for the LECs in \cite{Aebischer:2021hws}, we obtain
\bea
\left(\frac{\varepsilon'}{\varepsilon}\right)_{\bf 3\times 3} \simeq 62 \,{\rm TeV}^2 \,{\rm Im}\left(\left( L_{\bf 3\times 3}^{2311}\right)^*-L_{\bf 3\times 3}^{3211}\right)\,,
\eea 
implying that $\varepsilon'$ is sensitive to very high scales.
If we conservatively demand that this contribution is smaller than the measured value~\footnote{See e.g.\  \cite{Aebischer:2020jto,RBC:2020kdj,Cirigliano:2019cpi} for discussions on recent evaluations of the SM contribution based on lattice QCD, $\chi$PT, and dual QCD.}, $\left(\frac{\varepsilon'}{\varepsilon}\right)_{\rm expt.} = (16.6\pm 2.3)\ten{-4}$ \cite{Zyla:2020zbs}, we obtain $\Lambda\gtrsim 193$ TeV significantly more stringent than the $K^+\rightarrow \pi^+ +a$ limits, for $m_a\lesssim{\rm keV}$. 

\subsubsection*{Axion-nucleon couplings}

\renewcommand{\arraystretch}{1.8}
\begin{table}[t!]
\centering
\begin{tabular}{|c |c c|cc|}
\hline
 & $\alpha_{i}^{(0)}$  & $\bt_{i}^{(0)}$  & $\alpha_{i}^{(1)}$  & $\bt_{i}^{(1)}$ \\\hline
$L_{\substack{dG\\ uu}}$ &$\frac{\Delta m_N}{m_u}$ &$2 \frac{\bar B}{B m_u}\frac{\partial \Delta m_N}{\partial m_u}$ & 
$-\frac{\delta m_N}{2m_u}$ & $-\frac{\bar B}{2B m_d}\frac{\partial \delta m_N}{\partial (\bar m\epsilon)}$
\\
$L_{\substack{dG\\ dd}}$ &$\frac{\Delta m_N}{ m_d}$ &$2 \frac{\bar B}{B m_d}\frac{\partial \Delta m_N}{\partial m_d}$
& 
$-\frac{\delta m_N}{2m_d}$ & $\frac{\bar B}{2B m_u}\frac{\partial \delta m_N}{\partial (\bar m\epsilon)}$
\\
$L_{\substack{dG\\ ss\\ {\rm }}}$ &$\frac{\Delta m_N}{m_s}$ &$ 2\frac{\bar B}{B m_s}\frac{\partial \Delta m_N}{\partial  m_s}$
&$0$&$0$\\%
\hline
$L_{\bf 8\times 8}^{2112}$ &$\frac{m_u-m_d}{m_dm_u}\Delta m_N$ &$ \frac{F_0^2\mathcal A_{\bf 8\times 8}}{2B}\frac{m_u-m_d}{m_um_d}\frac{\partial \Delta m_N}{\partial m_u}$
&$0$ &$ -\frac{F_0^2\mathcal A_{\bf 8\times 8}}{8B}\frac{m_u+m_d}{m_um_d}\frac{\partial \delta m_N}{\partial (\bar m\epsilon)}$\\
$L_{\bf 8\times 8}^{3113}$ &$\frac{m_u-m_s}{m_um_s}\Delta m_N$ &$ \frac{F_0^2\mathcal A_{\bf 8\times 8}}{2B}\left[\frac{1}{m_s}\frac{\partial \Delta m_N}{\partial m_s}-\frac{1}{m_u}\frac{\partial \Delta m_N}{\partial m_u}\right]$
&$\frac{m_s-m_u}{2m_um_s}\delta m_N$ &$ -\frac{F_0^2\mathcal A_{\bf 8\times 8}}{8B}\frac{1}{m_u}\frac{\partial \delta m_N}{\partial (\bar m\epsilon)}$\\
$L_{\bf 8\times 8}^{3223}$ &$\frac{m_d-m_s}{m_dm_s}\Delta m_N$ &$ \frac{F_0^2\mathcal A_{\bf 8\times 8}}{2B}\left[\frac{1}{m_s}\frac{\partial \Delta m_N}{\partial m_s}-\frac{1}{m_d}\frac{\partial \Delta m_N}{\partial  m_d}\right]$
&$\frac{m_s-m_d}{2m_dm_s}\delta m_N$ & $ \frac{F_0^2\mathcal A_{\bf 8\times 8}}{8B}\frac{1}{m_d}\frac{\partial \delta m_N}{\partial (\bar m\epsilon)}$
\\
\hline
$L_{\bf 3\times 3}^{2211}$ &$\frac{m_d+m_u}{m_dm_u}\Delta m_N$ &$ \frac{F_0^2\mathcal A_{\bf 3\times 3}}{B}\frac{m_d+m_u}{m_um_d}\frac{\partial \Delta m_N}{m_u}$
&$0$ &$ \frac{F_0^2\mathcal A_{\bf 3\times 3}}{4B}\frac{m_d-m_u}{m_um_d}\frac{\partial \delta m_N}{\partial (\bar m\epsilon)}$\\
$L_{\bf 3\times 3}^{3311}$ &$\frac{m_s+m_u}{m_sm_u}\Delta m_N$ &$ \frac{F_0^2\mathcal A_{\bf 3\times 3}}{B}\left[\frac{1}{m_u}\frac{\partial \Delta m_N}{\partial  m_u}+\frac{1}{m_s}\frac{\partial \Delta m_N}{\partial m_s}\right]$
&$-\frac{m_s+m_u}{2m_sm_u}\delta m_N$ &$ \frac{F_0^2\mathcal A_{\bf 3\times 3}}{4B}\frac{1}{m_u}\frac{\partial \delta m_N}{\partial (\bar m\epsilon)}$\\
$L_{\bf 3\times 3}^{3322}$ &$\frac{m_d+m_s}{m_dm_s}\Delta m_N$ &$ \frac{F_0^2\mathcal A_{\bf 3\times 3}}{B}\left[\frac{1}{m_d}\frac{\partial \Delta m_N}{\partial  m_d}+\frac{1}{m_s}\frac{\partial \Delta m_N}{\partial m_s}\right]$
&$-\frac{m_s+m_d}{2m_sm_d}\delta m_N$ &$ -\frac{F_0^2\mathcal A_{\bf 3\times 3}}{4B}\frac{1}{m_d}\frac{\partial \delta m_N}{\partial (\bar m\epsilon)}$\\
\hline
$L_{\bf 6\times 6}^{2211}$ &$\frac{m_d+m_u}{m_dm_u}\Delta m_N$ &$ \frac{F_0^2\mathcal A_{\bf 6\times 6}}{B}\frac{m_d+m_u}{m_um_d}\frac{\partial \Delta m_N}{\partial m_u}$
 &$0$ &$ \frac{F_0^2\mathcal A_{\bf 6\times 6}}{4B}\frac{m_d-m_u}{m_um_d}\frac{\partial \delta m_N}{\partial (\bar m\epsilon)}$\\
$L_{\bf 6\times 6}^{3311}$ &$\frac{m_u+m_s}{m_um_s}\Delta m_N$ &$ \frac{F_0^2\mathcal A_{\bf 6\times 6}}{B}\left[\frac{1}{m_u}\frac{\partial \Delta m_N}{\partial m_d}+\frac{1}{m_s}\frac{\partial \Delta m_N}{\partial m_s}\right]$
&$-\frac{m_u+m_s}{2m_um_s}\delta m_N$ &$ \frac{F_0^2\mathcal A_{\bf 6\times 6}}{4B}\frac{1}{m_u}\frac{\partial \delta m_N}{\partial (\bar m\epsilon)}$\\
$L_{\bf 6\times 6}^{3322}$ &$\frac{m_d+m_s}{m_dm_s}\Delta m_N$ &$ \frac{F_0^2\mathcal A_{\bf 6\times 6}}{B}\left[\frac{1}{m_d}\frac{\partial \Delta m_N}{\partial  m_d}+\frac{1}{m_s}\frac{\partial \Delta m_N}{\partial m_s}\right]$
&$-\frac{m_d+m_s}{2m_dm_s}\delta m_N$ &$ -\frac{F_0^2\mathcal A_{\bf 6\times 6}}{4B}\frac{1}{m_d}\frac{\partial \delta m_N}{\partial (\bar m\epsilon)}$\\
$L_{\bf 6\times 6}^{1111}$ &$2\frac{\Delta m_N}{m_u}$ &$ \frac{F_0^2\mathcal A_{\bf 6\times 6}}{B}\frac{1}{m_u}\frac{\partial \Delta m_N}{\partial m_u}$
&$-\frac{\delta m_N}{m_u}$ &$ \frac{F_0^2\mathcal A_{\bf 6\times 6}}{4B}\frac{1}{m_u}\frac{\partial \delta m_N}{\partial (\bar m\epsilon)}$\\
$L_{\bf 6\times 6}^{2222}$ &$2\frac{\Delta m_N}{m_d}$ & $\frac{F_0^2\mathcal A_{\bf 6\times 6}}{B}\frac{1}{m_d}\frac{\partial \Delta m_N}{\partial m_d}$
&$-\frac{\delta m_N}{m_d}$ &$ -\frac{F_0^2\mathcal A_{\bf 6\times 6}}{4B}\frac{1}{m_d}\frac{\partial \delta m_N}{\partial (\bar m\epsilon)}$\\
$L_{\bf 6\times 6}^{3333}$ &$2\frac{\Delta  m_N}{m_s}$ &$  \frac{F_0^2\mathcal A_{\bf 6\times 6}}{B}\frac{1}{m_s}\frac{\partial \Delta m_N}{\partial  m_s}$&$0$&$0$\\
\hline
\end{tabular}
\caption{Coefficients determining the contributions to the axion-nucleon couplings. $g_S^{(0,1)} = \frac{m_*}{f_a}\sum_i\left[\frac{\partial}{\partial {\rm Re}\, L_i}\alpha_{i}^{(0,1)}+\bt_{i}^{(0,1)}
\right]{\rm Im}\, L_i
$, where $i$ runs over all the Wilson coefficients in the table.
Wilson coefficients that do not appear above have either been rewritten using the symmetry properties described below Eq.\ \eqref{eq:LagLEFT}, or do not contribute. Note that, we have the following relations at LO $2\frac{\partial \Delta m_N}{\partial m_{u,d}}=\frac{\Delta m_N(m_q)}{\bar m} = \frac{\sigma_{\pi N}}{\bar m}$.
}
\label{tab:aNN}
\end{table}
The axion couplings to the nucleons can be split into an isoscalar and isovector component,
\bea\label{eq:aNNlag}
\vL_{a\bar NN} = a\bar N \left[ g_S^{(0)}+g_S^{(1)}\tau_3\right]N\,,\qquad g_S^{(p,n)} = g_S^{(0)}\pm g_S^{(1)}\,.
\eea
These axion-nucleon couplings receive an `indirect' contribution from Eq.\ \eqref{eq:LagpiNindirect}, which appears after vacuum alignment, as well as a `direct' contribution that arises from Eq.\ \eqref{eq:LagpiNdirect}, which we write as $
g_\al = g_\al\vert_{\rm dir}+g_\al\vert_{\rm ind}$. The `direct' and `indirect' pieces can be written as
\bea
g_S^{(0,1)} \vert_{\rm dir}&=& \frac{m_*}{f_a}\sum_i \frac{\partial \alpha_{i}^{(0,1)}}{\partial {\rm Re}\, L_i}{\rm Im}\, L_i\,,\label{eq:aNNdir}\\
g_S^{(0,1)} \vert_{\rm ind}&=& \frac{m_*}{f_a}\sum_i\bt_{i}^{(0,1)}\label{eq:aNNind}
{\rm Im}\, L_i\,.
\eea
where the coefficients $\al^{(0,1)}$ and $\bt^{(0,1)}$ are given in Table \ref{tab:aNN}. Here the symmetry properties of the $L^{ijkl}_\al$, described below Eq.\ \eqref{eq:LagLEFT}, were used to rewrite all couplings in terms of those appearing in the table.
The expressions in Table \ref{tab:aNN} employ the following combinations of quark masses
\beq \bar m = \frac{m_u+m_d}{2}\,,\qquad \bar m \epsilon= \frac{m_d-m_u}{2}\,,
\eeq
 and combination of nucleon masses
 \beq
 \Delta m_N = \frac{m_n+m_p}{2}\,,\qquad \delta m_N = m_n-m_p\,.
 \eeq 
Couplings to baryons containing valence strange quarks involve mass combinations of the full baryon octet.
The expressions in Eqs.\ \eqref{eq:aNNdir} and \eqref{eq:aNNind} involve the derivatives of these baryon-mass combinations with respect to the quark masses, written in terms of $\bar m$, $\bar m \epsilon$, and $m_s$, as well as the real parts of Wilson coefficients, Re $L_\al$. The dependence on these quantities arises from contributions to the baryon masses of Eq.\ \eqref{eq:LagpiNindirect} and Eq.\ \eqref{eq:LagpiNdirect}, respectively. 

For comparison we briefly discuss the expected size of the scalar axion-nucleon coupling in the pure Standard Model. In this case, the CP-odd couplings arise from the CKM phase $\delta_{\mathrm{CKM}}$. Ref.~\cite{Georgi:1986kr} estimated
\begin{equation}\label{gsCKM}
g_S^{(0)}(\delta_{\mathrm{CKM}}) \simeq m_* J_{\mathrm{CP}} G_F^2 F_0^4/f_a \simeq 10^{-18}\,\mathrm{MeV}/f_a\,,
\end{equation}
with $J_{\mathrm{CP}} \simeq 3 \cdot 10^{-5}$ \cite{Jarlskog:1985ht}. Similar-sized contributions arise from CKM-induced contributions to the light-quark chromo-EDMs. In that case, the estimate can be enhanced by a power $m_c^2/F_0^2$ but additional loop suppression lead to a very similar estimate of $g_S^{(0)}(\delta_{\mathrm{CKM}})$ \cite{Okawa:2021fto}. Long-distance contributions involving hyperons also appear to be of similar size \cite{Okawa:2021fto}. We therefore take Eq.~\eqref{gsCKM} as a rough indication of the size of CP-odd axion-nucleon couplings within the Standard Model. 

\subsubsection*{Pion-nucleon couplings}

The CP-odd pion-nucleon couplings can be written as
\bea\label{g012}
\vL_{\pi \bar NN} = \bar N\left[\bar  g_0 \boldsymbol{\tau \cdot  \pi} +\bar g_1 \pi_0+ \bar g_2 \left(\tau_3 \pi_0-\frac{1}{3}\boldsymbol{\pi\cdot \tau}\right)\right] N\,,
\eea 
where $\bar g_{0,1,2}$ denote the isoscalar, isovector, and isotensor terms, respectively.
The direct pieces of these couplings are related to the axion-nucleon couplings as follows
\bea
\bar g_0\vert_{\rm dir}&=&\tilde g^{(1)}_S\vert_{\rm dir}-\frac{2}{3}\bar g_2\vert_{\rm dir}\,,\qquad
\bar g_1\vert_{\rm dir}=\tilde g^{(0)}_S\vert_{\rm dir}\,,
\nn\\
\bar g_2\vert_{\rm dir}&=&\frac{1}{F_0}\Bigg[{\rm Im}\, L_{\bf 6\times 6}^{2211}\frac{d}{d\, {\rm Re}\, L_{\bf 6\times 6}^{2211}}\left(
\Delta m_{\Sigma^0}+\Delta m_{\Sigma^-}-2\Delta m_{\Xi^-}
\right)\nn\\
&&+\sum_a t_3^a {\rm Im}\, L_{\bf 6\times 6}^{aaaa}\frac{d}{d\, {\rm Re}\, L_{\bf 6\times 6}^{aaaa}}\dt m_N
\Bigg]
\,,
\eea
where $t_3^a\equiv \left(t_3\right)_{aa}$ and $\tilde g_S^{(0,1)}\vert_{\rm dir} $ can be obtained from $g_S^{(0,1)}\vert_{\rm dir}$ in Eq.\ \eqref{eq:aNNdir} by using the following replacement rules on the imaginary parts of the appearing Wilson coefficients~\footnote{The partial derivatives with respect to the real parts of the Wilson coefficients should be left unchanged.}
\bea
L_{5}^{ab}&\to& -\frac{2f_a}{m_*F_0}\left(t_3^a+t_3^b\right)\left(\frac{1}{m_a}+\frac{1}{m_b}\right)^{-1}L_{5}^{ab}\,,\nn\\
L_{\bf r}^{abcd}&\to& -\frac{2f_a}{m_*F_0}\left(t_3^a+t_3^b+t_3^c+t_3^d\right)\left(\frac{1}{m_a}+\frac{1}{m_b}+\frac{1}{m_c}+\frac{1}{m_d}\right)^{-1}L_{\bf r}^{abcd}\,,\qquad {\bf r} \in \{{\bf 3\times 3\,,6\times 6}\}\,,\nn\\
L_{\bf r}^{abcd}&\to& -\frac{2f_a}{m_*F_0}\left(t_3^a-t_3^b-t_3^c+t_3^d\right)\left(\frac{1}{m_a}-\frac{1}{m_b}-\frac{1}{m_c}+\frac{1}{m_d}\right)^{-1}L_{\bf r}^{abcd}\,,\qquad {\bf r} ={\bf 8\times 8}\,,
\eea
where the repeated indices were not summed over. 
The indirect contributions can be written as
\bea
\bar g_0\vert_{\rm ind}&=&-\frac{F_0}{4B}\sum_{a,b}{\rm Im}\Bigg[
\left(|t_3^a|+|t_3^b|\right)\left(\mathcal A_{\bf 3\times 3}L_{\bf 3\times 3}^{aabb}+\mathcal A_{\bf 6\times 6}L_{\bf 6\times 6}^{aabb}
+2\dt^{ab}\frac{\bar B}{F_0^2} L_5^{aa}
\right)\nn\\
&&+\frac{1}{2}\left(|t_3^a|-|t_3^b|\right)\mathcal A_{\bf 8\times 8}L_{\bf 8\times 8}^{abba}
\Bigg]
\frac{\partial\delta m_N}{\partial (\bar m \epsilon)}
\,,\nn\\
\bar g_1\vert_{\rm ind}&=&-\frac{F_0}{2B}\sum_{a,b}{\rm Im}\Bigg[
\left(t_3^a+t_3^b\right)\left(\mathcal A_{\bf 3\times 3}L_{\bf 3\times 3}^{aabb}+\mathcal A_{\bf 6\times 6}L_{\bf 6\times 6}^{aabb}
+2\dt^{ab}\frac{\bar B}{F_0^2} L_5^{aa}
\right)\nn\\
&& + \frac{1}{2}\left(t_3^a-t_3^b\right)\mathcal A_{\bf 8\times 8}L_{\bf 8\times 8}^{abba}
\Bigg]
\frac{\partial\Delta m_N}{\partial\bar m}\,,\nn\\
\bar g_2\vert_{\rm ind}&=&0\,.
\eea
Values for the mesonic LECs are discussed in App.~\ref{app:LECs}. 

\section{Constraints from electric dipole moment experiments}\label{sec:EDMs}

The CP-odd electron-nucleon and pion-nucleon interactions in Eqs.~\eqref{CSP} and \eqref{g012} induce EDMs of various systems. We take the expressions from Ref.~\cite{Dekens:2018bci}. The semi-leptonic Wilson coefficients $C^{(0,1)}_S$ mainly contribute to CP-odd effects in polar molecules 
\cite{doi:10.1063/1.4968597,Fleig:2017mls,PhysRevA.93.042507}
\bea
\omega_{\text{YbF}} &=& -(17.6\pm2.0)(\mathrm{mrad}/\mathrm{s})\left(\frac{C_S }{10^{-7}}\right)\,,\\
\omega_{\text{HfF}} &=&+(32.0\pm1.3)(\mathrm{mrad}/\mathrm{s})\left(\frac{C_S }{10^{-7}}\right)\,,\\
\omega_{\text{ThO}} &=&+(181.6\pm7.3)(\mathrm{mrad}/\mathrm{s})\left(\frac{C_S }{10^{-7}}\right)\, ,
\label{eq:Molecules}
\eea
in terms of $C_S = C_S^{(0)}+\frac{Z-N}{Z+N} C_S^{(1)}$ where $Z$ and $N$ correspond to the number of protons and neutrons, respectively, of the heaviest atom of the molecule. 
In addition, the combination of CP-odd and CP-even axion couplings, $\sim g_S^{N}g_P^{(e)}$ or $\sim g_S^{(e)}g_P^{(e)}$, can give rise to CP-odd effects in nuclei, atoms, and molecules. Such contributions were considered in Ref.\ \cite{Stadnik:2017hpa}, which showed that the most stringent limits arise from the ThO measurement, giving
\bea \omega_{\rm ThO}(g_{S,P}) = \left[0.54\,g_S^{(e)}g_P^{(e)}+1.4\,g_S^{N}g_{P}^{(e)}\right]\cdot10^{19}\,,
\eea
where $g_P^{(f)}$ is connected to the CP-even couplings in Eq.\ \eqref{eq:Laxion} by $g^{f}_{P} = C^{f}_{P}  m_f/f_a$ and $g_P^{N} = \frac{A-Z}{A}g_{P}^n+\frac{Z}{A}g_{P}^p$.

The operators $C^{(0,1)}_P$ and $\bar g_{0,1,2}$ induce EDMs of nucleons, nuclei, and diamagnetic atoms. For the nucleon EDMs we use the results \cite{Seng:2014pba}
\begin{eqnarray}
d_n &=& - \frac{e g_A }{8 \pi ^2 F_\pi} \left[ \left(\bar g_0-\frac{\bar g_2}{3}\right) \left( \log \frac{m^2_\pi}{m_N^2} - \frac{\pi m_\pi}{2 m_N} \right)  + \frac{\bar g_1}{4 } \left( \kappa_1 - \kappa_0\right) \frac{m^2_\pi}{m_N^2} \log \frac{m^2_\pi}{m_N^2}  \right] \ ,
\label{eq:dn}
\\
d_p &=&  \frac{e g_A}{8 \pi^2 F_\pi} \Bigg[  \left(\bar g_0-\frac{\bar g_2}{3}\right) \left( \log \frac{m^2_\pi}{m_N^2} - \frac{2 \pi m_\pi}{m_N} \right)  
-\frac{\bar g_1}{4 } \left(  \frac{2 \pi m_\pi}{m_N} + \left( 5/2 + \kappa_1 + \kappa_0\right) \frac{m^2_\pi}{m_N^2} \log \frac{m^2_\pi}{m_N^2}  \right)  \Bigg]
\ ,\nonumber
\label{eq:dp}
\end{eqnarray}
where $g_A \simeq 1.27$ is the nucleon axial charge, and $\kappa_1 = 3.7$ and $\kappa_0 = -0.12$ are related to the nucleon magnetic moments. For $\bar g_1$ we kept the next-to-next-to-leading-order corrections as this is the first order where a neutron EDM is induced. We have set the renormalization scale to the nucleon mass $m_N$ in order to estimate the EDMs as function of pion-nucleon couplings. 

We should point out that the CP-odd LEFT operators induce additional CP-violating hadronic interactions that contribute to the nucleon EDMs as well. For example, a quark chromo-EDM operator $\sim \bar q \sigma^{\mu\nu} \gamma^5 T^A q\,G^A_{\mu\nu}$ leads to direct contributions to the neutron EDM in addition to the pion-nucleon terms. Such direct terms depend on hadronic matrix elements that do not appear in the CP-odd axion interactions given above (they would be connected to CP-odd axion-photon-nucleon terms instead). We therefore do not include these effects here, which leads to conservative limits\footnote{For certain LEFT operators such as the Weinberg operator, there are no contributions to $\bar g_{0,1,2}$ at the chiral order we work but terms do appear after a quark mass insertion leading to an additional suppression of $m_\pi^2/\Lambda_\chi^2$ \cite{deVries:2012ab}. In those cases, the direct contributions to the nucleon EDMs, which come with additional LECs, are a better estimate. To keep the discussion compact we do not further pursue this here.}, assuming there are no significant cancellations. Instead, we estimate the EDMs of neutrons (and protons) from their pion-loop contributions proportional to $\bar g_{0,1,2}$ in Eqs.~\eqref{eq:dn} and \eqref{eq:dp}.

\begin{table}[t]
\small\center
\renewcommand{\arraystretch}{1.2}
$\begin{array}{ccc|ccc}
\multicolumn{3}{c|}{{\rm neutron}\, {\rm and }\,{\rm atoms }  \,(e \, {\rm cm})} &\multicolumn{3}{c}{{\rm Molecules}  \,(\mathrm{mrad}/\mathrm{s}) }   \\\hline
d_n & d_{\rm Hg} & d_{\rm Ra} & \omega_{\text{YbF}}&\omega_{\text{HfF}} & \omega_{\text{ThO}}\\\hline 
1.8 \cdot 10^{-26} &6.3\cdot 10^{-30} & 1.2\cdot 10^{-23}&
23.5
&  4.6
&
1.3\\
  \end{array}$
\caption{Current experimental limits (at 90$\%$ C.L.) from measurements on the neutron \cite{Abel:2020pzs}, $^{199}$Hg \cite{Graner:2016ses}, $^{225}$Ra \cite{Bishof:2016uqx}, YbF
\cite{Hudson:2011zz}, HfF \cite{Cairncross:2017fip}, and ThO \cite{ACME:2018yjb}.}\label{tab:expt}
\end{table}
The expression for the Hg EDM becomes  
\cite{Engel:2013lsa,Fleig:2018bsf,Dzuba:2009kn,Latha:2009nq,Yamanaka:2017mef,Dmitriev:2003sc}
\bea\label{dHg} d_{\rm Hg}&=& -(2.1\pm0.5)
\Ex{-4}\bigg[(1.9\pm0.1)d_n +(0.20\pm 0.06)d_p\nn\\
&&+\bigg(0.13^{+0.5}_{-0.07}\,\bar g_0 +
0.25^{+0.89}_{-0.63}\,\bar g_1+0.09^{+0.17}_{-0.04}\,\bar g_2\bigg)e\, {\rm fm}\bigg]\nn\\
&& - \left[ (0.028\pm0.006) C_S- \frac{1}{3}(3.6\pm0.4) \left(\frac{Z\al}{5 m_N R}C_P\right)\right]\cdot
10^{-20}\, e\,\mathrm{cm}\,,
\eea
in terms of the nuclear radius $R\simeq 1.2\, A^{1/3}$ fm, and
$C_{P,T} = (C_{P,T}^{(n)}\langle \vec \sigma_n\rangle  +C_{P,T}^{(p)}\langle \vec \sigma_p\rangle
)/(\langle \vec \sigma_n\rangle +\langle \vec \sigma_p\rangle )$. Here we defined $C_{P,T}^{(n,p)}=C_{P,T}^{(0)}\mp C_{P,T}^{(1)}$. For $^{199}$Hg
we use the values \cite{Yanase:2018qqq} 
\bea
\langle \vec \sigma_n\rangle= -0.3249\pm0.0515\,,\qquad \langle \vec \sigma_p\rangle = 0.0031\pm 0.0118\,.
\eea
The expression for the octopole-deformed Ra is simpler as nuclear CP violation dominates the atomic EDM
\cite{Engel:2013lsa,Dobaczewski:2018nim}
\begin{eqnarray}\label{dRa}
d_{\mathrm{Ra}} &=& (7.7\cdot 10^{-4})\cdot\left[(2.5\pm 7.5)\,\bar g_0 - (65 \pm 40)\,\bar g_1+(14\pm 6.5)\bar g_2\right]e\, {\rm fm}\,.
\end{eqnarray}
Note that the radium and mercury EDMs are dominated by the contributions to the pion-nucleon couplings as long as $\bar g_{0,1,2}$ receive contributions at LO (which is the case for the LEFT operators under consideration here). The connection between the contributions to $d_{\rm Ra}$ and $d_{\rm Hg}$ and the axion-nucleon couplings is therefore more straightforward than was the case for the nucleon EDMs, where additional direct contributions appear at the same order.
The current best limits are collected in Table~\ref{tab:expt}.

\section{Fifth-force experiments}\label{sec:5thforce}

\begin{figure}[t]
    \centering
    \includegraphics[width=\textwidth]{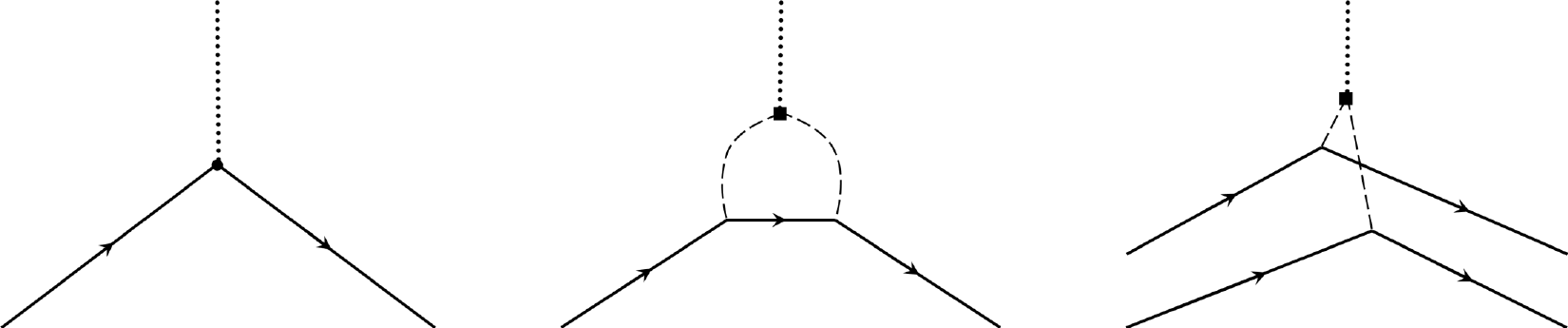}
    \caption{Diagrams contributing to the axion-nucleon interactions. The solid lines correspond to nucleons, the dotted lines to  axions, and the dashed lines to mesons.}
    \label{fig:feynman_diagrams}
\end{figure}

The CPV scalar axion-nucleon and axion-lepton couplings of Eqs.\ \eqref{eq:aNNlag} and \eqref{eq:lepAxion} lead to monopole-monopole forces, which would act like a `fifth force', thereby modifying Newton's inverse-square law (ISL) and violating the weak equivalence principle (WEP). The combination of the gravitational and axion potentials between two different bodies $I$ and $J$ then becomes
\be
V_{IJ}(r) = - \frac{G m_I m_J}{r_{IJ}} \left(1 + \al_{I}\al_{J}e^{-m_a r_{IJ}}\right),
\label{eq:fifth_force_pot}
\ee
where $m_a (=1/\lambda)$ is the axion mass, $r_{IJ}$ is the distance between $I$ and $J$, while $m_{I,J}$ and $\al_{I,J}$ are the masses and total `axion charges' of $I$ and $J$ (the latter are normalized to $m_{I,J}\sqrt{4\pi G}$). At leading order in $\chi$EFT this total charge is determined by the axion couplings to the electron, proton, and neutron, as well as the number of these particles present in the body under consideration. Explicitly, we have
\bea\label{alpha}
\al_I =\frac{1}{\sqrt{4\pi G}}\left[ g_S^{(n)}\frac{A-Z}{m_A}+g_S^{(p)}\frac{Z}{m_A}+g_S^{(e)}\frac{Z}{m_A}\right]\,,
\eea
where $m_{A}$ is the mass of the atom, which is most cases can be approximated by $m_A =A m_{\rm u}$, with $m_{\rm u}$ the atomic mass unit, while $Z$ and $A-Z$ are the number of protons and neutrons of the element that the body $I$ consists of. The leading-order contributions arise from the simple axion-nucleon diagram (left diagram of Fig.~\ref{fig:feynman_diagrams}).

 Axion-meson-meson couplings modify the axion-nucleon coupling at the one-loop level (middle diagram of Fig.~\ref{fig:feynman_diagrams}), in practice part of the one-loop contributions are automatically resummed by using the physical values for $\sigma_{\pi N}$ and $\delta m_N$ in Eq.~\eqref{sigma}. Graphs such as the one depicted in the right panel of Fig.~\ref{fig:feynman_diagrams} lead to two-nucleon contributions that cannot be captured by Eq.~\eqref{alpha}. In the analogous case of a dilaton ($\phi$) coupling to quarks, $\sim \phi \bar qq$, such contributions to the potential are related to the binding energy of the nuclei and were considered in Ref.\ \cite{alphaA}. Within $\chi$EFT, however, these two-body interactions appear at higher-order in the power counting. In particular, for scalar axion-quark interactions (e.g. $a \bar q q$) axion-nucleon-nucleon currents appear at next-to-leading-order in the power counting and could in principle be relevant \cite{Cirigliano:2012pq}. These currents were discussed in detail in light of WIMP scattering off atomic nuclei and were found to be somewhat smaller than power-counting predictions indicate and appear only at the few-percent level \cite{Korber:2017ery} although this could increase for larger nuclei \cite{Hoferichter:2018acd}. We neglect the subleading two-nucleon corrections in this work.
 
There are numerous experiments that search for the fifth force that would be induced by the $\al_I \al_J$ term in Eq.\ \eqref{eq:fifth_force_pot}. These experiments either look for violations of the WEP, which appear when $V_{IJ}$ is no longer proportional to $m_I m_J$, or departures from the inverse-square law due to deviations from the $1/r_{IJ}$ dependence of the usual gravitational potential.
In this section we summarize several of these experiments and discuss how they limit the axion couplings to nucleons and leptons.  

\subsection{MICROSCOPE mission}

The MICROSCOPE mission \cite{microscope} focuses on constraining the \eot which is a measure of WEP violations. The \eot is the normalized difference between the accelerations of two masses $I$ and $J$. In the case of Ref.\ \cite{microscope} these masses are made of  platinum and titanium and are in free fall aboard the MICROSCOPE satellite
\be
\eta = \left(\frac{\Delta a}{a}\right)_{IJ} = 2 \frac{\abs{\boldsymbol{a}_I - \boldsymbol{a}_J}}{\abs{\boldsymbol{a}_I + \boldsymbol{a}_J}}\,.
\ee
From Eq.\ \eqref{eq:fifth_force_pot} one finds that the 
\eot for two test masses in the external field of Earth ($E$) can be expressed as 
\be
\eta = \frac{\al_E|\al_I -\al_J| \left(1+x \right)e^{-x}}{1+\frac{1}{2}(\al_J +\al_J)\al_E \left(1+x \right)e^{-x}}
\simeq 
\al_E |\al_I - \al_J| (1+x )e^{-x}\,,
\ee
where $x= R m_a$, $R\approx 7000$ km is the distance from the center of the earth to the satellite, and $\al_E$ is the effective `axion charge' of the earth. Following \cite{microscope}, we model the earth as consisting of a core (which is taken to consist of iron) and the mantle (consisting of SiO$_2$), so that its charge takes the form \footnote{This result differs from the expression obtained in Ref.~\cite{microscope}.}
\bea\label{earth}
\al_E = \frac{m_C}{m_E}\al_{\rm Fe}\Phi\left(R_C m_a \right)+\frac{m_M}{m_E}\al_{\rm Si O_2}\frac{R_E^3\Phi(R_E m_a)-R_C^3\Phi(R_C m_a)}{R_E^3-R_C^3}\,,
\eea 
where $R_E\simeq 6371$ km and $R_C\simeq 3500$ km are the radii of the Earth and its core. $m_E, \,m_C,$ and $ m_M$ are, respectively, the masses of the Earth, its core, and its mantle, with $m_C/m_E\simeq 0.33$, while the function $\Phi (x) \equiv 3 (x \cosh x - \sinh x)/x^3$ describes the deviation from a simple Yukawa potential due to the finite size of the earth.

Combining these expressions with the experimental limit  \cite{microscope},
\be
\eta = (-1 \pm 27) \times 10^{-15}\,,
\ee
allows us to set constraints on $g_{S}^{(e,n,p)}$ as a function of $m_a$.

\subsection{E\"{o}t-Wash (WEP)}

The E\"{o}t-Wash experiment \cite{Smith:1999cr} constrained deviations from the WEP at distance scales $\gtrsim 0.1$ m. In this case two test bodies, made of Pb and Cu, were connected to a torsion balance around which a $^{238}$U attractor mass rotates. A difference in the accelerations of the two bodies would then show up as a torque, $\vec\tau = \frac{1}{2} \vec d \times (\vec F_{\rm Cu}-\vec F_{\rm Pb})$, where $\vec d$ is the distance between the two test bodies and $\vec F_{\rm Cu,Pb}$ are the forces that work on them, due to the earth and the attractor. The experiment looks for signals that vary as a function of the angle, $\phi$, between $\vec d$ and the vector from the test bodies to the attractor.
The fact that only accelerations orthogonal to $\vec d$ contribute to the torque implies $\tau$ is a measure of  deviations from the WEP, $\tau\sim \al_{\rm Cu} - \al_{\rm Pb}$, while looking for $\phi$-dependent signals means signals are due to the force (whether gravitational or axionic) exerted by the attractor. In total, the $\phi$-dependent part of the $z$-component of the torque can be written as~\footnote{Here we neglect a small correction to the contributions $\sim \al_A$ due to the centrifugal force induced by the earth's rotation.}
\bea
\frac{\tau_z\vert_{\rm varying}}{g' d m_{\rm Cu}/2} &=& \frac{|a_{\rm Pb}-a_{\rm Cu}|}{g'} \sin\phi\nn\\
&\simeq &
\left|  \al_{\rm Pb}- \al_{\rm Cu}\right|\left[\al_{ A} I_A(m_a)-\al_{E}(1+R_E m_a)e^{-R_Em_a}\right]\sin\phi\,,
\eea
where $g' = 9.2 \cdot 10^{-7}$ m/s$^2$ and $\al_{\rm A}$ are the gravitational acceleration and the axion charge of the element of the attractor, while $I_A$ is a function that captures the geometry of the attractor. 
The experimental limit
\bea
|a_{\rm Pb}-a_{\rm Cu}|\leq 5.7\cdot 10^{-15}\, {\rm m/s}^2\,, \qquad  95\% \,\,{\rm C.L.}
\eea
together with $I_A$, which we obtain from interpolating the numerical values in Table \ref{tab:IA}, again allows us to set constraints on $g_{S}^{(n,p,e)}$.

\begin{table}[h!]
\small\center
\renewcommand{\arraystretch}{1.2}
$\begin{array}{c|ccccccccccccccc}
\lambda \,\,{\rm (m)}& 0.01 & 0.014 & 0.020 & 0.028 & 0.05 & 0.07 & 0.1 & 0.2 & 0.5 & 1 & 2 & 5 
\\\hline 
 I_A(m_a) &1.3\cdot 10^{-5} &  1.8\cdot 10^{-4} & 0.0016 &0.0079& 0.057 & 0.13 & 0.26 &
0.59 & 0.89 &0.97 &0.99 &1.0 
\\
  \end{array}$
\caption{Numerical values for the function $I_A(m_a)$ that describes the geometry of the attractor in the E\"{o}t-Wash experiment \cite{Smith:1999cr} as a function of $\lambda = 1/m_a$.}\label{tab:IA}
\end{table}
Later work suspended the torsion pendulum from a rotating turntable, instead of using a rotating attractor, and used test bodies made of Be and Ti~\cite{Schlamminger:2007ht}. 
The role of source mass was dominated by features in the surrounding environment, local topography, and finally the earth, depending on the value of $\la$. Lacking knowledge of the size, density, and composition of these environmental features, we approximate the effective axion coupling of the source masses by $\al_{\rm source} = \al_{\rm SiO_2}$. The experimental results are shown in Table \ref{tab:WEPwash08}. 

\begin{table}[h!]
\small\center
\renewcommand{\arraystretch}{1.2}
$\begin{array}{c|ccccccccccccccc}
\lambda \,\,{\rm (m)}& 1 & 10 & 1\ten{2} & 1\ten{3} & 1\ten{4} & 1\ten{5} & 1\ten{6} & >1\ten{7}
\\\hline 
|\al_{\rm source}(\al_{\rm Be}-\al_{\rm Ti})| &8\ten{-6} & 1\ten{-6} & 2\ten{-7} &7\ten{-8} & 4\ten{-8} & 4\ten{-8} &5\ten{-9} & 2\ten{-10}
\\
  \end{array}$
\caption{Constraints set by E\"{o}t-Wash experiment \cite{Schlamminger:2007ht} as a function of $\lambda = 1/m_a$.}\label{tab:WEPwash08}
\end{table}

\subsection{Irvine}

This experiment \cite{Hoskins:1985tn} consists of two sets of measurements, searching for fifth forces at distances between $105$-$5$ cm and $5$-$2$ cm. Both measurements used a torsion balance to constrain the torque that would be induced by a Yukawa potential. The test and attractor masses used in these set-ups were designed so as to give rise to a net-zero torque if the force between them has a pure $1/r^2$ dependence. The measurements on smaller distance scales searched for fifth forces between a copper test mass and a stainless steel cylinder, while the test and attractor masses used in the measurements at larger distances were both made from copper. As the elements in these materials all have a similar $Z/A\simeq 0.45$, which determines $\al$ in Eq.\ \eqref{alpha}, we will approximate the effective axion charges $\al_{I,J}$ by $\al_{\rm Cu}$. 
We summarize the constraints on the combination $\al_I \al_J$ as a function of $\la$ in Table \ref{tab:Irvine}.

\begin{table}[h!]
\small\center
\renewcommand{\arraystretch}{1.2}
$\begin{array}{c|ccccccccccccccc}
\lambda \,\,{\rm (mm)}&  5 & 10 &  50 & 100 &  500 & 1.0\cdot 10^{3}& 5.0\cdot 10^{3}& 1.0\cdot 10^{4}
\\\hline 
|\al_{\rm Cu}\al_{\rm Cu}|&1.1\cdot 10^{-3}&2.1 \cdot 10^{-4}& 1.9\cdot 10^{-4} &4.2 \cdot 10^{-4} & 1.3\cdot 10^{-3} & 3.8 \cdot 10^{-3}& 0.088 & 0.46
\\
  \end{array}$
\caption{$95\%$ C.L.\ constraints on the combinations $|\al_{\rm Cu}\al_{\rm Cu}|$ as a function of $\lambda = 1/m_a$ set by the Irvine measurements \cite{Hoskins:1985tn}.}\label{tab:Irvine}
\end{table} 

\subsection{E\"{o}t-Wash (inverse-square law)}

Apart from searches for WEP violations there are experiments that look for deviations from the inverse-square law. The E\"{o}t-Wash experiments accurately measured the force between an attractor, made of Mo and Ta, and a Pt or Mo test body as a function of the distance between them \cite{Kapner:2006si,Lee:2020zjt}. 
The geometry of the experimental setup cancels the attraction due to the gravitational potential, $\sim 1/r^2$, allowing one to constrain Yukawa forces. Although the attractor was made of several materials, we approximate the probed combinations by $|\al_{\rm Mo}\al_{\rm Mo}|$ and $|\al_{\rm Pt}\al_{\rm Mo}|$, the resulting constraints on which are listed in Table \ref{tab:eotWash}. 

\begin{table}[h!]
\small\center
\renewcommand{\arraystretch}{1.2}
$\begin{array}{c|ccccccccccccccc}
\lambda \,\,{\rm (mm)}& 0.01 & 0.025 & 0.05 & 0.1 & 0.25 & 0.5 & 1  & 2.5 & 5
\\\hline 
|\al_{\rm Mo}\al_{\rm Mo}|&4.1\cdot 10^{4} & 43 & 1.4 & 0.1 &6.7\cdot 10^{-3} &2.4\cdot 10^{-3} &2.7\cdot 10^{-3} &7.2\cdot 10^{-3} &7.1\cdot 10^{-3}  
\\
|\al_{\rm Mo}\al_{\rm Pt}|&3.1\cdot 10^{3} & 6.4 & 0.42 & 0.077 &0.029 &0.025 & 0.019 & 0.013 & 0.012
\\
  \end{array}$
\caption{$95\%$ C.L.\ constraints on the combinations $|\al_{\rm Mo}\al_{\rm Mo}|$ \cite{Kapner:2006si} and $|\al_{\rm Pt}\al_{\rm Mo}|$ \cite{Lee:2020zjt} as a function of $\lambda = 1/m_a$ set by the E\"{o}t-Wash experiment.}\label{tab:eotWash}
\end{table}

\subsection{HUST}

The HUST experiment \cite{Tan:2020vpf,Tan:2016vwu,Yang:2012zzb,Tu:2007zz} searches for inverse-square law violations caused by a fifth force between two plane masses made of tungsten (W) using a torsion pendulum. 
The pendulum is suspended horizontally with a rectangular shaped test mass of tungsten at each end. The test masses are facing a rotating attractor made of rectangular tungsten source masses and compensating masses, designed to cancel out the torque due to Newtonian forces. This experiment is most sensitive in the range $\la = (40-350)\ \mu$m. The resulting constraints on $\al_I \al_J$ are collected in Table \ref{tab:HUST}. 

\begin{table}[h!]
\small\center
\renewcommand{\arraystretch}{1.2}
$\begin{array}{c|ccccccccccccccc}
\lambda \,\,{\rm (mm)}& 0.025 & 0.05 & 0.1 & 0.25 & 0.5 & 1 &  2.5 & 5
\\\hline 
|\al_{\rm W}\al_{\rm W}|&420 & 0.70 & 0.026 & 3.8\cdot 10^{-3} & 3.6 \cdot 10^{-3} & 1.4 \cdot 10^{-3} & 1.0 \cdot 10^{-3} & 2.2 \cdot 10^{-3}
\\
  \end{array}$
\caption{$95\%$ C.L.\ constraints on the combinations $|\al_{\rm W}\al_{\rm W}|$ as a function of $\lambda = 1/m_a$ set by the HUST experiment  \cite{Tan:2020vpf,Tan:2016vwu,Yang:2012zzb,Tu:2007zz}.
}
\label{tab:HUST}
\end{table} 

\subsection{Stanford}

The Stanford experiment \cite{Geraci:2008hb}  focused on axions in the range $\lambda \sim 5$ - $15\ \mu$m. A rectangular gold (Au) prism, located at the end of a cantilever, was used as a test mass, the force on which was determined through its displacement. A source mass, consisting of alternating gold and silicon (Si) bars, was then moved horizontally below the test mass. In the presence of a fifth force, the test mass experiences different forces depending on whether a Au or Si bar is located directly below it, resulting in a different displacement of the cantilever. This would induce an oscillating force when the source mass is displaced horizontally. Note that the background due to Newtonian forces is negligible at this level of precision. The amplitude of the induced force is proportional to $\rho_I \al_I$, with $\rho$ the mass density and $I=$  Au, Si. As $\rho_{\rm Au}\gg \rho_{\rm Si}$, we approximate the probed combination of axion charges by $\al_{\rm Au}^2$, the constraints on which are  shown in Table \ref{tab:stanford}. 

\begin{table}[h!]
\small\center
\renewcommand{\arraystretch}{1.2}
$\begin{array}{c|ccccccccccccccc}
\lambda \,\,{(\mu\rm m)}& 4& 6& 10& 18& 34&66
\\\hline 
|\al_{\rm Au}\al_{\rm Au}|&3.1 \cdot 10^{7}&4.6 \cdot 10^{5}&1.4 \cdot 10^{4} &1.1 \cdot 10^{3}&2.5 \cdot 10^{2}&1.5 \cdot 10^{2}   
\\
  \end{array}$
\caption{$95\%$ C.L.\ constraints on the combinations $|\al_I\al_J|$ as a function of $\lambda = 1/m_a$ set by the Stanford experiment  \cite{Geraci:2008hb}.}\label{tab:stanford}
\end{table}

\subsection{IUPUI}

This experiment searches for fifth forces by measuring the differential force on masses separated by distances in the nm range \cite{Chen:2014oda}, allowing it to probe axions in the range $\la \sim (40 - 8000)$ nm. The set up involves a spherical test mass, made in large part of sapphire (S), located above a rotating disk which serves as a source mass. The latter involves several rings, each with a number of alternating segments made of Au and Si.  The source mass is rotated at a constant frequency, so that a difference in force felt by the test mass due to Au and Si would show up as an oscillating signal. Such a difference would be a sign of a fifth force, while the Newtonian force for this design is below the experimental sensitivity. As both the attractor and the sources masses involve a number of materials, we approximate the effectively probed combination of couplings by $\al_{\rm S} (\al_{\rm Au}-\al_{\rm Si}) $, where $\al_{\rm S}$ is the effective axion charge of sapphire.
The resulting constraints on $\al_I \al_J$ are collected in Table \ref{tab:IUPUI}.

\begin{table}[h!]
\small\center
\renewcommand{\arraystretch}{1.2}
$\begin{array}{c|ccccccccccccccc}
\lambda \,\,{\rm (mm)}& 5\cdot 10^{-5} & 1\cdot 10^{-4} & 2.5 \cdot 10^{-4} & 5 \cdot 10^{-4} & 1 \cdot 10^{-3} &   2.5\cdot 10^{-3} & 5\cdot 10^{-3} & 1\cdot 10^{-2}
\\\hline 
|\al_{\rm S} (\al_{\rm Au}-\al_{\rm Si}) |&2.4\cdot 10^{13}&1.0\cdot 10^{11}&6.5\cdot 10^{8} &4.3\cdot 10^{7}&5.8\cdot 10^{6}&6.3\cdot 10^{5}&1.3\cdot 10^{5}&2.9\cdot 10^{4}
\\
  \end{array}$
\caption{$95\%$ C.L.\ constraints on the combinations $|\al_{\rm W}\al_{\rm W}|$ as a function of $\lambda = 1/m_a$ set by the IUPUI experiment  \cite{Chen:2014oda}.}\label{tab:IUPUI}
\end{table}
 
\subsection{Asteroids and planets}

It is possible to constrain a fifth force by measurements of the orbital trajectories of astronomical objects. In particular, Ref.~\cite{Tsai:2021irw} proposes to use the fifth-force-induced orbital precession of nine near-Earth asteroids, whose orbital trajectories are precisely tracked, to constrain a potential fifth force induced by the exchange of particles in the mass range $m_a \simeq 10^{-21}-10^{-15} \rm{eV}$.  The analysis of Ref.~\cite{Tsai:2021irw} assumed that the new scalar particles couple to the baryon charge, in which case the elemental composition of the sun and the asteroids is not relevant. To constrain the axion couplings, we assume that the asteroids consist mainly of iron ($\al_{\rm asteroid} = \al_{\rm Fe})$. Since the orbits can only be affected by axions with $\la \gg R_\odot$, we model the sun as a point particle, with
\be
\al_\odot 
=(0.75 \ \al_{\rm H} + 0.24 \ \al_{\rm He})\,.
\ee 
With these assumptions we convert the estimated sensitivity of Ref.~\cite{Tsai:2021irw} to limits on axion-nucleon and axion-electron scalar couplings.  

In a similar spirit, it is possible to constrain axion-induced fifth forces by measuring the perihelion procession of planetary orbits \cite{KumarPoddar:2020kdz}. The most stringent limits arise from the perihelion procession of Mars and Mercury. The analysis of Ref.~\cite{KumarPoddar:2020kdz} assumed a model where hypothetical ultralight $Z'$ bosons couple to electrons, which gives rise to a Yukawa potential similar to axions (but with opposite sign). We convert their limits by assuming Mars and Mercury have a similar composition to Earth with similar relative sizes of the mantle and core, which, for $\la\gg R_E$, gives $\al_{\rm planet}\simeq 0.33\al_{\rm Fe}+0.67\al_{\rm SiO_2}$. The constraints on $\al_I \al_J$ from the asteroids and planets as a function of $\la$ are collected in Table \ref{tab:asteroids}.

The resulting limits from the asteroid and planetary orbits are depicted by, respectively, black and gray lines in Figs.~\ref{figqCEDM}-\ref{fig:LR}. 

\begin{table}[h!]
\small\center
\renewcommand{\arraystretch}{1.2}
$\begin{array}{c|ccccccccccccccc}
\lambda \,\,{\rm (km)}& 1\cdot 10^6 & 5\cdot 10^6 & 1\cdot 10^7 & 5\cdot 10^7 & 1\cdot 10^8 & 5\cdot 10^8 & 1\cdot 10^9  & 5\cdot 10^9 
\\\hline 
|\al_{\rm Fe}\al_{\odot}|&2.5\cdot 10^{-6} & 9.8 \cdot 10^{-11} & 3.5 \cdot 10^{-11} & 7.5 \cdot 10^{-12} &1.1 \cdot 10^{-11} &1.1 \cdot 10^{-10} &4.2 \cdot 10^{-10}&1.1 \cdot 10^{-8} 
\\
|\al_{\rm planet}\al_{\odot}|&- & 3.4\cdot 10^{-9} & 1.8 \cdot 10^{-10} & 7.4 \cdot 10^{-12} & 4.3 \cdot 10^{-12} & 1.8 \cdot 10^{-11} & 4.9 \cdot 10^{-11} & 8.8 \cdot 10^{-10} 
\\
  \end{array}$
\caption{$95\%$ C.L.\ projected constraints on the combination $|\al_{\rm Fe}\al_{\odot}|$ \cite{Tsai:2021irw} and current limits on $|\al_{\rm planet}\al_{\odot}|$ \cite{KumarPoddar:2020kdz} as a function of $\lambda = 1/m_a$. The constraints due to the planetary orbits are dominated by Mars for $\la > 2.6\cdot 10^7$ km and by Mercury for smaller values of $\la$.
}\label{tab:asteroids}
\end{table} 

\subsection{Stellar Cooling}

Axions can be produced in the cores of stars. If they escape, this provides a new source of stellar cooling and leads to distinct astronomical signatures that can be searched for. Here we briefly discuss the most stringent limits arising from these searches. A recent more detailed discussion of these constraints can be found in Ref.~\cite{OHare:2020wah}.  

The pseudoscalar axion-electron interaction can generate axions through Compton scattering $\gamma  +e^- \rightarrow e^- + a $ and bremsstrahlung $e + Ze \rightarrow Ze + e + a$ ~\cite{Raffelt:1999tx,Raffelt:2006cw}. These cooling processes allow for heavier red giants as their cores now require more mass to reach the same temperature, thereby delaying helium ignition. The increase in mass then leads to a higher luminosity, so that measurements of the brightness of red giants allows one to constrain the cooling processes induced by axions. The resulting limit is given by~\cite{Capozzi:2020cbu} 
\be
g_{P}^e \lesssim 1.6 \ten{-13}\,.
\ee
These cooling processes are suppressed for heavier axions, as they cannot be produced once the mass becomes significantly heavier than the temperature in the core.  The limits in this section are valid for $m_a < 10$ keV.  

The scalar axion-electron interaction can be constrained by using the fact that it causes mixing of the axion with plasmons in stars ~\cite{Raffelt:1996wa}. This axion production is enhanced if the axion mass is below the plasmon frequency. The most stringent constraint comes from the resonant production in red giants ~\cite{Hardy:2016kme}
\be
g_S^e \lesssim 7.1 \ten{-16}\,.
\ee
The analogous resonant axion production in red giants, induced by scalar axion-nucleon interactions, gives the limit ~\cite{Hardy:2016kme}
\be
g_S^{(0)} \lesssim 1 \ten{-12}\,.
\ee

Finally, for the pseudoscalar coupling to neutrons, the most stringent limits arise from neutron stars. 
Young neutron stars that are formed from the collapsed star core after a supernova explosion can cool by emitting axions through bremsstrahlung, $n + n \rightarrow n + n +a$. Observation of a high surface temperature of neutron stars can then set a limit on the amount of axion emission. The resulting constraint is given by~\cite{Beznogov:2018fda} 
\be
g_P^{n} \lesssim 2.8 \ten{-10}\,.
\ee

\section{Searches for monopole-dipole interactions}\label{sec:ariadne}

In the previous section we discussed constraints on the product of two scalar axion couplings. In the presence of both CP-even and CP-odd interactions, axion exchange also leads to a monopole-dipole potential of the form $V\sim (\vec \sigma \cdot \hat r) e^{-m_a r}/r$, where $\sigma$ is the spin of the particle with a CP-even axion coupling. 
Potentials of this form are searched for by various experiments which we discuss in more detail below. Before doing so, we first introduce the CP-even axion interactions that make up the monopole-dipole potential. These CP-even couplings arise from the derivative terms $ \sim c_{L,R}^{f}\frac{\partial_\mu a}{f_a} \bar f \gamma^\mu\gamma_5 f$ in Eq.\ \eqref{eq:Laxion}. 
In addition, quark-axion interactions are generated by the axion-dependent part of the chiral rotation, $A$, 
which shifts the $\sim (c_R^{q}-c_L^{q})\partial_\mu a$ terms contained in $r_\mu-l_\mu$ in Eq.\ \eqref{eq:rotatedCouplings}. As a result, the quark interactions receive a model-independent contribution from the chiral rotation, while the lepton couplings only involve model-dependent terms since $c_{L,R}^{f}$ depend on the UV construction.

We write the final CP-conserving interactions as
\begin{equation}
\mathcal L_{aN} = \frac{\partial_\mu a}{2 f_a}\left( C^{p}_{P} \,\bar p\,\gamma^\mu \gamma^5\,p +C^{n}_{P}\, \bar n\,\gamma^\mu \gamma^5\,n + C^{e}_{P} \,\bar e\,\gamma^\mu \gamma^5\,e \right)\,,
\end{equation}
for couplings to protons, neutrons, and electrons, respectively. 
Within chiral EFT these nucleon interactions result from the $\sim S_\mu$ terms in the first line of Eq.\ \eqref{eq:LagpiNindirect} as the above Lorentz structure reduces to $\ga_\mu \ga_5\to 2S_\mu$ in the non-relativistic limit. 
For the axion-nucleon CP-even couplings we apply recent results from next-to-next-to-leading-order chiral perturbation theory \cite{Vonk:2020zfh}
\bea
C_P^p &=& -0.430(50)+0.862(75)X_u-0.417(66)X_d-0.035(54)X_s\,,\nn\\
C_P^n &=& 0.007(46)-0.417(66)X_u+0.862(75)X_d-0.035(54)X_s\,, \nn\\
C_P^e &=&X_e =  \left(\frac{c_R^{e}-c_L^{e}}{2}\right)_{11}\,, 
\eea
with $X_q = {\rm diag}(X_u,\,X_d,\, X_s) =\frac{1}{2}{\rm diag} \left(c_R^q - c_L^q\right)$.
These couplings are sometimes written in pseudoscalar form using the equations of motion for on-shell fermions $-g^{f}_{P}\,a\,\bar f i\gamma^5 f$ where $g^{f}_{P} = C^{f}_{P}  m_f/f_a$ for $f=\{p,n,e\}$. 
Using Eq.~\eqref{axionmass}  we write \cite{OHare:2020wah}
\begin{equation}
g^{f}_{P} = 
1.7\cdot 10^{-13}\, C^{f}_{P} \,\left(\frac{m_f}{1\,\mathrm{GeV}}\right) \left(\frac{m_a}{1\,\mu\mathrm{eV}}\right)\,.
\end{equation}

Two of the most popular UV constructions (see Refs.~\cite{DiLuzio:2016sbl,Plakkot:2021xyx
} for more general constructions), which determine the $c_{L,R}^{(f)}$ couplings, are the KSVZ \cite{Kim:1979if,Shifman:1979if} and DFSZ \cite{Zhitnitsky:1980tq,Dine:1981rt} models, see Ref.~\cite{DiLuzio:2020wdo} for a recent review. In these scenarios the couplings take the following values
\bea
 {\rm DFSZ:}\qquad X_{u} &=&  \frac{1}{3}\sin ^2\bt\,,\nn\\ 
X_{d} &=& X_s\, =\, X_e \,\,=\,\,\frac{1}{3}(1-\sin ^2\bt)\,,\nn\\
 {\rm KSVZ:}\qquad X_{q} &=& X_e \,\,=\,\, 0\,,
\eea
where $\tan \beta = v_d/v_u$ is the ratio of vacuum expectation values of scalar fields in the DFSZ model. Assuming perturbativity of the Yukawa couplings appearing in the model, $\bt$ lies in the range $\tan \beta \in [0.25,170]$ \cite{Bjorkeroth:2019jtx,DiLuzio:2020wdo}. In our analysis, the exact values of the CP-even couplings are not our main concern (although they play a role in setting limits). For simplicity, we will consider the DFSZ model and set $\tan \beta \simeq 1$ and pick the central values of the matrix elements. That is, we take
$C^{p}_{P}=-0.36$, $C^{n}_{P}=0.08$, and  $C^{e}_{P}= 0.17$. Using other values of $\tan \beta$ or the KSVZ couplings will not dramatically change our findings. 

\subsection{ARIADNE}

The Axion Resonant InterAction Detection Experiment (ARIADNE) aims to probe axion masses up to $10^{-3}$\eV \ by using methods based on nuclear magnetic resonance (NMR) \cite{Arvanitaki:2014dfa,ARIADNE:2017tdd,ARIADNE:2020wwm}. The experiment is sensitive to the axion-mediated monopole-dipole potential between two nuclei, 
\begin{align}
V_{SP} &= \frac{  g_P^{N}g_S^{N}}{8 \pi m_N} \left(\frac{1}{r \lambda}+\frac{1}{r^2}\right) e^{-\frac{r}{\lambda}} (\boldsigma \cdot \hat{r})
\equiv \boldsymbol{\mu}\cdot B_{\rm eff}\,,
 \label{eq:ARIADNE_beff}
\end{align}
where $\boldsymbol{\mu} = \frac{1}{2} g_N \mu_B \boldsigma$, with $\mu_B = \frac{e}{2m_p}$ and $g_N$ the nuclear magneton and $g$ factor, respectively. As implied by the second equality, the effects of this potential can be interpreted as an effective magnetic field $B_{\rm eff}$. 
 
The setup consists of a source mass made of unpolarized tungsten in the form of a rotating cylinder with teeth-like structures pointing radially outwards. These teeth pass by an NMR sample made of $^3$He gas, thereby inducing an oscillating $B_{\rm eff}$ field. As the NMR sample resides in a conventional external magnetic field as well, the $B_{\rm eff}$ field will drive spin precession in the $^3$He sample if it is chosen to oscillate at the nuclear Larmor frequency, determined by the external field. The resulting magnetization, which is proportional to $g_S^N g_P^N$, is precisely measured. 
These couplings can be written as $g_S^{N} = \frac{A-Z}{A}g_S^n+\frac{Z}{A}(g_S^p+g_S^{(e)})$ for the tungsten source mass and $g_P^{N} = 0.88 g_P^n-0.047g_P^p$ \cite{deVries:2011an} for the $^3$He sample.
Ref.\ \cite{Arvanitaki:2014dfa} considered the projected limits that would results from several setups. Table \ref{tab:ARIADNE} shows the projected limits from their initial setup (with $T_2=1$ s) as well as those from a scaled-up version of the the apparatus.

\begin{table}[h!]
\small\center
\renewcommand{\arraystretch}{1.2}
$\begin{array}{c|ccccc ccc}
\lambda \,\,{(\rm cm)}& 0.003& 0.01 &0.03 & 0.1 & 0.3 & 1 & 3 & 10
\\\hline 
|g_S^N g_P^N (\rm projected)|\ten{33}& 6\ten{4} \hspace{1cm}&500&60 & 10 & 6  & 4  & 4  &4
\\
\,\,\,|g_S^N g_P^N (\rm upgrade)|\ten{39}& 7 \ten{6} & 4\ten{5}& 2 \ten{4} & 1 \ten{3} & 200 & 40 & 6 & 1
\\
  \end{array}$
\caption{The projected limits on the strength of the axion mediated  monopole-dipole interaction from the ARIADNE experiment using the initial and upgraded setups of Ref.~\cite{Arvanitaki:2014dfa}.}
\label{tab:ARIADNE}
\end{table}
 
\subsection{QUAX} 

The QUest for AXions (QUAX-$g_Pg_S$) experiment ~\cite{Crescini:2016lwj,Crescini:2017uxs} is similar in setup to the ARIADNE experiment, as it also makes use of the fictitious magnetic field induced by the combination of CP-even and CP-odd axion couplings. In this case the source masses consist of lead, while the detector measures the magnetization of a sample of paramagnetic crystals that would be induced by the axionic potential. A key difference with the ARIADNE experiment is that the magnetization is induced by the coupling to electrons, rather than nucleons. The probed couplings are therefore given by, $ g_P = g_P^{(e)}$ and $g_S^N = \frac{A-Z}{A}g_S^n+\frac{Z}{A}(g_S^p+g_S^{(e)})$ for the case of lead. 
We show the current constraint \cite{Crescini:2017uxs} and projected limits \cite{Crescini:2016lwj} in Table \ref{tab:QUAX}.

\begin{table}[h!]
\small\center
\renewcommand{\arraystretch}{1.2}
$\begin{array}{c|ccccc ccc}
\lambda \,\,{(\rm m)}& 0.003& 0.01 &0.03 & 0.1 & 0.3 & 1 
\\\hline 
\,\,\,\,\,|g_S^N g_P^{(e)} (\rm current)|\ten{30}& - \hspace{1cm}&530 & 17 & 5  & 4  & 4
\\
|g_S^N g_P^{(e)} (\rm projected)|\ten{34}&  14\ten{4}& 120 & 15 & 8 & 6 &6
\\
  \end{array}$
\caption{The current \cite{Crescini:2017uxs} and projected limits \cite{Crescini:2016lwj} on the strength of the axion mediated  monopole-dipole interaction from the QUAX.}
\label{tab:QUAX}
\end{table}

\section{Applications}\label{sec:pheno}

\subsection{Chromo-electric dipole moments}

To illustrate the use of the EFT framework we revisit a well-studied scenario where BSM CP violation is dominated by the chromo-electric dipole moments (CEDMs) of first-generation quarks. We turn on the LEFT operators 
\begin{eqnarray}
\mathcal L_{CEDM} &=& L_5^u\,\bar u_L\,T^AG_{\mu\nu}^A \sigma^{\mu\nu} u_R +  L_5^d\,\bar u_L\,T^AG_{\mu\nu}^A \sigma^{\mu\nu}d_R +\mathrm{h.c.}\nonumber\\
&=& \mathrm{Re}\,( L_5^u)\,\bar u\,G \cdot \sigma\, u + \mathrm{Im}\,(L_5^u)\,\bar u\,G \cdot \sigma  i \gamma^5  u + \left( u\leftrightarrow d\right)\,,
\end{eqnarray}
where we introduced $G\cdot \sigma = T^A G_{\mu\nu}^A \sigma^{\mu\nu}$in the second line. The terms proportional to the imaginary part of $L_5^q$ are the CP-violating quark CEDMs. We read from Table \ref{tab:aNN} the induced isoscalar scalar axion-nucleon couplings
\begin{eqnarray}
g_S^{(0)} = \frac{m_*}{f_a}\frac{1}{m_u}\mathrm{Im}\,(L_5^u)\left(\frac{\partial}{\partial \mathrm{Re}\,(L_5^u)} + \frac{2 \bar B}{B} \frac{\partial}{\partial m_u} \right)\Delta m_N+(u\leftrightarrow d)\,.
\end{eqnarray}
We introduce the isoscalar and isovector combinations 
\begin{eqnarray}
L_5^0 = \frac{1}{2}(L_5^u + L_5^d)\,,\qquad
L_5^3 = \frac{1}{2}(L_5^u - L_5^d)\,,
\end{eqnarray}
to rewrite 
\begin{eqnarray}\label{g0sqCEDM}
g_S^{(0)} = \frac{m_*}{f_a}\left(\frac{\mathrm{Im}\,L_5^u}{m_u}+\frac{\mathrm{Im}\,L_5^d}{m_d}\right) \left[\frac{1}{2}\frac{\partial}{\partial \mathrm{Re}\, L_5^0} + \frac{\bar B}{B} \frac{\partial}{\partial \bar m}\right]\Delta m_N
\end{eqnarray}
in agreement with Ref.~\cite{Pospelov:1997uv}. 

Similarly, we can compute the CP-violating pion-nucleon couplings. We focus on the isovector coupling (the discussion for the isoscalar coupling goes along similar lines) and obtain 
\begin{equation}\label{g1qCEDM}
\bar g_1 = -2\frac{\mathrm{Im}\,L_5^3}{F_0} \left[\frac{1}{2}\frac{\partial}{\partial \mathrm{Re}\, L_5^0} + \frac{2\bar B}{B} \frac{\partial}{\partial \bar m}\right]\Delta m_N\,,
\end{equation}
in agreement with Refs.~\cite{Pospelov:2001ys,deVries:2016jox}. While the matrix elements appearing in Eqs.~\eqref{g0sqCEDM} and \eqref{g1qCEDM} are poorly known, we see that the matrix element drops out in the ratio
\begin{equation}\label{ratiochromo}
\frac{\bar g_1}{g_S^{(0)}} = - \left(\frac{ f_a}{F_0}\right)\frac{1}{m_*}
\frac{ m_u m_d \,(\mathrm{Im}\,L_5^u - \mathrm{Im}\,L_5^d)}{ m_d \,\mathrm{Im}\,L_5^u + m_u \,\mathrm{Im}\,L_5^d}\,.
\end{equation}
In this way experiments looking for EDMs and CP-odd axion couplings can be directly compared for a given value of the axion mass $m_a \sim 1/f_a$.  

To determine the absolute scale that various experiments are sensitive to, we do need to determine the matrix element 
\beq
\tilde \Delta m_N \equiv \left[\frac{1}{2}\frac{\partial}{\partial \mathrm{Re}\, L_5^0} + \frac{\bar B}{B} \frac{\partial}{\partial \bar m}\right]\Delta m_N\,.
 \eeq
The two terms in square brackets correspond to direct and indirect contributions respectively. The latter only depend on vacuum matrix elements and are relatively well known. The direct term depends on the nucleon matrix element of the chromomagnetic operator and is poorly understood. Recently, Seng \cite{Seng:2018wwp} argued that while the direct term is not well known, it is subleading with respect to the indirect term. The argument is based on a connection between chromomagnetic nucleon matrix elements and higher-twist distributions that can be measured in deep inelastic scattering, finding only $10$-$20$\% corrections from the direct piece. This result is at odds with the QCD sum rule results Ref.~\cite{Pospelov:2001ys} where both terms are of similar size. Lattice-QCD might provide a resolution to this discrepancy \cite{deVries:2016jox}. For now we follow Ref.~\cite{Seng:2018wwp} and set $\bar B/B \simeq 0.4\,\mathrm{GeV}^2/g_s(2\,\mathrm{GeV})$ \cite{Belyaev:1982sa, Seng:2018wwp}, $g_s(2\,\mathrm{GeV})\simeq 1.85$, and $\bar m = 3.4$ MeV \cite{Aoki:2021kgd} to obtain 
\beq
\tilde \Delta m_N \simeq 3.7 \,{\rm GeV}^2\,,
\eeq
which is what we use in our analysis. 

\begin{figure}[t!]\center
\includegraphics[width=0.99\textwidth]{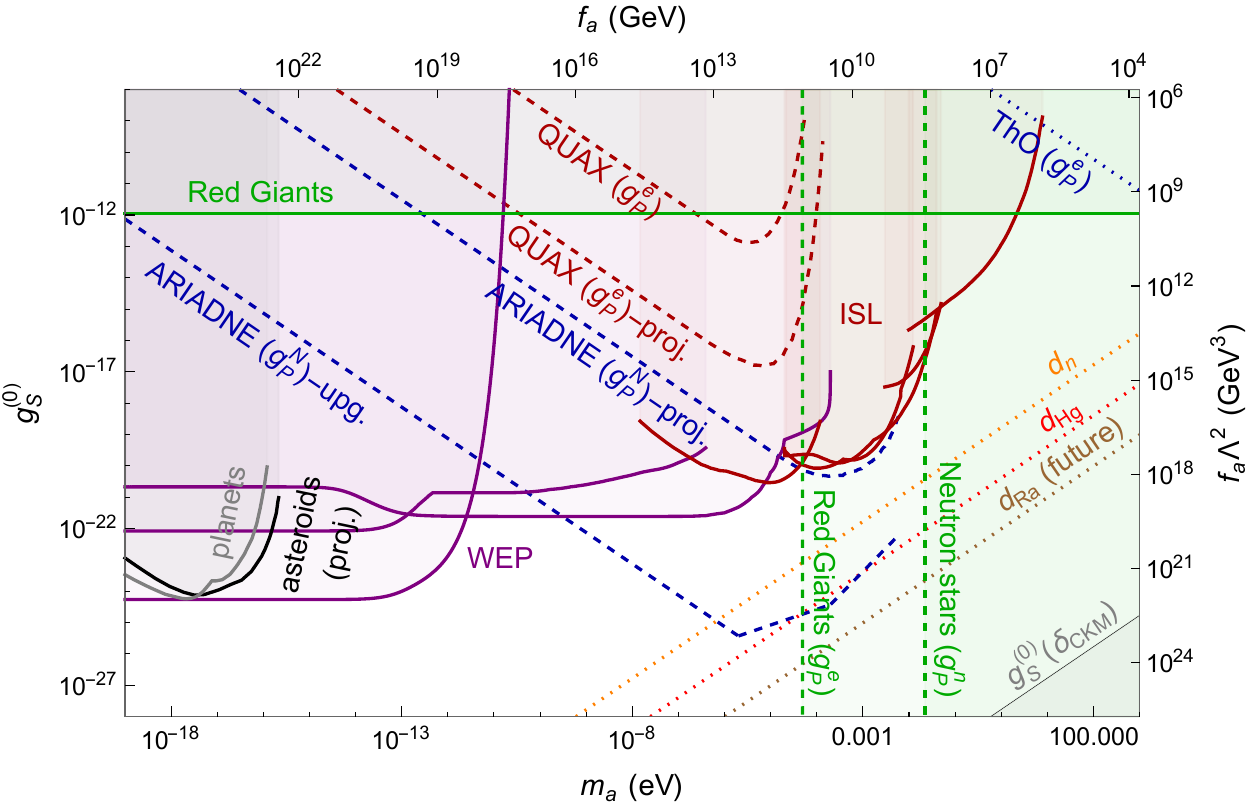}
 \caption{
 Constraints on the isoscalar scalar axion-nucleon coupling in case of a down-quark chromo-EDM. The CP-even axial axion-nucleon coupling is chosen as in the DFSZ model with $\tan \beta=1$.  The WEP experiment limits are shown in purple, from top to bottom on the low end of axion mass the lines denote E\"{o}t-Wash (2000) \cite{Smith:1999cr}, E\"{o}t-Wash (2008) \cite{Schlamminger:2007ht}, and the MICROSCOPE mission \cite{microscope}. The inverse-square law (ISL) experiments limits are shown in red, from left to right the lines denote Irvine \cite{Hoskins:1985tn}, HUST \cite{Tan:2020vpf,Tan:2016vwu,Yang:2012zzb,Tu:2007zz}, E\"{o}t-Wash \cite{Kapner:2006si,Lee:2020zjt}, Stanford \cite{Geraci:2008hb}, and IUPUI \cite{Chen:2014oda}. The astronomical bounds from planets \cite{KumarPoddar:2020kdz} are shown in gray and the projected limits from asteroids \cite{Tsai:2021irw} are shown black. The stellar cooling bounds are shown in green for Red Giants \cite{Capozzi:2020cbu,Hardy:2016kme} and neutron stars \cite{Beznogov:2018fda}. Note that the vertical line from Red Giants is a bound on $g_P^e$ which would be much weaker had we used the KSVZ model. The EDM limits are shown as dotted lines, where orange corresponds to the neutron \cite{Abel:2020pzs}, red to Hg\cite{Graner:2016ses}, brown to Ra \cite{Bishof:2016uqx}, and blue to ThO \cite{ACME:2018yjb}. The current and projected limits from QUAX \cite{Crescini:2017uxs}  are depicted by red-dashed lines. The ARIADNE limits are shown in blue-dashed for the initially envisioned setup (labeled `proj.') and a upgraded version (labeled `upg.')  \cite{Arvanitaki:2014dfa,ARIADNE:2017tdd,ARIADNE:2020wwm}. The estimated size of $g_S^{(0)}$ arising from the SM CKM phase is shown in gray at the bottom-right corner. 
 }\label{figqCEDM}
\end{figure}

\begin{figure}[t!]\center
\includegraphics[width=0.99\textwidth]{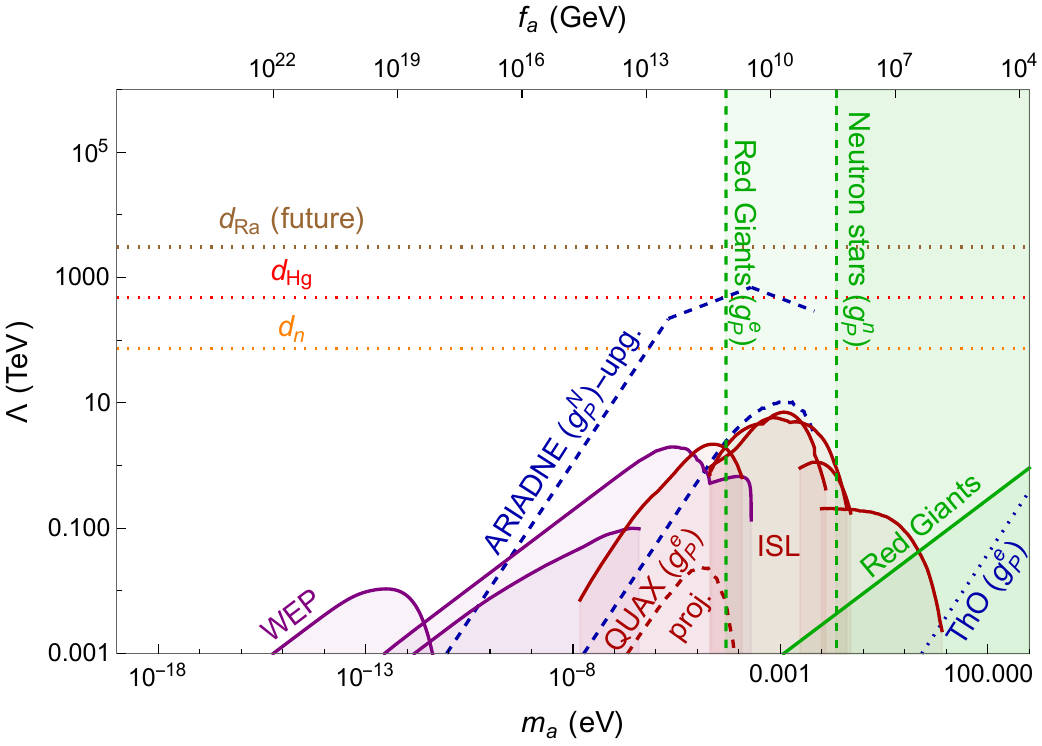}
 \caption{Constraints on $\Lambda$ from various experiments discussed in Section \ref{sec:pheno}. The labelling of the lines is explained in the caption of Fig.~\ref{figqCEDM}.
 }\label{fig:CEDM_lambda}
 \end{figure}
We can now compare the sensitivity of various experiments to the presence of the quark chromo-EDMs. 
For concreteness we turn on the down-quark chromo-EDM and assume ${\rm Im }\,L_5^d = \frac{m_d}{\Lambda^2}$. After the PQ mechanism is implemented, the presence of the CP-odd chromo-EDM leads to scalar axion-nucleon interactions which can be constrained by  fifth-force experiments. The isoscalar axion coupling is the most relevant for the present discussion and scales as $g_S^{(0)}\sim 1/(f_a \Lambda^2)$. 
Fig.~\ref{figqCEDM} shows the various constraints as a function of the axion mass (lower x-axis) or, equivalently, the axion decay constant (upper x-axis). For each observable we compute the limit on the CP-odd Wilson coefficient, which we translate into a limit on $f_a \Lambda^2$ (shown on the right y-axis) and the corresponding value of $g_S^{(0)}$ (depicted on the left y-axis). 
Showing $g_S^{(0)}$ is a somewhat arbitrary choice, as different experiments are in principle sensitive to different combinations of  $g_S^{(0,1)}$. However,
 once we assume a single Wilson coefficient,  plotting $g_S^{(1)}$ instead would simply result in a rescaling of the (left) y-axis.
The purple lines arise from searches for WEP violation, while the red lines are from tests of the inverse-square law.
The constraints are most stringent for axion masses below $10^{-13}$ eV, reaching a sensitivity of $g_S^{(0)} \lesssim 10^{-24}$ which stems from the MICROSCOPE experiment. The limits are weaker for larger axion masses and essentially disappear for $m_a > 1$ eV.

At the same time, the presence of a quark chromo-EDM can be looked for in EDM experiments. A down quark chromo-EDM induces the CP-odd pion-nucleon coupling $\bar g_1$ (among other CP-odd hadronic interactions) which leads to EDMs of nucleons and nuclei. In particular the limit on the EDM of the ${}^{199}$Hg atom sets a strong limit on $\bar g_1$ and thus $L_5^d \sim \Lambda^{-2}$. This limit can be converted into an indirect constraint on  $g_S^{(0)}$ using Eq.~\eqref{ratiochromo}. The corresponding constraints are depicted by dotted lines in Fig.~\ref{figqCEDM}. 
We observe that the indirect limits are at least several orders of magnitude stronger than the direct limits from fifth-force experiments, depending on $m_a$, in line with the conclusions of Ref.~\cite{Pospelov:1997uv}. We also depict the constraint from a prospected ${}^{225}$Ra EDM measurement at the level of $10^{-28}$ e cm \cite{Bishof:2016uqx}. This atom is particularly sensitive to $\bar g_1$ due to the octopole deformation of its nucleus and would improve upon the current $^{199}$Hg limit by one-to-two orders of magnitude at the projected sensitivity. 

Perhaps a more promising way to detect CP-violating axion interactions than the current fifth-force measurements are monopole-dipole searches. Under the reasonable assumption that axions also have a CP-even axion-nucleon interaction of typical size, the proposed ARIADNE experiment could come very close to the EDM sensitivity for axions in the $10^{-5}$ - $10^{-1}$ eV mass range. With the envisioned upgrade it could even overtake the EDM limits in the same mass window. However, the CKM-induced CP-odd axion-nucleon couplings would still be too small by many orders of magnitude to be detected.

Finally, in Fig.~\ref{fig:CEDM_lambda} we show the same information in a slightly different way by putting the BSM scale $\Lambda$ on the vertical axis. We observe that EDM limits reach scales of $10^2$ to $10^3$ TeV (note that this relies on the assumption $L_{5}^d = \frac{m_d}{\Lambda^2}$, while the quark chromo-EDMs are induced at one-loop order in many explicit BSM models), while fifth-force experiments only reach 10 TeV. The upgraded ARIADNE setup could compete with EDM experiments in reaching a scale around $10^3$ TeV. 

\subsection{A leptoquark extension} \label{sec:LQ}

Leptoquarks are hypothetical bosons that transform quarks into leptons and vice versa. They have become a popular model of BSM physics in light of various signals of lepton-flavor-universality violation, see e.g.\ \cite{Fajfer:2012jt,Buttazzo:2017ixm}. Leptoquarks generally have CP-violating interactions proportional to new, unconstrained phases. Unless these phases are chosen to be very small by hand, leptoquarks induce large radiative corrections to the QCD theta term and EDMs \cite{Dekens:2018bci}. To illustrate this we consider a simple scenario involving one scalar leptoquark. We pick the $S_1$ leptoquark that transforms as $(\bar 3,1,1/3 )$ under the SM gauge symmetries, $SU(3)_c\times SU(2)\times U(1)_Y$. These quantum numbers lead to four allowed dimension-four Yukawa-like interactions
 \bea
\vL_Y^{(S_1)}=S_1^\g\left[\bar Q^{c,I}_\g y_{LL} \epsilon_{IJ} L^J + \bar u_{R\, \g}^cy_{RR}e_R
-\epsilon^{\al\bt\g}\bar Q^I_\al z_{LL}^\dagger \epsilon_{IJ} Q_\bt^{c,J}+\epsilon^{\al\bt\g}\bar d_{R\, \al}z_{RR}^\dagger
u_{R\, \bt}^c\right]+{\rm h.c.}\,.\label{eq:S1Lag}
\eea
Here
$\al,\,\bt,\, \g$ are $SU(3)_c$ indices,  $y_{LL,RR}$ and $z_{RR}$ are generic $3\times 3$ matrices in flavor space, while $z_{LL}$ is
a symmetric $3\times 3$  matrix. 
$Q = (u_L\,, d_L)^T$ and $L = (\nu_L\,, e_L)^T$ denote the left-handed quark and lepton doublets in the weak eigenstate basis. We pick a basis in which the
up-type quark and charged-lepton mass matrices are diagonal, so that the translation from weak to mass basis is given by $d_L^{\rm weak}=Vd_L$, where $V$ is the CKM matrix. In principle, the interactions of $S_1$  lead to baryon-number-violating interactions which can be avoided if either  $y_{LL}=y_{RR}=0$ or  $z_{LL}=z_{RR}=0$. These two cases lead to rather different conclusions regarding which EDM experiments are relevant and the type of axion interactions that are induced, we therefore consider both cases separately.

In the absence of an IR solution to the strong CP problem, the above LQ interactions can generate dangerously large contributions to $\bar \theta$. Similar to the $R_2$ leptoquark \cite{deVries:2021sxz}, the $S_1$ interactions induce the following threshold correction at the scale $\mu = m_{S_1}$ (in the \textoverline{MS} scheme)
\bea
\delta \bar \theta &\simeq& \frac{1}{(4\pi)^2}\left(\ln \left(\frac{m_{S_1}^2}{\mu^2}\right)-1\right)
{\rm Im\, Tr\,}\Bigg[ Y_u^{-1}y_{LL}^*Y_e^*y_{RR}^T\nn\\
&&+4z_{LL}^\dagger\left(\left(Y_u^T\right)^{-1}z_{RR}Y_d^\dagger +Y_u^{*}z_{RR}Y_d^{-1}\right)
\Bigg]+\dots\,,
\eea
where the SM Yukawa couplings are defined through $-\vL_Y = \bar Q Y_u \tilde H u+ \bar Q Y_d H d +\bar L Y_e He+$h.c.
and the ellipses stand for terms requiring two insertions of the SM Yukawa couplings.  The correction to $\bar \theta$ is not suppressed by any high-energy scale. Thus, even when $\bar \theta=0$ at the high scale, significant tuning of the LQ couplings is needed to ensure it remains small at low energies unless an IR solution of the strong CP problem, such as a PQ mechanism, is implemented. In what follows we therefore consider the above LQ interactions, supplemented by a PQ mechanism. 

Integrating out the leptoquarks at tree-level leads to the following SM-EFT operators
\bea
\vL_{\psi^4} &=& C_{lequ}^{(1)\, abcd}(\bar L_a^I  e_{R_b}) \epsilon_{IJ}(\bar Q_c^J  u_{R_d})+ C_{lequ}^{(3)\, abcd}(\bar L_a^I  \simu e_{R_b})\epsilon_{IJ}\,(\bar Q_c^J  \sigma_{\mu\nu} u_{R_d}) \nn\\
&&+C_{quqd}^{(1)\, abcd}(\bar Q_a^I  u_{R_b}) \epsilon_{IJ}(\bar Q_c^J  d_{R_d})+C_{quqd}^{(8)\, abcd}(\bar Q_a^I T^A u_{R_b}) \epsilon_{IJ}(\bar Q_c^JT^A  d_{R_d})
+{\rm h.c.}\,,
\label{eq:EFT4fermion}
\eea
with Wilson coefficients evaluated at the leptoquark threshold
\bea\label{eq:MatchFQsemil} 
C^{(1)\, abcd }_{lequ}(m_{S_1})&=& - 4 C^{(3)\, abcd}_{lequ}(m_{S_1}) = \frac{1}{2}  \frac{\left(y^*_{LL}\right)^{ca}\left(y_{RR}\right)^{db}}{m_{S_1}\sq}\,,\nonumber\\
 C^{(1)\, abcd }_{quqd}(m_{S_1})&=& -\frac{1}{3} C^{(8)\, abcd }_{quqd}(m_{S_1})=  -\frac{2}{3} \frac{\left( z_{LL}^*\right)_{ac}\left( z_{RR}\right)_{bd}}{m_{S_1}\sq}\,.
\eea
The running of the induced SM-EFT operators as well as the subsequent matching onto the LEFT, is known one-loop order \cite{Jenkins:2013zja,Jenkins:2013wua,Alonso:2013hga,Dekens:2019ept,Jenkins:2017dyc}, however, as we mainly aim to illustrate the connection between EDMs and probes of axion couplings, we neglect these effects in what follows. Since the same dimension-six operators generate CP-odd effects with and without axions, their renormalization does not impact the comparison between the two types of experiments. The extraction of the bound on the BSM scale, $\Lambda\sim m_{S_1}$, however, can be affected by $\Or(1)$ factors.

After electroweak symmetry breaking, these SM-EFT coefficients generate the following LEFT interactions at tree level~\footnote{
Here we moved to the mass basis of the quarks and charged leptons, but left the neutrinos in the flavor eigenstates as they (and their masses) will not play a role in our analysis.
}
\bea
L_{\substack{eu\\prst}}^{S,RR}&=& -C_{\substack {lequ\\prst}}^{(1)}\,,\qquad L_{\substack{\nu edu\\prst}}^{S,RR} = C_{\substack {lequ\\prvt}}^{(1)}V_{vs}^*\,,\qquad
 L_{\substack{eu\\prst}}^{T,RR} = -C_{\substack{lequ\\prst}}^{(3)}\,,\qquad L_{\substack{\nu edu\\prst}}^{T,RR}= C_{\substack{lequ\\prvt}}^{(3)}V_{vs}^*\,,\nn\\
L_{\substack{ud\\prst}}^{S1,RR}&=&-L_{\substack{uddu\\prst}}^{S1,RR} = C_{\substack{quqd\\prvt}}^{(1)}V_{vs}^*\,,\qquad L_{\substack{ud\\prst}}^{S8,RR} =-L_{\substack{uddu\\prst}}^{S8,RR}= C_{\substack{quqd \\prvt}}^{(8)}V_{vs}^*\,.
\eea
Setting $V_{ud}\simeq 1$, these matching conditions then imply that $L_{\bf 3\times 3}^{2211} = \frac{1}{2}C_{\substack{quqd\\ 1111}}^{(1)}$ while $L_{\bf 6\times 6}^{2211}=0$ and similar for couplings with $C^{(1)}_{quqd}\to C^{(8)}_{quqd}$ and $L_i\to \bar L_i$.\\
\\
\noindent\textbf{Semi-leptonic CP violation.} We begin by setting $z_{LL}=z_{RR}=0$ such that only semi-leptonic operators are induced. For simplicity we consider couplings to first-generation fermions and set $a=b=c=d=1$ in Eq.\ \eqref{eq:MatchFQsemil} resulting in CP-odd interactions between electrons and first-generation quarks. In particular, we obtain
\begin{equation}
 C_S^{(0)} =-v_H^2\frac{\sigma_{\pi N}}{m_u+m_d}{\rm Im}\,L_{\substack{eu\\eeuu}}^{S,RR}\,,\qquad C_S^{(1)} = -\frac{v_H^2}{2}\frac{\dt m_N}{m_d-m_u}{\rm Im}\,L_{\substack{eu\\eeuu}}^{S,RR}\,,
\end{equation} 
which contribute to CP-odd effects in ThO through Eq.\ \eqref{eq:Molecules}.
Here $ C_S^{(1)} $ plays a marginal role and can, for the present discussion, be neglected. 
At the same time, the PQ mechanism leads to CP-odd axion-electron interactions. From Eq.~\eqref{eq:lepAxion} we read off 
\begin{equation}
g_S^{(e)} = \frac{m_*}{2f_a}\frac{m_\pi^2 F_0^2}{m_u+m_d}\frac{1}{m_u} {\rm Im}\, L_{\substack{eu\\eeuu}}^{S,RR}\,,
\end{equation} 
so that the ratio of these CP-odd interactions depends only on $f_a$ (and thus the axion mass) and known QCD matrix elements 
\begin{equation}
\frac{ C_S^{(0)}}{g_S^{(e)}} =-2 \left(\frac{f_a}{F_0}\right)\frac{\sigma_{\pi N} v_H^2 }{ m_\pi^2 F_0}\frac{m_u+m_d}{m_d}
\,,
\end{equation} 
in the limit $m_{u,d}/m_s\to 0$.

\begin{figure}\center 
\includegraphics[height=.58\linewidth,
                 keepaspectratio]{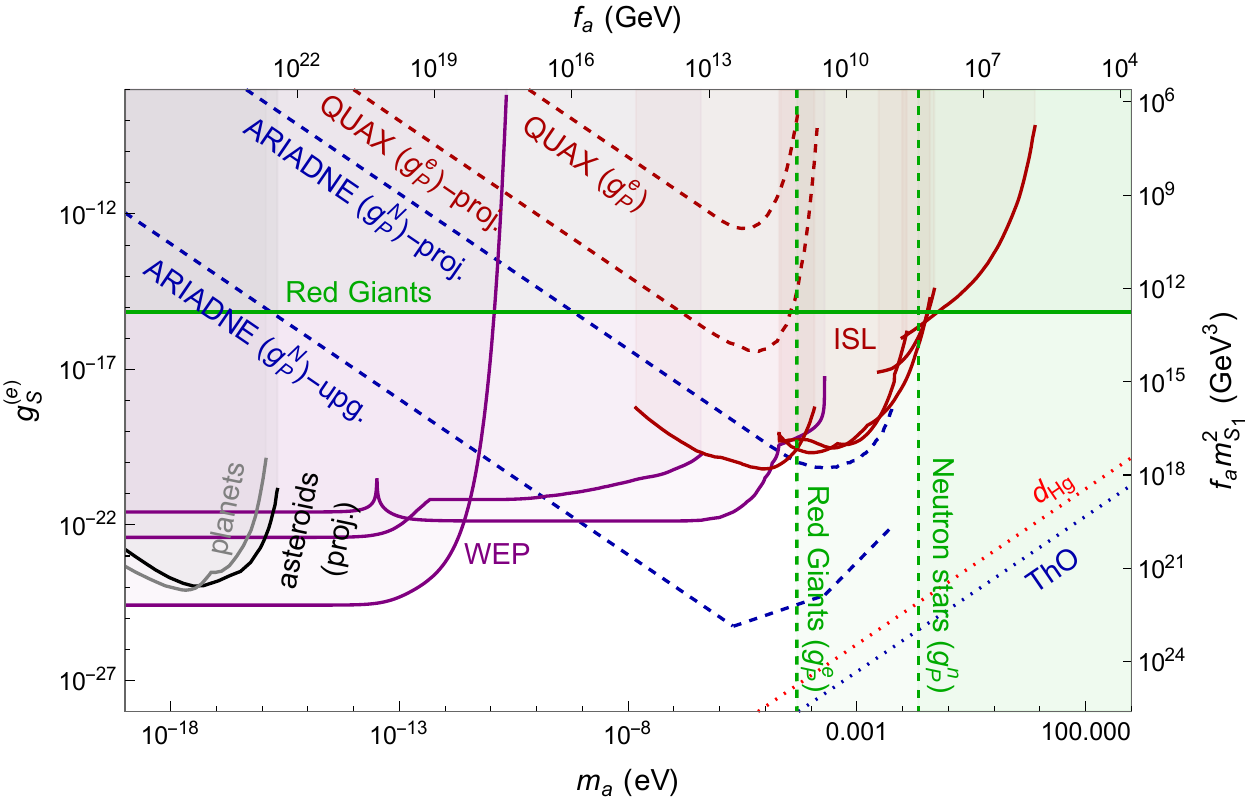}
\includegraphics[height=.58\linewidth,
                 keepaspectratio]{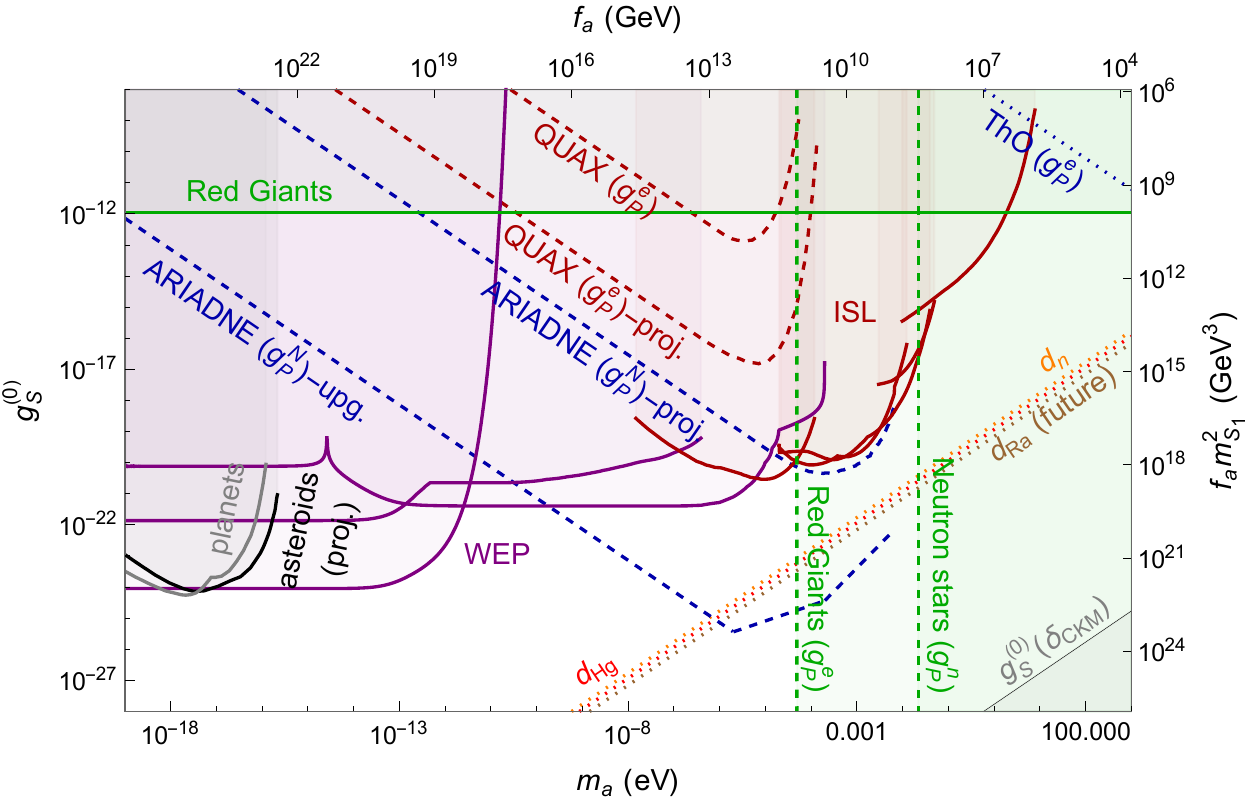}
 \caption{Constraints on the axion-electron and isoscalar axion-nucleon coupling in the leptoquark scenario discussed in Section \ref{sec:LQ}. The top panel depicts the case in which semileptonic operators are induced by $y_{LL,RR}\neq 0$ while we set $z_{LL,RR}=0$. The bottom panel shows the inverted scenario with $y_{LL,RR}= 0$ and $z_{LL,RR}\neq 0$, which leads to hadronic four-quark interactions. To obtain the mass scale, $m_{S_1}$, on right vertical axes we set the coupling constants to one, ${\rm Im}\left(y^*_{LL}\right)^{11}y_{RR}^{11} = 1$ and  ${\rm Im}\left(z^*_{LL}\right)^{11}z_{RR}^{11} = 1$, for the upper and lower panels, respectively.
 The CP-even axial axion-fermion couplings are chosen as in the DFSZ model with $\tan \beta=1$. For an explanation of the various lines we refer to the caption of Fig. \ref{figqCEDM}. }\label{fig:LQ}
\end{figure}
We compare the resulting constraints from EDM experiments and fifth-force searches in the $g_S^{(e)}$ versus $m_a$ plane in the top panel of Fig.~\ref{fig:LQ}. Fifth-force experiments constrain $g_S^{(e)} \lesssim 10^{-24}$ for axion masses below $10^{-13}$ eV. The indirect constraints from EDM experiments are many orders of magnitude more stringent over the entire axion mass range. This gap is larger than was the case for the purely hadronic chromo-EDM operator because of the extremely tight limit from the ACME ThO experiment (Hg is slightly less constraining). The envisioned sensitivity of the ARIADNE experiment is no longer competitive with the EDM experiments in this case. Note that the SM contribution to $g_S^{(e)}$ is also harder to observe than the SM value of $g_S^{(0)}$, as can be seen from the fact that our estimate in Eq.\ \eqref{eq:gSeSM} is too small to appear in Fig.~\ref{fig:LQ}.
\\
\\
\noindent\textbf{Hadronic CP violation.} We now set $y_{LL}=y_{RR}=0$ and focus on the resulting CP-odd four-quark interactions. In this case, we obtain CP-odd pion-nucleon couplings
\begin{equation}
\bar g_{0}=- \frac{F_0 \mathcal A_{\bf 3\times 3}}{4B}\frac{\partial \dt m_N}{\partial(\bar m\epsilon)}{\rm Im} \left(C_{quqd}^{(1)}\right)
\,,\qquad \bar g_1=\bar g_2=0\,,
\end{equation} 
and axion-nucleon interactions 
\bea
g_S^{(0)} &=& 
\frac{1}{2}\frac{m_*}{f_a}\frac{m_u+m_d}{m_u m_d}{\rm Im} \left(C_{quqd}^{(1)}\right)\left[\frac{1}{2}\frac{F_0^2\mathcal A_{\bf 3\times 3}}{B}\frac{\partial}{\partial\bar m}+\frac{\partial}{\partial{\rm Re}\,L_{\bf 3\times 3}^{2211}}\right]\Delta m_N
\,,\nn\\
 g_S^{(1)} &=& 
 \frac{m_*}{8f_a}\frac{m_d-m_u}{m_u m_d}{\rm Im} \left(C_{quqd}^{(1)}\right)\frac{F_0^2\mathcal A_{\bf 3\times 3}}{B}\frac{\partial \dt m_N}{\partial(\bar m\epsilon)}\,,
\eea
with similar contributions from color-octet operators, $C_{quqd}^{(1)}\to C_{quqd}^{(8)}$, $L_i\to \bar L_i$, $\mathcal A_{\bf 3\times 3}\to\bar{ \mathcal A}_{\bf 3\times 3}$.

The ratios of these CP-odd interactions depend on the QCD matrix element $\partial \Delta m_N/\partial {\rm Re}\,L_{\bf 3\times 3}^{2211}$ which is not known. For our analysis we consider the indirect pieces only for which we do control the matrix elements. Under this assumption, we obtain 
\bea
\frac{\bar g_0}{g_S^{(0)}} &=& -\frac{1}{2}\frac{f_a}{F_0}\frac{m_um_d}{m_*(m_u+m_d)}\frac{\partial \dt m_N}{\partial(\bar m\epsilon)}\left(\frac{\partial \Delta m_N}{\partial\bar m}\right)^{-1}
\simeq-\frac{1}{2}\frac{f_a}{F_0}\frac{\partial \dt m_N}{\partial(\bar m\epsilon)}\left(\frac{\partial \Delta m_N}{\partial\bar m}\right)^{-1}+\Or\left(\frac{m_{u,d}}{m_s}\right)
\,,\nn\\
\frac{\bar g_0}{g_S^{(1)}} &=&-2\frac{f_a}{F_0}\frac{m_um_d}{m_*(m_d-m_u)}
\simeq -2\frac{f_a}{F_0}\frac{m_u+m_d}{m_d-m_u}+\Or\left(\frac{m_{u,d}}{m_s}\right)
\,,
\eea
where the first ratio could be affected by $\mathcal O(1)$ factors due to unknown the direct contribution. 

In the bottom panel of Fig.~\ref{fig:LQ} we compare limits from fifth-force experiments, EDM searches, and monopole-dipole experiments in the $g_S^{(0)}-m_a$ plane. Here the mass scale on the right vertical axis is obtained by setting ${\rm Im}\left(z^*_{LL}\right)^{11}z_{RR}^{11} = 1$ and using the color-singlet contributions as an estimate of the complete effect, with  $\mathcal A_{\bf 3\times 3} = \Lambda_{\chi}^2$, see App.\ \ref{app:LECs}.
As in Fig.~\ref{figqCEDM}, EDM experiments are more stringent than fifth-force searches. In this case, the projected $^{225}$Ra measurement provides a smaller improvement on the $^{199}$Hg limit than was the case for the down-quark CEDM.  This is because the generated four-quark operator, $L_{\bf 3\times 3}^{2211}$, only induces $\bar g_0$ instead of $\bar g_1$, which comes with a smaller LEC (proportional to $\sim \dt m_N$ compared to $\sim \sigma_{\pi N}$). We again find that an upgraded version of ARIADNE could overtake the EDM limits in a small window of axion masses. 

\subsection{Left-right symmetric models}\label{sec:LR}

Left--right symmetric models are based on an extended gauge symmetry $SU(3)_c \times SU(2)_L \times SU(2)_R\times U(1)_{B-L}$ \cite{Pati:1974yy, Mohapatra:1974hk, Senjanovic1975}. Minimal versions contain an enlarged scalar sector with one scalar bidoublet and two triplets whose vevs spontaneously break the extended gauge symmetry. Variants of the model with generalized parity forbid a QCD theta term at high scales where the discrete symmetry is exact. While this naively solves the strong CP problem, dangerous contributions to $\bar \theta$ are induced once parity is spontaneously broken by the vevs of the scalars. The leading correction arises from the phases in the quark mass matrices which can be calculated explicitly \cite{Maiezza:2014ala}
\begin{equation}
\bar \theta \sim {\rm Arg\, Det\,}\left( M_u M_d\right)\sim \frac{m_t}{ m_b} \frac{\xi}{1-\xi^2}\sin \alpha \,,
\end{equation}
where $\alpha$ is a phase related to CP violation in the scalar sector and $\xi$ is related to the ratio of vacuum expectation values. This implies there is still a strong CP problem in the model: the question why $\bar \theta$ is much smaller than the naive $\Or(1)$ expectation is then essentially translated to the question of why $\al\ll 1$  in the mLRSM. 

\begin{figure}[t!]\center 
\includegraphics[width=0.99\textwidth]{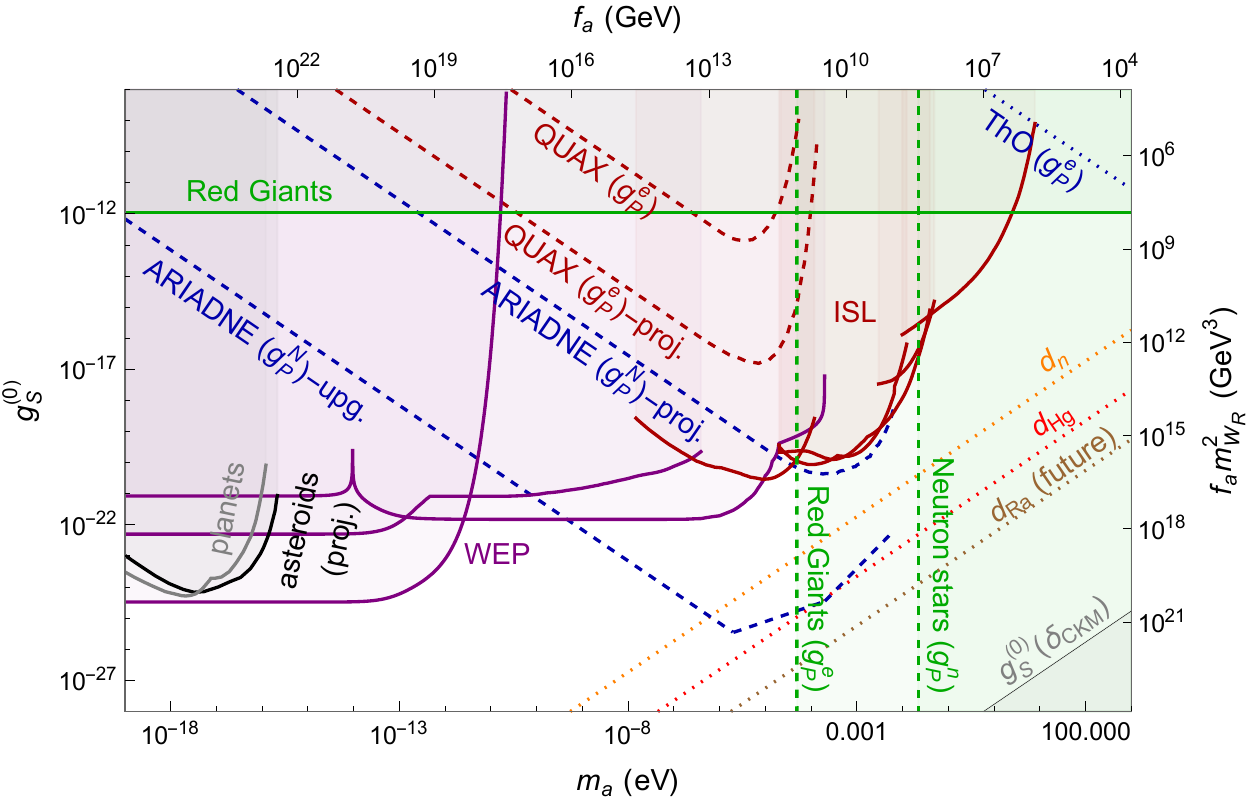}\\
 \caption{Constraints on the isoscalar axion-nucleon coupling in the left-right model discussed in Section \ref{sec:LR}. 
 The CP-even axial axion-fermion couplings are chosen as in the DFSZ model with $\tan \beta=1$.  To obtain the mass scale, $m_{W_R}$, on the right vertical axis, we assumed $\frac{\xi \sin\al}{1+\xi^2} = m_b/m_t\,\sin \alpha$ (see Refs.~\cite{Senjanovic:2015yea,Dekens:2021bro}) and pick $\sin \alpha =1$. The labelling of the lines is explained in the caption of Fig.~\ref{figqCEDM}.}\label{fig:LR}
\end{figure}

\begin{figure}[t!]\center
\includegraphics[width=0.99\textwidth]{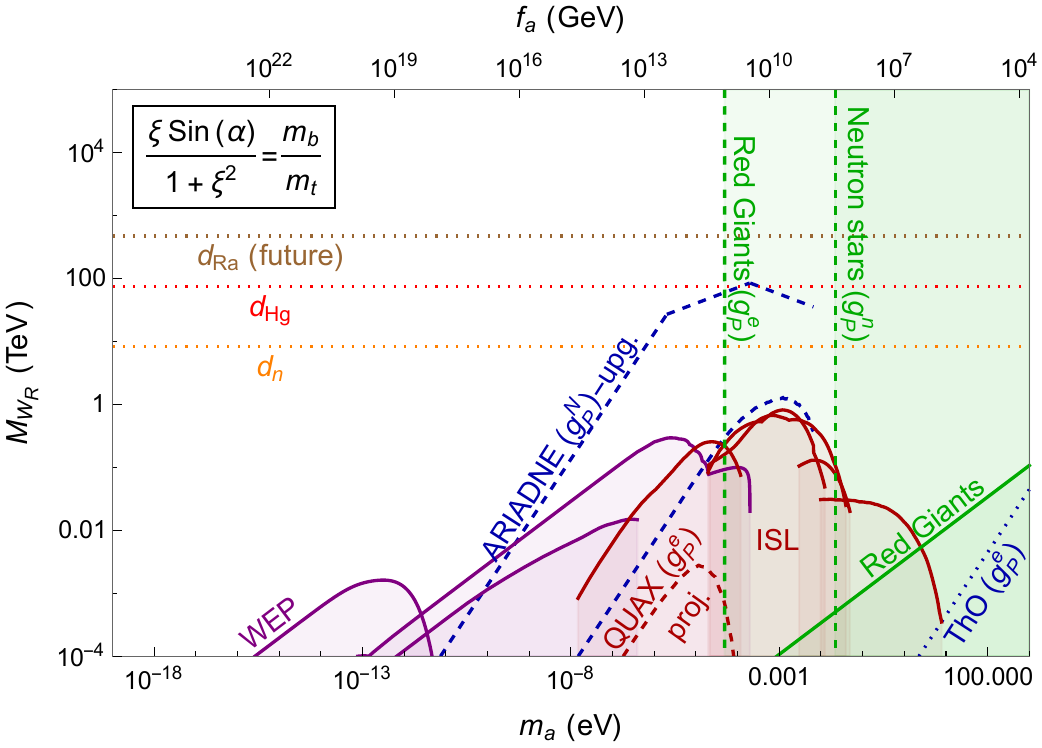}
 \caption{Limits  on the mass of right-handed gauge bosons from various experiments discussed in Section \ref{sec:pheno}. The labelling of the lines is explained in the caption of Fig.~\ref{figqCEDM}.
 }\label{fig:LR_lambda}
 \end{figure} 
If one extends the mLRSM with a PQ mechanism, the above contribution to $\bar \theta$ is relaxed to $0$, but other contributions to $\langle \theta_a\rangle= \bar \theta + \frac{\langle a\rangle}{f_a}$ arise from CP-odd higher-dimensional operators that appear in the left-right model. In particular, integrating out the right-handed W boson leads to a dimension-six SM-EFT operator \cite{Dekens:2014jka}
\bea
\vL_{6,\rm{mLRSM} } &=&   C^{ij}_{Hud} i \tilde{\vp}^{\dagger} D_{\mu} \vp \, \bar{u}^i_R \gamma^\mu  d^j_R 
+  \mathrm{h.c.}\, ,
\label{dim6edms}
\eea
where 
\begin{eqnarray}\label{eq:mHmatching}
C^{ij}_{Hud}  &=&\frac{g_R^2 }{ m_{W_R}^2} \frac{\xi e^{i\alpha}}{1+\xi^2}V_{R,\,ij}\,,
\end{eqnarray}
in terms of the $SU(2)_R$ gauge coupling, $g_R=g$, and the mass of the right-handed gauge boson $m_{W_R}$. The matrix $V_R$ is the right-handed version of the CKM matrix, and under a generalized parity symmetry we have $V_R = V_{\rm }$ in the limit  $\xi\sin\alpha\to 0$  \cite{Senjanovic:2014pva}. For more details we refer to Ref.~\cite{Dekens:2021bro}. 

Again neglecting renormalization-group effects, we match to the LEFT operators below the electroweak  scale
\begin{equation}
L_{\substack{\nu e du\\prst}}^{V,LR}= - \dt_{pr}\left(C_{Hud}^{ts}\right)^*\,,\qquad L_{\substack{uddu\\prst}}^{V1,LR}= - V_{pr}\left(C_{Hud}^{ts}\right)^*\,,\qquad L_{\substack{uddu\\prst}}^{V8,LR}=0\,.
\end{equation}
Focussing on the couplings to the first generation, we have $L_{\bf 8\times 8}^{1221} = - V_{ud}\left(C_{Hud}^{ud}\right)^*$ and we read off the CP-odd pion-nucleon and axion-nucleon interactions  
\bea
g_S^{(0)} &=& \frac{m_*}{f_a}\frac{m_d-m_u}{m_u m_d}{\rm Im\,} L_{\bf 8\times 8}^{1221}\left[
\frac{\partial}{\partial {\rm  Re\,} L_{\bf 8\times 8}^{1221}}+\frac{1}{2}\frac{F_0^2\mathcal A_{\bf 8\times 8}}{2B}\frac{\partial}{\partial \bar m}
\right]\Delta m_N\,,\nn\\
g_S^{(1)} &=& \frac{m_*}{f_a}\frac{m_d+m_u}{m_u m_d}{\rm Im\,} L_{\bf 8\times 8}^{1221}\frac{F_0^2\mathcal A_{\bf 8\times 8}}{8B}\frac{\partial\delta m_N}{\partial (\bar m\epsilon)}\,.
\eea
The CP-odd pion-nucleon couplings are
\bea
\bar g_0 &=&\bar g_2=0\,,\qquad
\bar g_1 =
-\frac{2}{F_0}{\rm Im\,} L_{\bf 8\times 8}^{1221} \left[\frac{\partial }{\partial {\rm  Re\,} L_{\bf 8\times 8}^{1221}} +\frac{1}{2}\frac{F_0^2\mathcal A_{\bf 8\times 8}}{2B}
\frac{\partial }{\partial \bar m}
\right]\Delta m_N
\,.\eea
While the matrix element $\partial \Delta m_N/\partial  {\rm  Re\,} L_{\bf 8\times 8}^{1221}$ is not known, it drops out in the ratio of the isovector CP-odd pion-nucleon coupling to the isoscalar axion-nucleon coupling 
\bea
\frac{\bar g_1}{g_S^{(0)}} =  - 2 \frac{f_a}{F_0}\frac{m_u m_d}{(m_d-m_u)m_*}\,,
\eea
while the less relevant ratio $\bar g_1/g_S^{(s)}$ depends on the unknown matrix element. If we consider only the indirect pieces we get a rough estimate  
\bea
g_S^{(0)}\simeq 0.11 \,{\rm GeV}^2\,\frac{F_0}{f_a}{\rm Im\,} L_{\bf 8\times 8}^{1221}\,,\qquad
\bar g_1\simeq -0.62 \,{\rm GeV}^2\,{\rm Im\,} L_{\bf 8\times 8}^{1221}\,,
\eea
which we use to generate the lines in Fig.~\ref{fig:LR}. In Fig.~\ref{fig:LR_lambda} we show the same information but now interpreted in terms of a limit on the mass of the right-handed gauge bosons $M_{W_R}$. EDM experiments set stringent limits on the CP-odd axion-nucleon coupling and on the mass of right-handed gauge bosons of around $M_{W_R}>100$ TeV for reasonable choices of $\xi$ and $\sin \alpha$. The neutron EDM limit was recently discussed in a similar LR scenario \cite{Bertolini:2020hjc}, with which we find general agreement. In addition, we observe that the Hg EDM sets an even more stringent constraint, that can only be overtaken by the upgraded set-up of ARIADNE in a small window of axion masses around $10^{-4}$ eV. 
Note that the projected ${}^{225}$Ra EDM measurement would provide a significant improvement on the current $^{199}$Hg limits, as the generated four-quark operator, $L_{\bf 8\times 8}^{1221}$, contributes to $\bar g_1$ in contrast to $L_{\bf 3\times 3}^{2211}$ discussed in the previous section.

\section{Conclusions and outlook}\label{sec:conclusions}

Axions provide a compelling solution to the strong CP problem that has lead to a tremendous amount of theoretical and experimental effort towards their first detection. 	In the presence of CP-violating sources beyond the QCD $\bar \theta$-term, the axion develops interactions with SM fields that violate CP symmetry. Within the Standard Model, the CKM phase leads to small CP-odd axion-lepton and axion-hadron interactions that seem impossible to detect with foreseeable technology. However, in the presence of BSM sources of CP violation, motivated for instance by the matter-antimatter asymmetry, CP-odd axion interactions can be much larger. If such BSM sources emerge at energies well above the electroweak scale they can be captured by local effective operators. In this work, we have performed a systematic study of the form and size of CP-violating axion interactions that are induced by CP-violating dimension-six interactions. We list here the main results of our analysis:	
\begin{itemize}	
\item We have implemented a Peccei-Quinn mechanism in the presence of a general set of CP-violating EFT operators built from elementary Standard Model fields. The CP-odd interactions involving quarks and gluons shift the minimum of the axion potential away from that of the pure QCD-axion case, leaving a remnant of CP violation behind. 
In addition, hadronic CP-odd operators can cause a misalignment of the vacuum, allowing for meson-vacuum transitions. We determined the chiral rotations that are needed in order to align the vacuum in Section \ref{sec:vacAlign}, with explicit expressions in App.~\ref{app:rotations}. 
The main consequence is that electric dipole moments and other CP-violating observables can be larger than the predictions from the CKM phase of the Standard Model, and that axions can obtain CP-odd Lorentz-scalar interactions with nucleons and leptons in addition to the usual (derivative) axial-vector couplings. 

\item We have used chiral perturbation theory to derive CP-violating axion-lepton, axion-meson, and axion-baryon interactions, for dimension-five and -six CP-violating operators involving light quarks and gluons in LEFT. The axion-lepton interactions arise from CP-violating lepton-quark interactions that, for example, appear in leptoquark models. Because of spontaneous chiral symmetry breaking and the appearance of a nonzero quark condensate, these interactions allow the axion to couple to leptons. Hadronic CP violation leads to couplings between axions and two pseudoscalar Goldstone bosons without derivatives. Axion couplings to pions lead to corrections to axion-nucleus interactions through loop diagrams and two-nucleon currents. Flavor-changing axion-meson-meson couplings can lead to rare decays such as $K\rightarrow \pi +a$. 

The most important hadronic interactions however are axion-nucleon couplings. We determined which low-energy constants, or QCD matrix elements, are required to calculate the coupling strengths. In general, these axion-nucleon couplings obtain indirect contributions from purely mesonic matrix elements after vacuum alignment. Most of these matrix elements are relatively well known, for instance from lattice-QCD calculations for neutrinoless double beta decay \cite{Nicholson:2018mwc} or the bag factors that enter $B-\bar B$ oscillations \cite{FlavourLatticeAveragingGroup:2019iem}. In addition, there are direct contributions at the same order, involving baryonic matrix elements, about which much less is known. An advantage is that the same matrix elements appear in the study of EDMs and this has led to large-scale efforts to determine them from lattice-QCD calculations, see Ref.~\cite{Shindler:2021bcx} for a recent review. 

\item The CP-odd LEFT operators not only induce CP-odd couplings of the axion, but they also generate CP-violating  nucleon-lepton and nucleon-pion couplings. We show that these CP-odd couplings depend on the same QCD matrix elements that enter the expressions for the axion-lepton and axion-nucleon interactions. As a result, there exists a clean relation between the CP-odd couplings with and without axions.
\item We have collected direct and indirect constraints on CP-odd axion-nucleon and axion-electron couplings from a broad range of experiments. The direct constraints are set by searches for fifth-forces that are proportional to the product of two CP-odd couplings, and by astrophysical processes. Indirect constraints arise from the product of a CP-even and CP-odd coupling. This leads to some model dependence as the CP-even couplings depend on the UV implementation of the Peccei-Quinn mechanism. Indirect constraints also arise from experiments probing beyond-the-Standard-Model CP violation, in particular, electric dipole moment searches. 

For a given source of CP violation and a given axion mass, there exists a direct connection between CP-odd axion-nucleon and axion-electron couplings and electric dipole moments of nucleons, atoms, and molecules. In general, we find that EDMs set the 
most stringent constraints, but the prospects for the ARIADNE experiment are sufficiently strong to overtake EDM limits in a window of axion masses ($10^{-5}$-$10^{-1}$ eV). 
This implies that if ARIADNE measures a nonzero signal, it would point to a fairly specific range of axion masses. Similarly, if future fifth-force experiments do find evidence for axions, it will point to a rather non-generic beyond-the-Standard-Model scenario. One option is that the dimension-six operators arose in a very specific combination, such that their contributions to EDMs are negligible, while still generating sizable CP-odd axion couplings. The second possibility would be that the effective-field-theory framework set-up in this work does not apply, implying the existence of additional light degrees of freedom.
\end{itemize}

The framework we have constructed can be further developed into several directions. 
\begin{itemize}	
\item First of all, we have only computed the leading-order axion-nucleus interactions. In principle, the chiral Lagrangian leads to a richer structure at higher orders where, in addition to axion-nucleon effects, there also appear axion-nucleon-nucleon interactions (see Fig.~\ref{fig:feynman_diagrams}). The power counting for nuclear currents is not fully understood and it would be interesting to further study such contributions, in analogy to similar studies for WIMP-nucleus scattering \cite{Cirigliano:2012pq,Korber:2017ery,Hoferichter:2018acd}. 

\item Another direction would involve the study of EFT interactions with heavier leptons. We have focused on axion-electron couplings, which are probed by fifth-force searches and for which strong indirect constraints exist from EDM experiments. By turning on effective interactions between quarks and muons (motivated for instance by the muon $g-2$ discrepancy), CP-odd axion-muon interactions can appear which are not as stringently constrained directly and for which no indirect EDM constraints exist. Such axion-muon interactions can potentially be constrained by supernovae cooling rates in analogy to the analysis in Ref.~\cite{Bollig:2020xdr} for CP-even axial vector couplings. 

\item Our work has mainly focused on flavor-conserving CP-violating LEFT operators involving light quarks and electrons. While we have briefly discussed more general couplings, for example the CP-odd $\Delta S=1$ couplings that lead to $K\rightarrow \pi +a$ transitions or lepton-flavor-violating operators that could lead to $\mu + N \rightarrow e +N$, a more thorough analysis of such interactions would be very interesting. We have also not discussed operators involving bottom or charm quarks or SM-EFT operators containing electroweak gauge and Higgs bosons, or top quarks, that are integrated out at the level of our LEFT analysis. It might be that CP-odd axion couplings to heavier fields could lead to interesting phenomenology at higher energies not discussed in this work. Of course, the CP-odd operators are still stringently constrained by low-energy experiments such as EDMs or probes of lepton number violation and it remains to be seen how much room there is for axionic couplings.

\item If axions form the dark matter in our universe this would lead to additional tests. For instance, with just CP-even interactions the axion dark matter background would lead to an oscillating neutron EDM \cite{Graham:2013gfa} with a frequency set by the axion mass. Similarly the CP-odd couplings to electrons and nucleons would lead to time-varying electron and nucleon masses whose signal can be searched for \cite{ Arvanitaki:2014faa, Arvanitaki:2016fyj}.

\end{itemize}

\section*{Acknowledgments} 
We thank Emanuele Mereghetti and Yu-Dai Tsai for useful discussions.
JdV acknowledges support from the Dutch Research Council (NWO) in the form of a VIDI grant. 
\newpage
\appendix

\section{LEFT operators}\label{app:LEFT}

\subsection{Notation and selection of operators}\label{app:operators}

Here we clarify some of the notation related to the LEFT  as well as the selection of the operators in Table  \ref{tab:oplist1}.

As mentioned, the sum in Eq.\ \eqref{eq:lagLEFT} extends over all the operators in Table  \ref{tab:oplist1}, as well as their hermitian conjugates, and flavor indices.
For example, the Lagrangian due to $\Or_{dd}^{\rm S1,RR}$ and its hermitian conjugate is written as $\vL\supset \sum_{p,r,s,t }L_{\substack{dd\\prst}}^{\rm S1,RR} (\bar q^p_L q_R^r)\,(\bar q^s_L q_R^t)+{\rm h.c.}$, where we explicitly sum over all possible flavor combinations. Note that this means that several different coefficients multiply the same operator since, e.g.\ $\Or_{\substack{dd\\prst}}^{\rm S1,RR}=\Or_{\substack{dd\\stpr}}^{\rm S1,RR}$. Without loss of generality this allows us to assume similar relations between the Wilson coefficients. The relevant cases for the operators of interest are
\bea
&L_{\substack{dd\\prst}}^{\rm S1,RR}=L_{\substack{dd\\stpr}}^{\rm S1,RR}\,,\qquad L_{\substack{dd\\prst}}^{\rm S8,RR}=L_{\substack{dd\\stpr}}^{\rm S8,RR}\,,\nn\\
&L_{\substack{ud\\uu st}}^{\rm V1,LR}=\left(L_{\substack{ud\\uu ts}}^{\rm V1,LR}\right)^*\,,\qquad L_{\substack{du\\ st uu}}^{\rm V1,LR}=\left(L_{\substack{du\\ ts uu}}^{\rm V1,LR}\right)^*\,,\qquad L_{\substack{dd\\pr st}}^{\rm V1,LR}=\left(L_{\substack{dd\\rp ts}}^{\rm V1,LR}\right)^*\,,
\eea
and similar for the $L^{\rm V8,LR}$ couplings.

As we focus on the dimension-five and -six operators that can be written as external sources or give rise to leading-order chiral interactions, it will be useful to write them in terms of the $SU(3)_L\times SU(3)_R$ representations they belong to. Below we consider the different classes of operators and the chiral representations they lead to. It turns out that all the operators in Table \ref{tab:oplist1} only lead to a limited set of irreducible representations 
\begin{itemize}
\item $X^3$\\
This class only includes $SU(3)_L\times SU(3)_R$ singlets and the CP-odd term does not appear in the chiral Lagrangian at LO.
\item $(\bar L R)X$\\
This class contains the quark EM dipole moments that can be captured by the tensor sources, $t_R^{\mu\nu}$, described in App.\ \ref{app:sources}. In addition, there are the chromo-magnetic dipole moments which transform in the same way as the quark mass terms, $\bf \bar 3_L\times  3_R$, leading to analogous interactions in the chiral Lagrangian. These cannot be captured by the external sources and are discussed in App.\ \ref{app:dim56}.
\item $(\bar LL)(\bar LL)$ and $(\bar RR)(\bar RR)$
\\
The semi-leptonic operators in these classes contribute to the external currents, $l_\mu$ and $r_\mu$ given in App.\ \ref{app:sources}. The four-quark interactions transform as $(\bf{\bar3}\times 3\times \bf{\bar3}\times 3)_{L,R}$, which lead to singlet terms that appear at LO but are CP even, or representations that can be CPV but require derivatives in the chiral Lagrangian. We thus only take into account the semi-leptonic terms. 
\item $(\bar LL)(\bar RR)$\\
The semi-leptonic operators can again be captured by $l_\mu$ and $r_\mu$ discussed in App.\ \ref{app:sources}. The four-quark interactions now transform as  $(\bf{\bar3}\times 3)_L\times (\bf{\bar3}\times 3)_R$. This again includes singlet terms that are CP even, as well as non-singlet pieces, $\sim\bf{1_{L}}\times 8_{R}$, that come with derivatives in the chiral Lagrangian. The remaining representation, $\bf{8_{L}}\times 8_{R}$, can lead to CPV terms that appear at LO in the chiral Lagrangian and is discussed further in App.\ \ref{app:dim56}.
\item $(\bar LR)(\bar LR)$\\
The semi-leptonic operators in this class can be described by the scalar and tensor currents $s, p,$ and $t_R^{\mu\nu}$. The four-quark operators now lead to two representations that can violate CP and appear at LO in the chiral Lagrangian, $(\bf{\bar3}\times \bar3)_L\times (\bf{3}\times 3)_R\supset 3_L\times \bf{\bar 3}_R \oplus \bar6 _L\times \bf{ 6}_R$.
\item $(\bar LR)(\bar RL)$\\
This class only involves semi-leptonic operators which are captured by the scalar currents $s$ and $p$.
\end{itemize}

\subsection{External sources}\label{app:sources}

Here we list the explicit expressions of the external sources appearing in Eq.\ \eqref{eq:sourcesLag}. The vector sources can be written as,
\bea
l_\mu &=& -e Q A_\mu+\frac{\partial_\mu a}{2f_a}c^q_L\nn\\
&&+\bar e_L^p\ga_\mu e_L^r \bma
L_{\substack{e u\\pruu}}^{\rm V,LL} &0& 0\\
0& L_{\substack{e d\\prdd}}^{\rm V,LL}&L_{\substack{e d\\prds}}^{\rm V,LL}\\
0& L_{\substack{e d\\prsd}}^{\rm V,LL}&L_{\substack{e d\\prss}}^{\rm V,LL}\ema
+\bar \nu_L^p\ga_\mu \nu_L^r \bma
L_{\substack{\nu u\\pruu}}^{\rm V,LL} &0& 0\\
0& L_{\substack{\nu d\\prdd}}^{\rm V,LL}&L_{\substack{\nu d\\prds}}^{\rm V,LL}\\
0& L_{\substack{\nu d\\prsd}}^{\rm V,LL}&L_{\substack{\nu d\\prss}}^{\rm V,LL}
\ema\nn\\
&&+\bar e_R^p\ga_\mu e_R^r \bma
L_{\substack{ue\\uupr}}^{\rm V,LR} &0& 0\\
0& L_{\substack{de\\ddpr}}^{\rm V,LR}&L_{\substack{de\\dspr}}^{\rm V,LR}\\
0& L_{\substack{de\\sdpr}}^{\rm V,LR}&L_{\substack{de\\sspr}}^{\rm V,LR}\ema
+\Bigg[\bar \nu_L^p\ga_\mu e_L^r \bma
0 &0& 0\\
L_{\substack{\nu edu\\ pr du}}^{\rm V,LL}& 0&0\\
L_{\substack{\nu edu\\ pr su}}^{\rm V,LL}&0 & 0
\ema + {\rm h.c.}\Bigg]\,,\nn\\
r_\mu &=&   -e Q A_\mu+\frac{\partial_\mu a}{2f_a}c^q_R\nn\\
&&+\bar e_L^p\ga_\mu e_L^r \bma
L_{\substack{e u\\pruu}}^{\rm V,LR} &0& 0\\
0& L_{\substack{e d\\prdd}}^{\rm V,LR}&L_{\substack{e d\\prds}}^{\rm V,LR}\\
0& L_{\substack{e d\\prsd}}^{\rm V,LR}&L_{\substack{e d\\prss}}^{\rm V,LR}\ema
+\bar \nu_L^p\ga_\mu \nu_L^r \bma
L_{\substack{\nu u\\pruu}}^{\rm V,LR} &0& 0\\
0& L_{\substack{\nu d\\prdd}}^{\rm V,LR}&L_{\substack{\nu d\\prds}}^{\rm V,LR}\\
0& L_{\substack{\nu d\\prsd}}^{\rm V,LR}&L_{\substack{\nu d\\prss}}^{\rm V,LR}
\ema\nn\\
&&+\bar e_R^p\ga_\mu e_R^r \bma
L_{\substack{e u\\pruu}}^{\rm V,RR} &0& 0\\
0& L_{\substack{e d\\prdd}}^{\rm V,RR}&L_{\substack{e d\\prds}}^{\rm V,RR}\\
0& L_{\substack{e d\\prsd}}^{\rm V,RR}&L_{\substack{e d\\prss}}^{\rm V,RR}\ema
+\Bigg[\bar \nu_L^p\ga_\mu e_L^r \bma
0 &0& 0\\
L_{\substack{\nu edu\\ pr du}}^{\rm V,LR}& 0&0\\
L_{\substack{\nu esu\\ pr su}}^{\rm V,LR}&0 & 0
\ema + {\rm h.c.}\Bigg]\,,
\eea
while the scalar, pseudoscalar, and tensor terms are given by
\bea
-(s+ip) &=& \bar e_L^p e_R^r \bma
L_{\substack{e u\\pruu}}^{\rm S,RL} &0& 0\\
0& L_{\substack{e d\\prdd}}^{\rm S,RL}&L_{\substack{e d\\prds}}^{\rm S,RL}\\
0& L_{\substack{e d\\prsd}}^{\rm S,RL}&L_{\substack{e d\\prss}}^{\rm S,RL}\ema
+\bar e_R^p e_L^r \bma
L_{\substack{e u\\rpuu}}^{\rm S,RR} &0& 0\\
0& L_{\substack{e d\\rpdd}}^{\rm S,RR}&L_{\substack{e d\\rpsd}}^{\rm S,RR}\\
0& L_{\substack{e d\\rpds}}^{\rm S,RR}&L_{\substack{e d\\rpss}}^{\rm S,RR}
\ema^*\nn\\
&&+\bar \nu_L^pe_R^r \bma
0 &0& 0\\
L_{\substack{\nu edu\\ pr du}}^{\rm S,RL}& 0&0\\
L_{\substack{\nu edu\\ pr su}}^{\rm S,RL}&0 & 0
\ema+
\bar e_R^p \nu_L^r \bma
0 &L_{\substack{\nu edu\\ rp du}}^{\rm S,RR}& L_{\substack{\nu edu\\ rp su}}^{\rm S,RR}\\
0&0 & 0\\
0&0 & 0
\ema^*\,,\nn\\
t_R^{\mu\nu}&=& 
F^{\mu\nu}\bma
L_{\substack{u\ga\\uu}}^{} &0& 0\\
0& L_{\substack{d\ga\\dd}}^{}& L_{\substack{d\ga\\ds}}^{}\\
0& L_{\substack{d\ga\\sd}}^{}& L_{\substack{d\ga\\ss}}^{}\ema\nn\\
&&+\bar e_L^p \sigma^{\mu\nu}e_R^r \bma
L_{\substack{e u\\pruu}}^{\rm T,RR} &0& 0\\
0& L_{\substack{e d\\prdd}}^{\rm T,RR}&L_{\substack{e d\\prds}}^{\rm T,RR}\\
0& L_{\substack{e d\\prsd}}^{\rm T,RR}&L_{\substack{e d\\prss}}^{\rm T,RR}\ema
+\bar \nu_L^p\sigma^{\mu\nu} e_R^r \bma
0 &0& 0\\
L_{\substack{\nu edu\\ pr du}}^{\rm T,RR}& 0&0\\
L_{\substack{\nu edu\\ pr su}}^{\rm T,RR}&0 & 0
\ema\,.
\eea

\subsection{Hadronic dimension-five and -six terms}\label{app:dim56}

The expressions for the couplings that are induced by the purely hadronic operators, appearing in Eq.\ \eqref{eq:LagLEFT} are given by
\bea
L_5 = \bma
L_{\substack{uG\\uu}}^{} &0& 0\\
0& L_{\substack{dG\\dd}}^{}& L_{\substack{dG\\ds}}^{}\\
0& L_{\substack{dG\\sd}}^{}& L_{\substack{dG\\ss}}^{}\ema\,,
\eea
\bea
L_{\bf 8\times 8}^{ijkl} &=& L_{\rm LLRR}^{ijkl}-\frac{\dt^{ij}}{3} L_{\rm LLRR}^{nnkl}-\frac{\dt^{kl}}{3} L_{\rm LLRR}^{ijnn}+\frac{\dt^{ij}\dt^{kl}}{9} L_{\rm LLRR}^{nn mm}\,,
\eea
where
\bea\label{eq:L8x8}
L_{\rm LLRR}^{ijkl}&=&
L_{\substack{ud\\uuds}}^{\rm V1,LR}\dt^{i}_1\dt^{j}_1\dt^{k}_2\dt^{l}_3+L_{\substack{ud\\uusd}}^{\rm V1,LR}\dt^{i}_1\dt^{j}_1\dt^{k}_3\dt^{l}_2
+L_{\substack{du\\dsuu}}^{\rm V1,LR}\dt^{k}_1\dt^{l}_1\dt^{i}_2\dt^{j}_3+L_{\substack{du\\sduu}}^{\rm V1,LR}\dt^{k}_1\dt^{l}_1\dt^{i}_3\dt^{j}_2\nn\\
&&+L_{\substack{dd\\sddd}}^{\rm V1,LR}\dt^{i}_3\dt^{j}_2\dt^{k}_2\dt^{l}_2+L_{\substack{dd\\dsdd}}^{\rm V1,LR}\dt^{i}_2\dt^{j}_3\dt^{k}_2\dt^{l}_2+L_{\substack{dd\\ddsd}}^{\rm V1,LR}\dt^{i}_2\dt^{j}_2\dt^{k}_3\dt^{l}_2+L_{\substack{dd\\ddds}}^{\rm V1,LR}\dt^{i}_2\dt^{j}_2\dt^{k}_2\dt^{l}_3\nn\\
&&+L_{\substack{dd\\dsss}}^{\rm V1,LR}\dt^{i}_2\dt^{j}_3\dt^{k}_3\dt^{l}_3+L_{\substack{dd\\sdss}}^{\rm V1,LR}\dt^{i}_3\dt^{j}_2\dt^{k}_3\dt^{l}_3+L_{\substack{dd\\ssds}}^{\rm V1,LR}\dt^{i}_3\dt^{j}_3\dt^{k}_2\dt^{l}_3+L_{\substack{dd\\sssd}}^{\rm V1,LR}\dt^{i}_3\dt^{j}_3\dt^{k}_3\dt^{l}_2\nn\\
&&+L_{\substack{dd\\dssd}}^{\rm V1,LR}\dt^{i}_2\dt^{j}_3\dt^{k}_3\dt^{l}_2+L_{\substack{dd\\sdds}}^{\rm V1,LR}\dt^{i}_3\dt^{j}_2\dt^{k}_2\dt^{l}_3
+L_{\substack{dd\\dsds}}^{\rm V1,LR}\dt^{i}_2\dt^{j}_3\dt^{k}_2\dt^{l}_3+L_{\substack{dd\\sdsd}}^{\rm V1,LR}\dt^{i}_3\dt^{j}_2\dt^{k}_3\dt^{l}_2\nn\\
&&+L_{\substack{uddu\\uddu}}^{\rm V1,LR}\dt^{i}_1\dt^{j}_2\dt^{k}_2\dt^{l}_1+L_{\substack{uddu\\usdu}}^{\rm V1,LR}\dt^{i}_1\dt^{j}_3\dt^{k}_2\dt^{l}_1+L_{\substack{uddu\\udsu}}^{\rm V1,LR}\dt^{i}_1\dt^{j}_2\dt^{k}_3\dt^{l}_1+L_{\substack{uddu\\ussu}}^{\rm V1,LR}\dt^{i}_1\dt^{j}_3\dt^{k}_3\dt^{l}_1\nn\\
&&+\left[L_{\substack{uddu\\uddu}}^{\rm V1,LR}\dt^{j}_1\dt^{i}_2\dt^{l}_2\dt^{k}_1+L_{\substack{uddu\\usdu}}^{\rm V1,LR}\dt^{j}_1\dt^{i}_3\dt^{l}_2\dt^{k}_1+L_{\substack{uddu\\udsu}}^{\rm V1,LR}\dt^{j}_1\dt^{i}_2\dt^{l}_3\dt^{k}_1+L_{\substack{uddu\\ussu}}^{\rm V1,LR}\dt^{j}_1\dt^{i}_3\dt^{l}_3\dt^{k}_1\right]^* \,,\nn\\
\eea
where we neglected CP even terms such as $L_{\substack{uu\\uuuu}}^{V1,LR}$.
$\bar L_{\bf 8\times 8}^{ijkl} $ can be obtained from the expressions above by replacing, $L_{\al}^{\rm V1,LR}\to L_{\al}^{\rm V8,LR}$ in Eq.\ \eqref{eq:L8x8}.

For the operators in the LRLR class, we can define
\bea
L_{\rm LRLR}^{ijkl} &=&L_{\substack{uu\\uuuu}}^{\rm S1,RR}\dt^{i}_1\dt^{j}_1\dt^{k}_1\dt^{l}_1
+L_{\substack{dd\\dddd}}^{\rm S1,RR}\dt^{i}_2\dt^{j}_2\dt^{k}_2\dt^{l}_2+L_{\substack{dd\\ssss}}^{\rm S1,RR}\dt^{i}_3\dt^{j}_3\dt^{k}_3\dt^{l}_3\nn\\
&&+L_{\substack{dd\\ssdd}}^{\rm S1,RR}\dt^{i}_3\dt^{j}_3\dt^{k}_2\dt^{l}_2+L_{\substack{dd\\sdsd}}^{\rm S1,RR}\dt^{i}_3\dt^{j}_2\dt^{k}_3\dt^{l}_2+L_{\substack{dd\\sdds}}^{\rm S1,RR}\dt^{i}_3\dt^{j}_2\dt^{k}_2\dt^{l}_3\nn\\
&&+L_{\substack{dd\\dssd}}^{\rm S1,RR}\dt^{i}_2\dt^{j}_3\dt^{k}_3\dt^{l}_2+L_{\substack{dd\\dsds}}^{\rm S1,RR}\dt^{i}_2\dt^{j}_3\dt^{k}_2\dt^{l}_3+L_{\substack{dd\\ddss}}^{\rm S1,RR}\dt^{i}_2\dt^{j}_2\dt^{k}_3\dt^{l}_3\nn\\
&&+L_{\substack{dd\\sddd}}^{\rm S1,RR}\dt^{i}_3\dt^{j}_2\dt^{k}_2\dt^{l}_2+L_{\substack{dd\\dsdd}}^{\rm S1,RR}\dt^{i}_2\dt^{j}_3\dt^{k}_2\dt^{l}_2+L_{\substack{dd\\ddsd}}^{\rm S1,RR}\dt^{i}_2\dt^{j}_2\dt^{k}_3\dt^{l}_2+L_{\substack{dd\\ddds}}^{\rm S1,RR}\dt^{i}_2\dt^{j}_2\dt^{k}_2\dt^{l}_3\nn\\
&&+L_{\substack{dd\\dsss}}^{\rm S1,RR}\dt^{i}_2\dt^{j}_3\dt^{k}_3\dt^{l}_3+L_{\substack{dd\\sdss}}^{\rm S1,RR}\dt^{i}_3\dt^{j}_2\dt^{k}_3\dt^{l}_3+L_{\substack{dd\\ssds}}^{\rm S1,RR}\dt^{i}_3\dt^{j}_3\dt^{k}_2\dt^{l}_3+L_{\substack{dd\\sssd}}^{\rm S1,RR}\dt^{i}_3\dt^{j}_3\dt^{k}_3\dt^{l}_2\nn\\
&&+L_{\substack{ud\\uudd}}^{\rm S1,RR}\dt^{i}_1\dt^{j}_1\dt^{k}_2\dt^{l}_2
+L_{\substack{ud\\uuds}}^{\rm S1,RR}\dt^{i}_1\dt^{j}_1\dt^{k}_2\dt^{l}_3+L_{\substack{ud\\uusd}}^{\rm S1,RR}\dt^{i}_1\dt^{j}_1\dt^{k}_3\dt^{l}_2+L_{\substack{ud\\uuss}}^{\rm S1,RR}\dt^{i}_1\dt^{j}_1\dt^{k}_3\dt^{l}_3\nn\\
&&+L_{\substack{uddu\\uddu}}^{\rm S1,RR}\dt^{i}_1\dt^{j}_2\dt^{k}_2\dt^{l}_1
+L_{\substack{uddu\\udsu}}^{\rm S1,RR}\dt^{i}_1\dt^{j}_2\dt^{k}_3\dt^{l}_1+L_{\substack{uddu\\usdu}}^{\rm S1,RR}\dt^{i}_1\dt^{j}_3\dt^{k}_2\dt^{l}_1+L_{\substack{uddu\\ussu}}^{\rm S1,RR}\dt^{i}_1\dt^{j}_3\dt^{k}_3\dt^{l}_1\,,
\eea
so that 
\bea
L_{\bf 6\times 6}^{ijkl} &=&\frac{1}{4}\left[L_{\rm LRLR}^{ijkl} +L_{\rm LRLR}^{ilkj} +L_{\rm LRLR}^{kjil}+L_{\rm LRLR}^{klij} \right]\,,\nn\\
L_{\bf 3\times 3}^{ijkl} &=&\frac{1}{4}\left[L_{\rm LRLR}^{ijkl} -L_{\rm LRLR}^{ilkj} -L_{\rm LRLR}^{kjil}+L_{\rm LRLR}^{klij} \right]\,.
\eea
The couplings with different color structures, $\bar L_{\bf 6\times 6}$ and $\bar L_{\bf 3\times 3}$, can be obtained from the expressions for $ L_{\bf 6\times 6}$ and $ L_{\bf 3\times 3}$ with the replacement $L_{\al}^{\rm S1,RR}\to L_{\al}^{\rm S8,RR}$.

The four-quark operators in Eq.\ \eqref{eq:LagIrreps} then reproduce those in the original Lagrangian, Eq.\ \eqref{eq:lag0}, after one takes into account the relations
\bea
L_{\substack{dd\\prst}}^{\rm S1,RR}=L_{\substack{dd\\stpr}}^{\rm S1,RR}\,, \qquad L_{\substack{dd\\prst}}^{\rm S8,RR}=L_{\substack{dd\\stpr}}^{\rm S8,RR}\,.
\eea

\subsection{Hadronic dimension-five and -six terms}\label{app:LECs}

The matrix elements that enter in the rotation angles $\al_i$ discussed in Sect.\ \ref{sec:vacAlign} and their relations to the LECs discussed in Sect.\ \ref{sec:ChiLag}, can be written as
\bea\label{eq:LECs}
&&\langle 0| \bar q_p q_r|0\rangle = -F_0^2 B\dt_{pr}\,,\qquad \langle 0| \bar q_p \sigma^{\mu\nu}G_{\mu\nu}^At^Aq_r|0\rangle = -2 F_0^2 \bar B\dt_{pr}\,,\nn\\
&&\langle0|(\bar q_{L\,p}\ga_\mu q_{L\,r})\,(\bar q_{R\,s}\ga^\mu q_{R\,t})_{\bf 8_L\times 8_R}|0\rangle = -\frac{F_0^4}{4}\mathcal A_{\bf 8\times 8}\left(\dt_{pt}\dt_{rs}-\frac{1}{3}\dt_{pr}\dt_{st}\right)\,,\nn\\
&&\langle0|(\bar q_{L\,p}\ga_\mu t^A q_{L\,r})\,(\bar q_{R\,s}\ga^\mu t^A q_{R\,t})_{\bf 8_L\times 8_R}|0\rangle = -\frac{F_0^4}{4} \mathcal{\bar A}_{\bf 8\times 8}\left(\dt_{pt}\dt_{rs}-\frac{1}{3}\dt_{pr}\dt_{st}\right)\,,\nn\\
&&\langle0|(\bar q_{L\,p} q_{R\,r})\,(\bar q_{L\,s} q_{R\,t})_{\bf 3\times 3}|0\rangle = -\frac{F_0^4}{4}\mathcal A_{\bf 3\times 3}\frac{\dt_{pr}\dt_{st} - \dt_{pt}\dt_{rs}}{2}\,,\nn\\
&&\langle0|(\bar q_{L\,p} q_{R\,r})\,(\bar q_{L\,s} q_{R\,t})_{\bf 6\times 6}|0\rangle = -\frac{F_0^4}{4}\mathcal A_{\bf 6\times 6}\frac{\dt_{pr}\dt_{st} + \dt_{pt}\dt_{rs}}{2}\,,\nn\\
&&\langle0|(\bar q_{L\,p}t^A q_{R\,r})\,(\bar q_{L\,s} t^Aq_{R\,t})_{\bf 3\times 3}|0\rangle = -\frac{F_0^4}{4}\mathcal {\bar A}_{\bf 3\times 3}\frac{\dt_{pr}\dt_{st} - \dt_{pt}\dt_{rs}}{2}\,,\nn\\
&&\langle0|(\bar q_{L\,p}t^A q_{R\,r})\,(\bar q_{L\,s} t^Aq_{R\,t})_{\bf 6\times 6}|0\rangle = -\frac{F_0^4}{4}\mathcal{\bar A}_{\bf 6\times 6}\frac{\dt_{pr}\dt_{st} + \dt_{pt}\dt_{rs}}{2}\,.
\eea
These relations hold after performing the basis transformation in Eq.\ \eqref{eq:transf} has been performed. 
Here $B$, $\bar B$, $\mathcal A_{i}$, and $\mathcal{\bar A}_{i}$ are low-energy constants which appear in the chiral Lagrangian. Comparing with \cite{Cirigliano:2018yza,Dekens:2021bro} we have,
\bea
\bar B &=& -\tilde B/g_s\,,\nn\\
\mathcal { A}_{\bf 8\times 8}  &=& \mathcal A_{1\, LR}=-g_{4}^{\pi\pi}  \,, \qquad \mathcal { \bar A}_{\bf 8\times 8}  =\frac{1}{2} \mathcal A_{2\, LR}  -\frac{1}{2N_c} \mathcal A_{1\, LR}=-\left[\frac{1}{2}g_{5}^{\pi\pi}-\frac{1}{2N_c} g_{4}^{\pi\pi}\right]  \,,\nn\\
\mathcal { A}_{\bf 6\times 6}  &=& -g_2^{\pi\pi}  \,, \qquad \mathcal { \bar A}_{\bf 6\times 6}  =-\left[\frac{1}{2} g_3^{\pi\pi}  -\frac{1}{2N_c} g_2^{\pi\pi}\right]  \,.
\eea

The LECs of these four-quark operators can be determined from matrix elements of the form $\langle (\pi\pi)_{I=0,2}| O_i |K^0\rangle$ which have been computed on the lattice \cite{RBC:2015gro,Blum:2012uk,RBC:2020kdj}. Using chiral symmetry, the same LECs can be related to matrix elements that play a role in neutrinoless double beta decay \cite{Nicholson:2018mwc} or to the bag factors appearing in $ K-\bar K$ oscillations \cite{FlavourLatticeAveragingGroup:2019iem}, up to $SU(3)$ corrections  \cite{Cirigliano:2017ymo}.
Using the results of Ref.\ \cite{Nicholson:2018mwc} , we have
\bea\label{eq:LECsKpp}
g_2^{\pi\pi} &=& 2.0(0.2){\rm GeV}^2 \,,\qquad g_3^{\pi\pi} = -0.62(0.06){\rm GeV}^2\,,\nn\\
g_4^{\pi\pi} &=& -1.9(0.2){\rm GeV}^2 \,,\qquad g_5^{\pi\pi} = -8.0(0.6){\rm GeV}^2\,,
\eea
while $\mathcal { A}_{\bf 3\times 3}$ and $\mathcal { \bar A}_{\bf 3\times 3}$ have not been computed. They are expected to be $\Or(\Lambda_\chi^2)$ by naive dimensional analysis \cite{Manohar:1983md,Gavela:2016bzc} and we use the estimate $\mathcal { A}_{\bf 3\times 3}=\mathcal { \bar A}_{\bf 3\times 3}\simeq 1\, {\rm GeV}^2$ in numerical evaluations.

\subsection{Chiral rotations}\label{app:rotations}

Here we give explicit expressions for the rotations discussed in Section \ref{sec:vacAlign}. Starting with the vev of the axion, we have $a = \langle a\rangle +a_{\rm ph}$ with
\bea
\frac{\langle a\rangle}{f_a} &=& -\frac{\bar B}{B}\langle i L_5M_0^{-1}+{\rm h.c.}\rangle+
\frac{F_0^2}{B}\sum_{i\geq j}(1-\frac{1}{2}\dt_{ij})\left(\frac{1}{m_i}+\frac{1}{m_j}\right)\left( \mathcal A_{\bf 3\times 3} \,{\rm Im}\,L_{\bf 3\times 3}^{iijj}+\mathcal A_{\bf 6\times 6} \,{\rm Im}\,L_{\bf 6\times 6}^{iijj}\right)\nn\\
&&
+
\frac{F_0^2\mathcal A_{\bf 8\times 8}}{2B}\sum_{i> j}\left(\frac{1}{m_i}-\frac{1}{m_j}\right) \mathcal A_{\bf 8\times 8} \,{\rm Im}\,L_{\bf 8\times 8}^{ijji}\,
+\bma 
L_\al \to \bar L_\al\\
\mathcal{ A}_{\al}\to \mathcal{\bar A}_\al
\ema\,.
\eea
The axion-independent parts of the rotation angles needed to align the vacuum are given by
\bea 
\al_3 &=& \frac{m_* F_0^2}{2Bm_u m_d m_s}\Bigg\{\mathcal A_{\bf 3\times 3}
{\rm Im}\left[(m_d-m_u)L_{\bf 3\times 3}^{2211}+(2m_s+m_u)L_{\bf 3\times 3}^{3311}-(2m_s+m_d)L_{\bf 3\times 3}^{3322}\right]\nn\\
&&+\mathcal A_{\bf 6\times 6}
{\rm Im}\Big[(2m_s+m_d)L_{\bf 6\times 6}^{1111}-(2m_s+m_u)L_{\bf 6\times 6}^{2222}+(m_u-m_d)L_{\bf 6\times 6}^{3333}\nn\\
&&+(m_d-m_u)L_{\bf 6\times 6}^{2211}+(2m_s+m_u)L_{\bf 6\times 6}^{3311}-(2m_s+m_d)L_{\bf 6\times 6}^{3322}\Big]\nn\\
&&+\frac{1}{2}\mathcal A_{\bf 8\times 8}
{\rm Im}\Big[(2m_s+2m_u-m_d)L_{\bf 8\times 8}^{3223}-(4m_s+m_d+m_u)L_{\bf 8\times 8}^{2112}-(2m_s-m_u+2m_d)L_{\bf 8\times 8}^{3113}\Big]\nn\\
&&+\frac{2\bar B}{F_0^2}{\rm Im}\Big[(2m_s+m_d)L_5^{11}-(2m_s+m_u)L_5^{22}+(m_u-m_d)L_5^{33}\Big]
\Bigg\}+\dots\,,\nn\\
\al_7+i \al_6 &=& \frac{F_0^2}{B( m_d +m_s)}\Bigg\{
\sum_i\Big[\mathcal A_{\bf 3\times 3}\left(L_{\bf 3\times 3}^{23ii}-\left(L_{\bf 3\times 3}^{32ii}\right)^*\right)+\mathcal A_{\bf 6\times 6}\Big(L_{\bf 6\times 6}^{23ii}-\left(L_{\bf 6\times 6}^{32ii}\right)^*
\Big)\Big]
\nn\\
&&+\frac{\mathcal A_{\bf 8\times 8}}{2}\sum_i \Big(L_{\bf 8\times 8}^{2ii3}-\left(L_{\bf 8\times 8}^{3ii2}\right)^*
\Big)
+\frac{2\bar B}{F_0^2}\Big[L_5^{23}-\left(L_5^{32}\right)^*\Big]
\Bigg\}+\dots\,,\nn\\
\al_8 &=& \frac{\sqrt{3}m_* F_0^2}{2Bm_u m_d m_s}\Bigg\{\mathcal A_{\bf 3\times 3}
{\rm Im}\left[(m_d+m_u)L_{\bf 3\times 3}^{2211}-m_uL_{\bf 3\times 3}^{3311}-m_dL_{\bf 3\times 3}^{3322}\right]\nn\\
&&+\mathcal A_{\bf 6\times 6}
{\rm Im}\Big[m_dL_{\bf 6\times 6}^{1111}+m_uL_{\bf 6\times 6}^{2222}-(m_u+m_d)L_{\bf 6\times 6}^{3333}\nn\\
&&+(m_d+m_u)L_{\bf 6\times 6}^{2211}-m_uL_{\bf 6\times 6}^{3311}-m_dL_{\bf 6\times 6}^{3322}\Big]\nn\\
&&+\frac{1}{2}\mathcal A_{\bf 8\times 8}
{\rm Im}\Big[(m_u-m_d)L_{\bf 8\times 8}^{2112}-(2m_u+m_d)L_{\bf 8\times 8}^{3223}-(2m_d+m_u)L_{\bf 8\times 8}^{3113}\Big]\nn\\
&&+\frac{2\bar B}{F_0^2}{\rm Im}\Big[m_dL_5^{11}+m_uL_5^{22}-(m_u+m_d)L_5^{33}\Big]
\Bigg\}+\dots\,,
\eea
where the dots stand for analogous terms for the color-octet operators, with $ L_\al \to \bar L_\al,\,
\mathcal{ A}_{\al}\to \mathcal{\bar A}_\al$, as well as axion-dependent contributions and higher-order terms in $1/\Lambda$. 
The remaining angles vanish, $\al_{1,2,4,5}=0$.

\subsection{Meson-meson-axion couplings}\label{app:pipia}

As discussed in Section \ref{sec:HadrAxionCouplings}, the hadronic LEFT operators can induce CP-odd interactions between the axions and mesons through the chiral Lagrangian of Eq.\ \eqref{eq:mesonLag}. These interactions can arise both from the terms in Eq.\ \eqref{eq:mesonLag} involving the LEFT Wilson coefficients, as well as from those involving $\sim \chi$. The former arise after performing the $U(1)_A$ rotation that removes the $aG\tilde G$ term from the Lagrangian, while the latter are induced once the $\al_i$ rotations, needed for vacuum alignment, have been performed as well. Although the general expressions are fairly lengthy, the axion-meson interactions can be related to the somewhat simpler contributions to the meson masses. These contributions to the meson masses can be written as,
\bea
\vL_{m}&=&F_0^2 \Bigg\{
(t\cdot \pi)_{ij}(t\cdot \pi)_{kl}\Bigg[\left(
{\cal A}_{\bf 3\times 3} \frac{L_{\bf 3\times 3}^{jilk}-L_{\bf 3\times 3}^{lijk}}{2}+{\cal A}_{\bf 6\times 6} \frac{L_{\bf 6\times 6}^{jilk}+L_{\bf 3\times 3}^{lijk}}{2}+{\rm h.c.}\right)-{\cal A}_{\bf 8\times 8} L_{\bf 8\times 8}^{lijk}\Bigg]\nn\\
&&
+(t\cdot \pi \,t\cdot \pi)_{ij}\Bigg[
\left(
{\cal A}_{\bf 3\times 3} \frac{L_{\bf 3\times 3}^{aaji}-L_{\bf 3\times 3}^{jaai}}{2}+{\cal A}_{\bf 6\times 6} \frac{L_{\bf 6\times 6}^{aaji}+L_{\bf 3\times 3}^{jaai}}{2}+{\rm h.c.}\right)\nn\\
&&+\frac{{\cal A}_{\bf 8\times 8}}{2} \left(L_{\bf 8\times 8}^{aija}+L_{\bf 8\times 8}^{jaai}\right)+\frac{1}{F_0^2}\left(- B\left(M_q+M_q^\dagger\right)+2\bar B\left(L_5+L_5^\dagger\right)\right)_{ji}
\Bigg]\Bigg\}\,,
\eea 
with analogous terms for the $\bar L_{\bf 3\times 3,6\times 6,8\times8}$ Wilson coefficients. 
Using the fact that 
\bea
\pi\cdot t = \frac{1}{\sqrt{2}}
\bma
\frac{\pi_3}{\sqrt{2}}+\frac{\pi_8}{\sqrt{6}} & \pi^+ & K^+ \\
\pi^- & -\frac{\pi_3}{\sqrt{2}}+\frac{\pi_8}{\sqrt{6}} &  K^0 \\
K^-&\overline{K}^0 & -2\frac{\pi_8}{\sqrt{6}} 
\ema\,,
\eea
the contributions from a specific a given Wilson coefficient can then be read off. 
These mass terms also allow one to obtain the axion-meson-meson interactions after making the following replacements
\bea
\vL_{a\pi^2} &=& \frac{m_*}{2}\frac{a}{f_a}\vL_m\left( M_q\to \tilde M_q\,,
L_\al \to \tilde L_\al\right)\,,\nn\\
\left(\tilde M_q\right)_{ij} &=& i\frac{m_i^2+6m_i m_j +m_j^2}{m_i m_j (m_i+m_j)}\Bigg[
\frac{\bar B}{B}L_5^{ij}+\frac{F_0^2\cal A_{\bf 3\times 3}}{2B}L_{\bf 3\times 3}^{aa ij}+\frac{F_0^2\cal A_{\bf 6\times 6}}{2B}L_{\bf 6\times 6}^{aa ij}+\frac{F_0^2\cal A_{\bf 8\times 8}}{4B}L_{\bf 8\times 8}^{iaaj}
\Bigg]\,,\nn\\
\tilde L_{5}^{ij} &=& i\left(\frac{1}{m_i}+\frac{1}{m_j}\right)L_5^{ij}\,,\nn\\
\tilde L_{\substack{\bf 3\times 3\\
\bf 6\times 6}
}^{ijkl}&=&i\left(\frac{1}{m_i}+\frac{1}{m_j}+\frac{1}{m_k}+\frac{1}{m_l}\right)L_{\substack{\bf 3\times 3\\
\bf 6\times 6}
}^{ijkl}\,,\nn\\
\tilde L_{\bf 8\times 8
}^{ijkl}&=&i\left(\frac{1}{m_i}-\frac{1}{m_j}-\frac{1}{m_k}+\frac{1}{m_l}\right)L_{\bf 8\times 8
}^{ijkl}\,.
\eea
Note that these replacements do not change the flavor structure, so that the contribution from a
specific Wilson coefficient to the $\pi^2 a$ interactions is determined purely from its contribution to the meson masses and $\tilde M_q$.

\bibliographystyle{utphysmod}
{\small
\bibliography{bibliography}
}

\end{document}